\documentclass[a4paper,fleqn,usenatbib]{mnras}
\usepackage[T1]{fontenc}
\usepackage{ae,aecompl}
\usepackage{graphicx}	
\usepackage{amsmath}	
\usepackage{amssymb}

\newcommand{\be}{\begin{equation}}
\newcommand{\ba}{\begin{eqnarray}}
\newcommand{\ee}{\end{equation}}
\newcommand{\ea}{\end{eqnarray}}

\newcommand{\gtsima}{$\; \buildrel > \over \sim \;$}
\newcommand{\ltsima}{$\; \buildrel < \over \sim \;$}
\newcommand{\gsim}{\lower.5ex\hbox{\gtsima}}
\newcommand{\lsim}{\lower.5ex\hbox{\ltsima}}
\newcommand{\Lya}{Ly$\alpha$~}
\newcommand{\msun}{M_\odot}

\begin{document}

\title[Large-Scale Observational Signatures of EoR]{The Large-Scale Observational Signatures of Low-Mass Galaxies During Reionization}
\author[K.~L.~Dixon, et al.]
{Keri L. Dixon$^{1}$\thanks{e-mail: K.Dixon@sussex.ac.uk},
Ilian T. Iliev$^{1}$,
Garrelt Mellema$^{2}$,
Kyungjin~Ahn$^3$,
Paul~R.~Shapiro$^4$
\\
$^1$ Astronomy Centre, Department of Physics \& Astronomy, Pevensey II Building, University of Sussex, Falmer, Brighton, BN1 9QH, \\United Kingdom
\\
$^2$ Stockholm Observatory, AlbaNova University Center, Stockholm University, SE-106 91 Stockholm, Sweden\\
$^3$ Department of Earth Sciences, Chosun University, Gwangju, Korea\\
$^4$ Department of Astronomy, University of Texas, Austin, TX 78712-1083
}
\date{\today}
\pubyear{2015} \volume{000} \pagerange{1}

\voffset-.6in
\maketitle\label{firstpage}

\begin{abstract}
Observations of the epoch of reionization give us clues about the nature and evolution of the sources of ionizing photons, or early stars and galaxies. We present a new suite of structure formation and radiative transfer simulations from the PRACE4LOFAR project designed to investigate whether the mechanism of radiative feedback, or the suppression of star formation in ionized regions from UV radiation, can be inferred from these observations. Our source halo mass extends down to $10^8\,\msun$, with sources in the mass range $10^8$ to $10^9\,\msun$ expected to be particularly susceptible to feedback from ionizing radiation, and we vary the aggressiveness and nature of this suppression. Not only do we have four distinct source models, we also include two box sizes (67~Mpc and 349~Mpc), each with two grid resolutions. This suite of simulations allows us to investigate the robustness of our results. All of our simulations are broadly consistent with the observed electron-scattering optical depth of the cosmic microwave background and the neutral fraction and photoionization rate of hydrogen at $z\sim6$. In particular, we investigate the redshifted 21-cm emission in anticipation of upcoming radio interferometer observations. We find that the overall shape of the 21-cm signal and various statistics are robust to the exact nature of source suppression, the box size, and the resolution. There are some promising model discriminators in the non-Gaussianity and small-scale power spectrum of the 21-cm signal.
\end{abstract}


\section{Introduction}

The epoch of reionization (EoR) is a major transition period for the Universe. During this time, the first luminous sources form, which begin to reionize the intergalactic medium (IGM) that is mostly neutral hydrogen as a result of recombination. Eventually, as dark matter haloes grow, galaxies begin to form, completing reionization. Clues from recent indirect measures have constrained the EoR to likely be an extended process, the bulk of which spans the range $6\lesssim z \lesssim 10$ \citep[e.g.][]{Robe15a,Bouw15a,Mitr15a}. These observations include high-redshift quasar spectra \citep[e.g.][]{Fan06, Mort11a, Bolt11a,McGr15a}, the cosmic microwave background (CMB) polarization \citep{Koma11a,Plan15a}, IGM temperature measurements \citep[e.g.][]{Theu02,Rask12a,Bolt12a}, and the decline of Lyman-$\alpha$ (Ly$\alpha$) emission in high-redshift galaxies\citep[e.g.][]{Star10a,Sche12a,Pent14a,Tilv14a}. Direct observations of the main sources of reionization (presumably star-forming galaxies) remain elusive, but the boundaries are being pushed to ever higher redshift \citep[e.g.][]{Bouw15b}.

The best constraints are likely to result from redshifted 21-cm emission from neutral hydrogen present in the IGM. Current experiments with the low-frequency radio interferometers, including the Giant Metrewave Radio Telescope (GMRT)\footnote{\url{http://gmrt.ncra.tifr.res.in/}} \citep{Paci11a}, the Low Frequency Array (LOFAR)\footnote{\url{http://www.lofar.org/}} \citep[e.g.][]{Hark10a}, the Murchison Widefield Array (MWA)\footnote{\url{http://www.mwatelescope.org/}} \citep{Lons09a}, and the Precision Array for Probing the Epoch of Reionization (PAPER)\footnote{\url{http://eor.berkeley.edu/}} \citep{Pars10a}, are attempting to measure the 21-cm radiation from the EoR. The next generation of telescopes, such as the Square Kilometre Array (SKA)\footnote{\url{http://www.skatelescope.org/}}, will have higher sensitivity and will measure further back in time. The goal of these experiments is to produce three-dimensional information about the morphology and evolution of reionization. 

Much theoretical work has gone into understanding how the underlying physical processes will shape the 21-cm signal \citep[see e.g.][]{Prit12a}. A number of analytic methods \citep[e.g][]{Furl04}, semi-numerical models \citep[e.g.][]{Mesi07}, and numerical simulations \citep[e.g.][]{Ilie06a, McQu07a,Ilie14a} have been developed for to model the 21-cm signal, but capturing the small-scale physics and large-scale structure simultaneously is difficult. There is significant disconnect between the large scales -- from  few up to tens, even hundreds of Mpc -- at which the reionization is patchy \citep{Ilie14a} and at which most observations are done, and the much smaller scales at which galaxy formation and radiative feedback occurs \citep[e.g.][]{Wise14a}. Therefore, detailed modelling is required to connect these very disparate scales and to gain a better understanding of the early galaxies based on the large-scale observational signatures.

In this paper, we focus on modelling the sources and evolution thereof during the EoR. By straightforward theoretical arguments, heating the IGM reduces the cooling necessary to form stars, and recent hydrodynamical simulations show that radiative feedback from ionizing sources suppresses star formation in dwarf galaxies \citep[e.g.][]{Simp13a,Ocvi15a}. No general consensus exists in the literature on what this may mean for reionization or, more generally, the escape of ionizing radiation into the IGM, especially when all the complicated processes of star and galaxy formation are considered. On scales less than 10~Mpc, some recent studies find that large galaxies may not be the dominant contributor to the ionizing photon budget as often assumed reionization models where every dark matter halo emits radiation proportional to its mass \citep{Wise14a,Paar15a}. This information paints a complicated picture that indicates the need for sophisticated treatments of ionizing sources. In our previous work, we have implemented some simplified models for suppression, mostly instantaneous full suppression of star formation for dwarf galaxies in ionized regions, and found significant differences from a model with no suppression present \citep{Frie11a,Ilie12a}. \citet{McQu07a}, using a $N$-body and radiative transfer code for $<100$~Mpc box, and \citet{Soba13a}, applying a semi-numeric approach to reionization informed by 1D collapse simulations, find radiative feedback to have a minimal effect on the progress of reionization, though we are primarily interested in observational signatures and not just the timing of reionization.

We apply detailed radiative transfer (RT) modelling to track the ionized regions and their evolution in cosmological volumes, with structures provided by large-scale $N$-body simulations (up to $369$~Mpc on a side) to make statistically meaningful predictions of observable signatures. In this work we are interested in what imprints the radiative feedback on low-mass galaxies might have left, and what can we learn about the high-redshift galaxies. The results from detailed radiative hydrodynamical simulations and theoretical considerations suggest that radiative feedback from photoionizing radiation, which heats the gas to at most few tens of thousands of degrees, affects mostly smaller galaxies and leaves larger ones unchanged. We, therefore, separate the ionizing sources into two distinct populations, high-mass ones, with masses above $10^9\,\msun$ (high-mass, atomic-cooling haloes, or `HMACHs') and those between $10^8$ and $10^9\,\msun$ (low-mass, atomic-cooling haloes, or `LMACHs'). The $10^8\,\msun$ mass limit roughly corresponds to a virial temperature of $10^4$\,K, below which the halo gas is unable to radiatively cool through hydrogen and helium atomic lines. haloes with virial temperature less than $10^4$ K, or minihaloes, collapse much earlier than HMACHs and even LMACHs. Because of their early formation epoch, many of these haloes have zero or very low metallicity, and stars inside them can form only through H$_2$ molecular cooling. Even though its cooling rate is slower than the atomic cooling, some fraction of these haloes can host very metal-poor stars, which can yield a non-negligible impact on the early stages of cosmic reionization through photo-ionization \citep{Wyit03b, Haim06a, Ahn12a} or X-ray heating from their by-products \citep{Mira11a, Fial14a, Xu14a, Jeon14a, Chen15a}.

We do not consider such small haloes in this work, because we mainly focus on relatively late stages of reionization, believed to be dominated by LMACHs and HMACHs. HMACHs are above the Jeans mass in the ionized and heated medium and, thus, are assumed to be unaffected by radiative feedback. Given our current incomplete understanding of the effects of radiative feedback on the star formation in early galaxies, we employ several physically motivated models for the suppression of LMACHs, as discussed in detail in Section~\ref{sec:supp} below. In this work, we present two new models that build on our previous efforts. These models can be characterised by how aggressively we suppress star formation in LMACHs, from complete suppression at all times, to full suppression in ionized regions, to partial suppression, where LMACHs remain active sites of star formation with a diminished efficiency. These cases, therefore, sample much of the available parameter space and provide clues on the observational signatures to be expected.

The outline of the paper is as follows. In Section~\ref{sec:supp}, we outline in detail the theoretical underpinnings of our source models. We present our suite of simulations, including $N$-body and RT, in Section~\ref{sec:sims}. Section~\ref{sec:results} contains our results, which include the reionization history and the morphology and various statistics of the 21-cm signal. We then conclude in Section~\ref{sec:summary}. The background cosmology is based on \emph{Wilkinson Microwave Anisotropy Probe} (\emph{WMAP}) 5-year data combined with constraints from baryonic acoustic oscillations and high-redshift supernovae ($\Omega_{\rm m} = 0.27, \Omega_\Lambda=0.73, h=0.701, \Omega_{\rm b}=0.044, \sigma_8 =0.8, n=0.96$). 

\section{Theory of reionization sources}
\label{sec:theory}

The exact nature of the sources of ionizing radiation during the EoR is still quite uncertain, although most likely the majority of the ionizing radiation was produced by massive stars in galaxies. In this section, we outline the physical processes we consider in our source modelling.

\subsection{Source suppression by Jeans-mass filtering}
\label{sec:supp}

During photoionization, the excess photon energy above the Lyman limit heats the gas to temperatures above $\sim\!10^4$\,K. The exact value of the temperature reached varies and depends on the local level of the ionizing flux, its spectrum, and the relevant cooling mechanism \citep[see e.g.][for detailed numerical calculations]{Shap04a}. Typical values are $T_{\rm IGM}=10,000-20,000$~K, but could be as high as $\sim\!40,000$~K for a hot black-body spectrum, such as could be found in Population~III (Pop.~III) stars present in the early Universe. However, hydrogen-line radiative cooling is highly efficient for $T_{\rm IGM}>8,000$~K, particularly at high redshift where the gas is denser on average. This cooling would typically bring the temperature down to $T_{\rm IGM}\sim10^4$~K, possibly somewhat lower due to the adiabatic cooling from the expansion of the Universe.

The increase of the IGM temperature caused by its photoheating results in a corresponding increase in the Jeans mass. In linear theory, the instantaneous Jeans mass is given by
\ba
M_{\rm J}&=&4.1\times10^9\,M_\odot\left(\frac{T_{\rm IGM}}{10^4\,K}\right)^{3/2}\left(\frac{\Omega_{\rm m} h^2}{0.1327}\right)^{-1/2}\nonumber\\
& &\times\left(\frac{\Omega_{\rm b}h^2}{0.02162}\right)^{-3/5}\left(\frac{1+z}{10}\right)^{3/2}
\ea
or roughly $M_{\rm J} \sim 10^9 \msun$ \citep[e.g.][and references therein]{Shap94a,Ilie02a,Ilie08a}. Even in linear theory, the actual filter mass differs somewhat from this instantaneous Jeans mass, since the mass scale at which baryons successfully collapse out of the IGM is determined by integrating the differential equation for perturbation growth over time for the evolving IGM \citep{Shap94a,Gned98a,Gned00a}. In full, non-linear cosmological simulations, the situation is still more complicated. A halo collapsing inside an ionized and heated region can only acquire enough gas to form stars if it is sufficiently massive. The minimum mass depends on the detailed gas dynamics of the process and on radiative heating and cooling. No sharp cutoff exists above which a collapsing halo retains all its gas and below which the gas does not collapse with the dark matter. Instead, simulations show that in haloes with mass $M_{\rm halo} \lesssim 10^9\msun$ the cooled gas fraction decreases gradually with decreasing halo mass \citep{Efst92a,Thou96a,Nava97a, Dijk04a, Shap04a, Okam08a, Finl11a, Hase13a}. The exact halo mass threshold for the onset of suppression from photoionization heating and the dependence of suppression on the halo mass below that threshold depends on the assumed physical processes. For simplicity, we assume that star formation is suppressed in haloes with masses below $10^9\,\msun$ and not suppressed in larger haloes, in rough agreement with the linear Jeans mass estimate for $10^4$\,K gas and the above dynamical studies.

\subsection{Source efficiencies and the Pop. III to Pop. II transition}
\label{sec:eff}

For the majority of our source models, we assume that the source emissivities are proportional to the host halo mass with an effective mass-to-light ratio, with different values adopted for LMACHs and HMACHs. For all haloes in the simulation volume, each halo that is not suppressed by Jeans-mass filtering is an ionizing source. For a source with halo mass, $M_{\rm halo}$, and lifetime, $t_{\rm s}$, we assign ionizing photon emissivity, $\dot{N}_\gamma$, according to
\be
\dot{N}_\gamma=g_\gamma\frac{M_{\rm halo}\Omega_{\rm b}}{m_{\rm p}(10\,\rm Myr)\Omega_0},
\ee
where $m_{\rm p}$ is the proton mass and the proportionality coefficient, $g_{\gamma}$, reflects the ionizing photon production efficiency of the stars per stellar atom, $N_{\rm i}$, the star formation efficiency, $f_*$, and the escape fraction, $f_{\rm esc}$:
\be
g_\gamma=f_*f_{\rm esc}N_{\rm i}\left(\frac{10 \;\mathrm{Myr}}{t_{\rm s}}\right).
\ee
\citep[e.g.][]{Haim03a,Ilie12a}. The factor $g_\gamma$, as defined, has the advantage that it is independent of the length of the source lifetime as long as the ionizing luminosity ($N_{\rm i}/t_{\rm s}$) is a constant and, as such, allows a direct comparison between different runs with varying source luminosities. All quantities determining the source efficiencies remain quite uncertain, especially at high redshift, see e.g. \citet{Ilie05a} for discussion. Recent theoretical studies have indicated that the first, metal-free (Pop.~III) stars might have been quite massive \citep[e.g.][]{Abel00a,Brom02a,OShe07a}, even when these stars are formed as multiples inside minihaloes \citep{Turk09a,Grei12a,Hira14a}. Massive stars are more efficient producers of ionizing photons, emitting up to $N_{\rm i}\sim10^5$ ionizing photons per stellar atom \citep{Brom01a,Scha02a,Venk03a}. Integrating over a top-heavy IMF for Pop.~III stars leads to estimates of $N_{\rm i}\sim25,000-90,000$ \citep{Scha02a}. As supernovae enrich the Universe with metals, Population~II (Pop.~II) stars form and become dominant, and the Salpeter IMF for these stars gives $N_{\rm i}=3,000-10,000$ \citep{Leit99a}. The values of $f_*$ and $f_{\rm esc}$ are even less certain, ranging from $\sim\!0.01$ to 1 for each of these quantities. Several recent studies have found that the photon escape fraction is mass-dependent and significantly higher for small galaxies that are more typical at high redshift than for large galaxies that form at later times \citep{Kita04a,Alva06a,Wise14a,Paar15a}, though \citet{Yaji14a} finds the ionizing radiation escape fraction to be $\sim\!0.2$ and be independent or redshift and galaxy property. Although the details are complicated, we include reionization scenarios with a higher $g_{\gamma}$ assigned to smaller haloes than the larger ones to capture the basic consensus.

\subsection{Mass-dependent feedback}
\label{sec:grad_theory}

Recent high-resolution, cosmological hydrodynamics simulations of galaxy formation suggest that source emissivities are mass-dependent with smaller haloes being more susceptible to radiative feedback \citep[][in prep.] {Wise09a,Ocvi15a,Sull15a}. The sharp distinction between LMACHs and HMACHs described above is, therefore, a simplified picture. In particular, the largest LMACHs behave nearly as HMACHs, while the smallest LMACHs have highly suppressed star formation in ionized regions. The transition between unsuppressible and highly suppressible is likely to be gradual and proportional to the mass of the halo.

Loosely following \citet{Wise09a} and \citet{Sull15a} (in prep.), we assume that the mass-dependence of our efficiency in ionized regions is:
\be
 g_\gamma = g_{\gamma,\rm HMACH} \times \left[\frac{M_{\rm halo}}{9\times10^8 \msun}-\frac{1}{9}\right],
 \label{eq:grad_eff}
\ee
essentially linear in logarithmic units of halo mass with $g_\gamma = g_{\gamma,\rm HMACH}$ at $10^9\,\msun$ and $g_\gamma = 0$ at $10^8\,\msun$. The precise formula for the suppression of ionizing photon production in smaller haloes is not important to our conclusions, since we are comparing \emph{methods} of suppression. Our main motivations are a simple relation and mass boundaries to match our other source models. The important characteristics here are that star formation in ionized regions is suppressed in a mass-dependent manner and that the smallest haloes are affected the most. Although such a simplified model is unable to capture all the expected halo-to-halo variation in physical quantities, we aim to capture the general behavior of ionizing radiation escaping haloes.

\section{The Simulations}
\label{sec:sims}

Our basic simulation methodology has been previously described in \citet{Ilie06a}, \citet{Mell06b}, and \citet{Ilie07a}, with the current, massively paralleled generation of the codes used here described in \citep{Ilie12a}. Hence, we will mainly focus on the new features we introduce, as well as outline the main simulation parameters.

\subsection{$N$-body simulations}

\begin{table*}
\caption{$N$-body simulation parameters. Background cosmology is based on the \emph{WMAP} 5-year results and constraints from baryonic acoustic oscillations and high-redshift supernovae and is consistent with the \citet{Plan15a} results.
}
\label{summary_N-body_table}
\begin{center}
\begin{tabular}{@{}llllll}
\hline
box size & $N_{\rm part}$   & mesh   & force softening & $m_{\rm particle}$ & $M_{\rm halo,min}$ 
\\[2mm]\hline
47$\,h^{-1}$~Mpc & $1728^3$ & $3456^3$ & $1.36\,h^{-1}$kpc & $2.153\times10^6\,\msun$ & $1.076\times10^8\,\msun$
\\[2mm]
244$\,h^{-1}$~Mpc & $4000^3$ & $8000^3$ & $3.05\,h^{-1}$kpc & $2.429\times10^7\,\msun$ & $0.971\times10^9\,\msun$
\\
\hline
\end{tabular}
\end{center}
\end{table*}

We start by performing very high-resolution $N$-body simulations of the formation of high-redshift structures. We use the \textsc{\small CubeP$^3$M} $N$-body code \citep{Harn13a}\footnote{\url{http://wiki.cita.utoronto.ca/mediawiki/index.php/CubePM} \url{https://github.com/jharno/cubep3m}}. This code uses a two-level, particle-mesh grid to calculate the long-range gravitational forces, kernel-matched to a local direct particle-particle interaction. The distance from a given particle up to which the direct forces are calculated is a code parameter. In the current simulations, we set this to eight fine grid cells, or two coarse grid cells, which we found provides the best tradeoff between precision and speed. Extending this further makes the calculations much more expensive, while providing little additional accuracy. The basic $N$-body simulation parameters are listed in Table~\ref{summary_N-body_table}. The force-smoothing length in both cases is set at 1/20th of the mean interparticle spacing. The larger computational volume, with a box size $L_{\rm box}=244\,h^{-1}=349~$Mpc, is chosen to recreate the large-scale reionization patchiness \citep{Ilie14a}. The smaller volume, $L_{\rm box}=47\,h^{-1}=67~$Mpc, provides significantly better resolution, which is useful for method validation purposes and provides faster radiative transfer simulation runtimes. The corresponding particle numbers, at $4000^3$ for the large box and $1728^3$ for the small box, are chosen to ensure reliable halo identification down to $10^9\,\msun$ (with 40~particles) and $10^8\,\msun$ (with 50~particles), respectively. As discussed above, $M_{\rm halo}\sim10^8\,\msun$ roughly corresponds to the atomically cooling limit, while $M_{\rm halo}\sim10^9 \msun$ is roughly the mass below which Jeans-filtering occurs in intergalactic gas at a temperature of $10^4$\,K, which is typical for the post-reionization IGM. The unresolved haloes are added using a sub-grid model, as discussed in detail in \citet{Ahn15a}. This model provides the mean local halo abundance based on the cell density and, here, is used to include haloes with masses $10^8\,\msun<M_{\rm halo}<10^9\,\msun$ in the larger-volume ($244\,h^{-1}$Mpc) simulation. Even though the correlation between the halo abundance and the cell density is stochastic, we do not include such an effect here. In the current simulations, we also do not include the effects of minihaloes, with masses below $M_{\rm halo}<10^8\msun$. These sources could be included using the same sub-grid model coupled with radiative transfer for the H$_2$ molecule-destroying Lyman-Werner band photons \citep{Ahn09a,Ahn12a}. However, while these sources drive the early reionization process and can contribute significantly to the integrated electron-scattering optical depth derived from the CMB, $\tau_{\rm es}$, their contribution at the later times of interest here is more limited, thus we leave this for future work.

The linear power spectrum of density fluctuations was calculated with the code \textsc{\small CAMB} \citep{Lewi00a}. Initial conditions were generated using the Zel'dovich approximation at redshifts high enough to employ linear theory and low enough to ensure against numerical artefacts, where $z_{\rm i}=150$ for the $244\,h^{-1}$Mpc volume and $z_{\rm i}=300$ for $47\,h^{-1}$Mpc \citep{Croc06a}.

\subsection{Radiative transfer simulations}

The radiative transfer simulations are performed with our code \textsc{\small C$^2$-Ray} (Conservative Causal Ray-Tracing) \citep{Mell06a}. The method is explicitly photon-conserving in both space and time for individual sources and, to a good approximation, for multiple sources. This method ensures the tracking of ionization fronts without loss of accuracy, independent of the spatial and time resolution, with corresponding gains in efficiency. The code has been tested in detail against a number of exact analytical solutions \citep{Mell06a}, as well as in direct comparison with a number of other independent radiative transfer methods on a standardised set of benchmark problems \citep{Ilie06b,Ilie09a}. The ionizing radiation is ray-traced from every source to every grid cell using the short characteristics method, whereby the neutral column density between the source and a given cell is found by interpolating the column densities of the intervening cells, in addition to the neutral column density through the cell itself. The contribution of each source to the local photoionization rate of a given cell is first calculated independently. Then, all contributions are added together, and a nonequilibrium chemistry solver is used to calculate the resultant ionization state. Typically, multiple sources contribute to the local photoionization rate of each cell. Changes in the rate from additional sources modify the neutral fraction and, therefore, the neutral column density, which in turn changes the local photoionization rates themselves. Consequently, an iteration procedure is required in order to converge -- with certain tolerance -- to the correct, self-consistent solution.

The $N$-body simulations discussed above provide us with the spatial distribution of cosmological structures and their evolution in time, including the locations and masses of galactic haloes, lists of the $N$-body particles which belong to each halo, and the intergalactic gas density field. We then use this information as the input to a full, 3D radiative transfer simulation of the reionization history, as follows. We have saved a series of slices, including halo particle lists, halo catalogues, and the density field smoothed to a grid of the intended resolution of the \textsc{\small C$^2$-Ray} simulation, from redshift 50 down to 6. These time-slices are uniformly spaced in time, every $\Delta t=11.53$~Myr, for a total of 82~slices. Simulating the transfer of ionizing radiation with the same spatial resolution as the underlying $N$-body (fine grid of $8000^3$, dynamic range $\sim\!10^5$) is still not feasible with current computational capacity. We, therefore, use an SPH-style smoothing scheme using nearest neighbours to transform the data to lower resolution, with $306^3$ or $612^3$ cells for 47\,$h^{-1}$~Mpc and $250^3$ or $500^3$ cells for 244\,$h^{-1}$~Mpc, for the radiative transfer simulations. We combine sources which fall into the same coarse cell, which slightly reduces the number of sources to be considered compared to the total number of haloes.

All simulations presented here include an approximate treatment of Lyman-limit systems (LLS), which are small, dense neutral regions that act as absorbers. During the early stages of reionization, the photon mean free path is set by the large neutral patches, making LLS unimportant; while at late times, they set a mean free path of several tens of Mpc \citep{Song02a}. In the current simulations, we roughly model this mean free path by imposing a hard limit on the distance an ionizing photon can travel, set at 40 comoving Mpc. We consider more detailed LLS models and their effects on reionization in \citet{Shuk15a}.

All identified haloes are potential sources of ionizing radiation, with different suppression criteria and ionizing photon production efficiencies imposed depending on the source model. We present a series of radiative transfer simulations with varying source models, summarized in Table~\ref{tab:summary}, as follows:
\begin{itemize}
\item{\raggedright{}\textit{HMACHs only:}}

In this scenario, we assume that only large haloes produce ionizing photons, corresponding to reionization being driven exclusively by relatively large galaxies. In terms of source suppression, this model could be considered an extreme case where all LMACHs are fully suppressed (or never formed) at all times. This situation may also arise when mechanical feedback from supernovae quickly (on scales smaller than our time-step) quenches the star formation in low-mass haloes. Though not considered the most realistic option, this model provides a good baseline to gauge the absolute contributions to observables from HMACHs alone, as well as facilitating comparison to older simulations with lower resolution \citep[e.g.][]{Ilie06a,Seme07a}. All HMACHs have a source efficiency of $g_\gamma = 1.7$. 
\\
 
\item{\raggedright{}\textit{Fully suppressed LMACHs (S):}}

This model was proposed previously in \citep{Ilie07a,Ilie12a}. HMACHs are once again assigned $g_\gamma = 1.7$, while LMACHs are assigned a higher efficiency $g_\gamma = 7.1$ in neutral regions to mimic the properties of early galaxies. Likely, these galaxies had higher photon production from massive, Pop.~III stars and/or higher escape fractions, and therefore higher photon production efficiencies overall, as detailed in Section~\ref{sec:eff}. We assume that LMACHs in ionized regions are completely suppressed, producing no ionizing photons. This scenario corresponds to the case of aggressive suppression of LMACHs from either mechanical or radiative feedback or a combination thereof.\\

\item{\raggedright{}\textit{Partially suppressed LMACHs (pS):}} 

For this model, first introduced in this paper, LMACHs are assumed to contribute to reionization at all times. In neutral regions, we assign LMACHs a higher efficiency as in the previous model, $g_\gamma = 7.1$. In ionized regions, these small galaxies are suppressed, resulting in diminished efficiency, and we set the efficiency to the same as the HMACHs, $g_\gamma = 1.7$. Here, star formation remains ongoing, but at a lower rate. Physically, this situation could arise if the fresh gas supply is cut off or diminished by the photoheating of surrounding gas, but a gas reservoir within the galaxy itself remains available for star formation. In this model, HMACHs are again given $g_\gamma = 1.7$.\\

\item{\raggedright{}\textit{Mass-dependent suppression of LMACHs (gS):}}

This model is also introduced in this paper for the first time. Instead of a sharp decrease in ionizing efficiency, as in the previous two cases, we also consider the gradual, mass-dependent suppression of sources in ionized regions. As before, HMACHs are assigned $g_\gamma = 1.7$ everywhere, and LMACHs have that same efficiency when residing in neutral regions. In ionized patches, LMACHs are suppressed in a mass-dependent manner, described by equation~(\ref{eq:grad_eff}), where larger galaxies are less susceptible to any kind of suppression.
\end{itemize}

As discussed above, there are two series of radiative transfer simulations based on the structure formation data from the 244$h^{-1}=349$~Mpc and the 47$h^{-1}=67$~Mpc volumes, with the first having sufficiently large volume to faithfully represent the reionization observables and the second affording better mass resolution. These two very different computational volumes also allow us to investigate the effects of resolution and sub-grid model and to evaluate which features of reionization and observable signatures are predicted robustly. We label all runs by a short label (listed in the first column of Table~\ref{tab:summary}) for more compact notation. Large-box runs are labelled LB1-LB4, while small-box ones are labelled SB1-SB4. The radiative transfer grid resolutions are $250^3$ and $306^3$ for the large and small volumes, with LB1, LB3, and SB2 also run at higher grid resolutions of $500^3$, $500^3$, and $612^3$ (LB1\_HR, LB3\_HR, and SB2\_HR), respectively.  Note, that LB3\_HR was not run through the end of reionization, but still provides a useful comparison.

Our full simulation notation reads $Lbox\_gI\_(J)(Supp)$ (the bracketed quantities are listed only when needed), where $'Lbox'$ is the simulation box size in Mpc, $'I'$ and $'J'$ are the values of the $g_{\gamma}$ factor for HMACHs and LMACHs, respectively. The symbol `Supp' indicates the suppression model S (fully suppressed), pS (partially suppressed), or gS (mass-dependent suppression) with no symbol meaning HMACHs only. For example, 63Mpc\_g1.7\_7.1pS indicates that large sources have an efficiency $g_\gamma=1.7$, while small sources have an efficiency $g_\gamma=7.1$ in neutral regions and are suppressed to $g_\gamma=1.7$ in ionized regions.

Most of these simulations were run on Curie at GENCI, France under the PRACE4LOFAR Tier-0 (Petascale) project, which was awarded time under the $5^{th}$ and $9^{th}$ Partnership for Advanced Computing in Europe (PRACE) calls. The rest of the simulations were run on computers in Germany (SuperMUC at LRZ, Hermit and Hornet at HLRS), Sweden (Triolith at NSC and Abisco at HPC2N), Finland (SISU), United States (Lonestar at TACC), and UK (Archer at EPCC and Apollo at the University of Sussex). The $N$-body simulations were run on 864 cores (47\,$h^{-1}$~Mpc) and 8,000 cores (244\,$h^{-1}$~Mpc) and required 89k and 456k core-hours respectively to complete. The radiative transfer simulations were run on a variable number of computing cores, up to 32,000. The lower-resolution runs required between 100k and 1M core-hours (47\,$h^{-1}$~Mpc volume) and between 256k and 3M core-hours (244\,$h^{-1}$~Mpc volume). The high-resolution runs required 4M (47\,$h^{-1}$~Mpc volume) and 2M (244\,$h^{-1}$~Mpc volume), respectively. The resources required for each radiative transfer run are dependent on the grid resolution used \emph{and} the number of active sources, with the latter varying significantly depending on the source suppression model -- from relatively low (LB1, LB2, SB1, SB2) to extremely high (LB3, LB4). In the latter cases, all grid cells contain active sources at late times.

\begin{table*}
\caption{Reionization simulation parameters and global reionization history results.}
\label{tab:summary}
\begin{center}
\begin{tabular}{@{}llllllllllllll}\hline
\hline
label & run & box size & $g_{\gamma}$ & $g_{\gamma}$ & $g_{\gamma}$ &mesh  &  $\tau_{\rm es}$ & $z_{10\%}$&$z_{50\%}$&$z_{90\%}$&$z_{\rm reion}$ \\
&      & \tiny{[$h^{-1}$Mpc]}  & \tiny{HMACH} & \tiny{LMACH} & \tiny{LMACH$_{\rm supp}$} & &  & &&&
\\[1.5mm]
\hline
LB1 & 349Mpc\_g1.7\_0 & 244 & 1.7 & 0 & 0 & $250^3$ & 0.049 & 8.515 & 7.059 & 6.483 & 6.231
\\[2mm]
LB1\_HR & 349Mpc\_g1.7\_0\_HR & 244 & 1.7 & 0 & 0 & $500^3$ & 0.049 & 8.456 & 7.059 & 6.483 & 6.201
\\[2mm]
LB2 & 349Mpc\_g1.7\_7.1S & 244 & 1.7 & 7.1 & 0 & $250^3$ & 0.055 & 10.290 & 7.263 & 6.617 & 6.323
\\[2mm] 
LB3 & 349Mpc\_g1.7\_7.1pS & 244 & 1.7 & 7.1 & 1.7 & $250^3$ & 0.068 & 11.200 & 8.636 & 7.859 & 7.525
\\[2mm]
LB3\_HR$^1$ & 349Mpc\_g1.7\_7.1pS & 244 & 1.7 & 7.1 & 1.7 & $500^3$ &  & 10.673 & 8.515 &  & 
\\[2mm]
LB4 & 349Mpc\_g1.7\_gS & 244 & 1.7 & 1.7 & eqn.~\ref{eq:grad_eff} & $250^3$ & 0.057 & 9.938 & 7.712 & 6.981 & 6.721
\\[2mm] 
\\
SB1 & 67Mpc\_g1.7\_0 & 47 & 1.7  & 0 & 0 & $306^3$ & 0.052 & 8.762 & 7.348 & 6.721 & 6.418
\\[2mm]
SB2 & 67Mpc\_g1.7\_7.1S & 47 & 1.7 & 7.1  & 0 &$306^3$ & 0.054 & 9.308 & 7.480 & 6.793 & 6.483
\\[2mm]
SB2\_HR & 67Mpc\_g1.7\_7.1S\_HR & 47 & 1.7 & 7.1  & 0 & $612^3$ & 0.053 & 9.235 & 7.435 & 6.757 & 6.418
\\[2mm]
SB3 & 67Mpc\_g1.7\_7.1pS & 47 & 1.7 & 7.1 & 1.7 & $306^3$ & 0.064 & 10.383 & 8.515 & 7.809 & 7.480
\\[2mm]
SB4 & 67Mpc\_g1.7\_gS & 47 & 1.7 & 1.7 & eqn.~\ref{eq:grad_eff} & $306^3$ & 0.058 & 9.382 & 7.760 & 7.099 & 6.793
\\[2mm]
\hline
\end{tabular}
\end{center}
\begin{flushleft}
$^1$ Simulation not run beyond $x_{\rm m} = 0.78$.
\end{flushleft}
\end{table*}

\section{Results}
\label{sec:results}

\subsection{Comparison to observations}
\label{sec:obs_comp}

\begin{figure*}
\begin{center} 
\hspace{-0.4in} 
\includegraphics[height=1.8in]{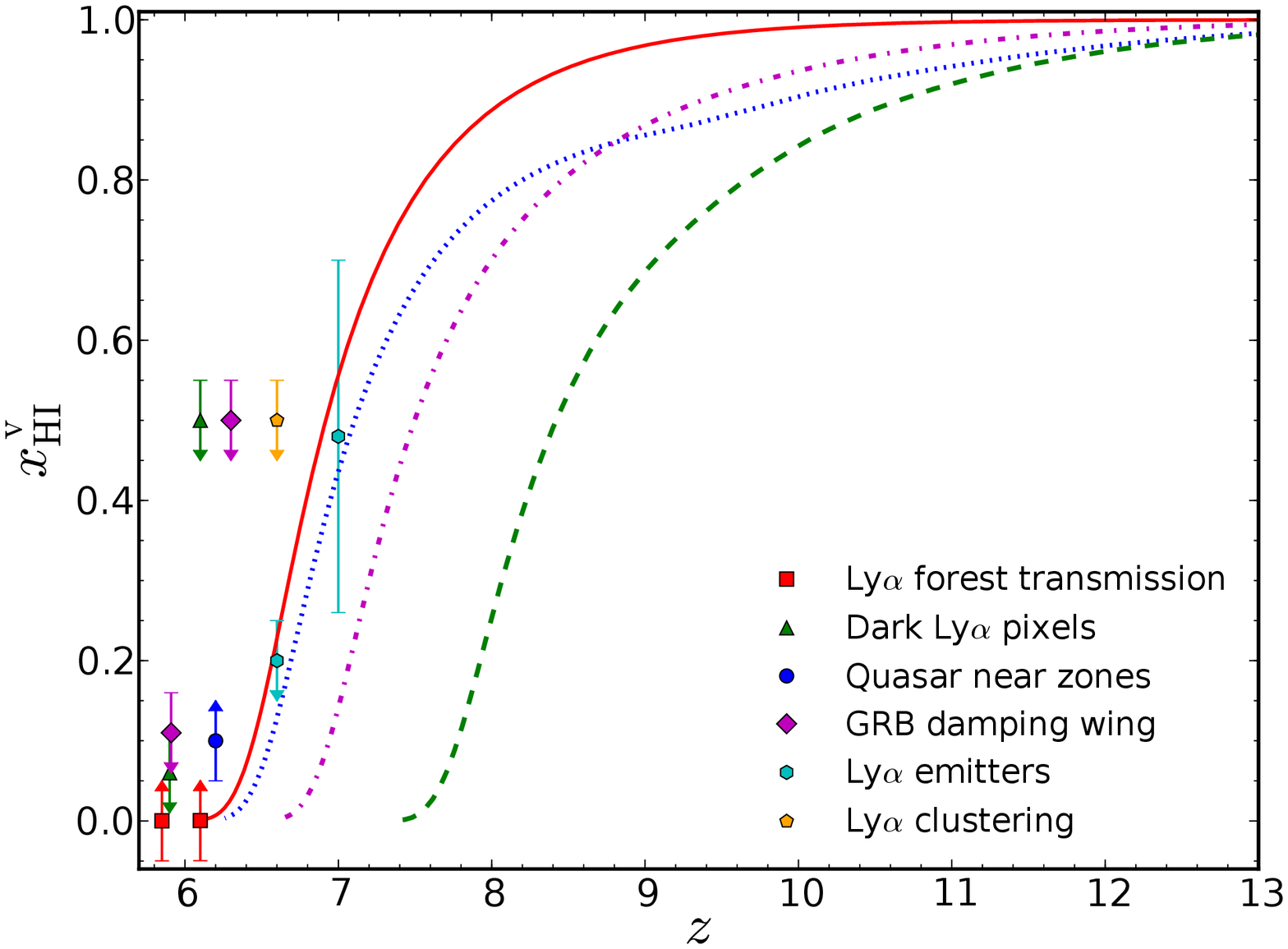}
\vspace{-0.2in} 
\hspace{-0.09in} 
\includegraphics[height=1.8in]{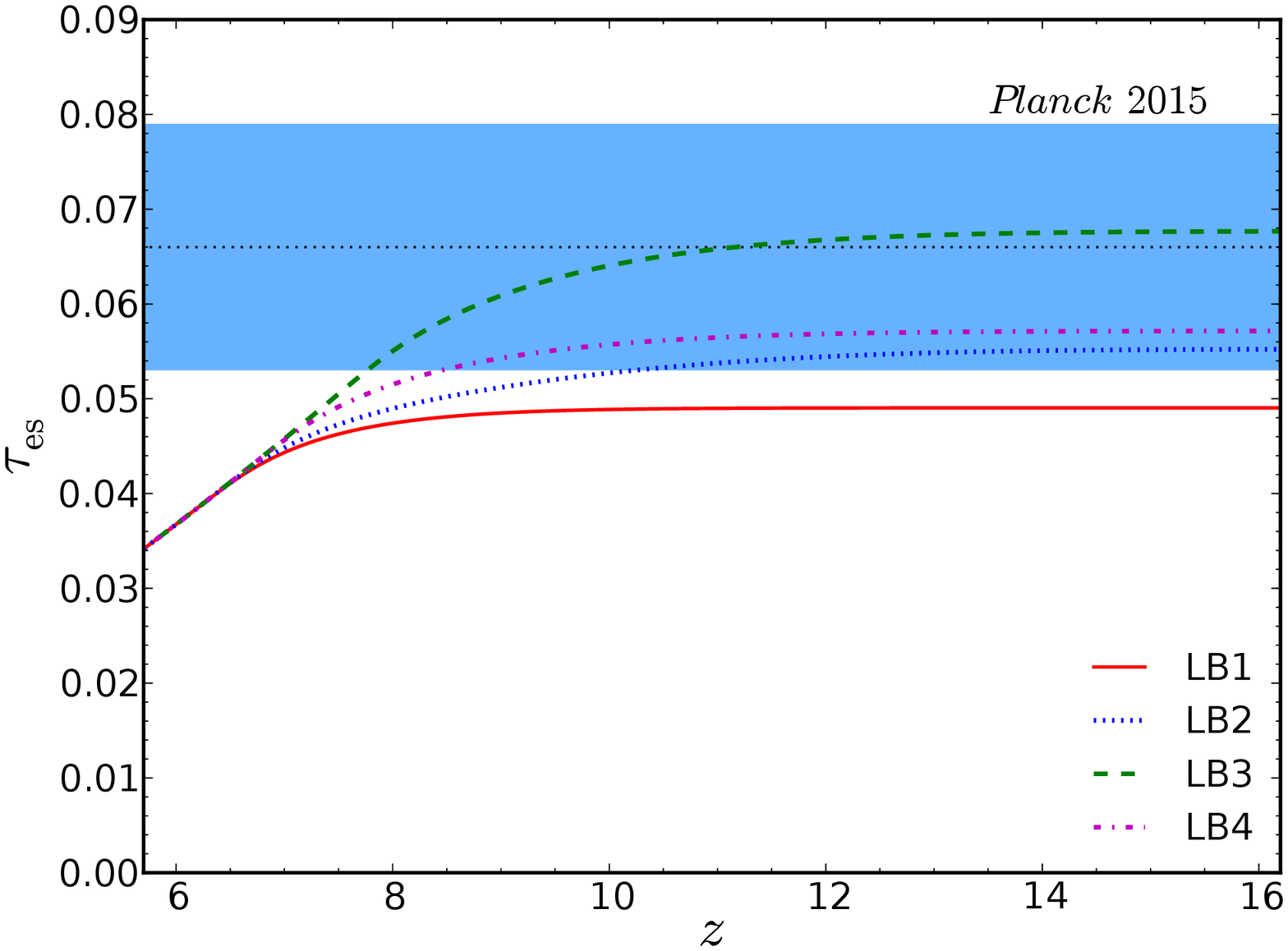}
\vspace{-0.2in}
\hspace{-0.082in} 
\includegraphics[height=1.8in]{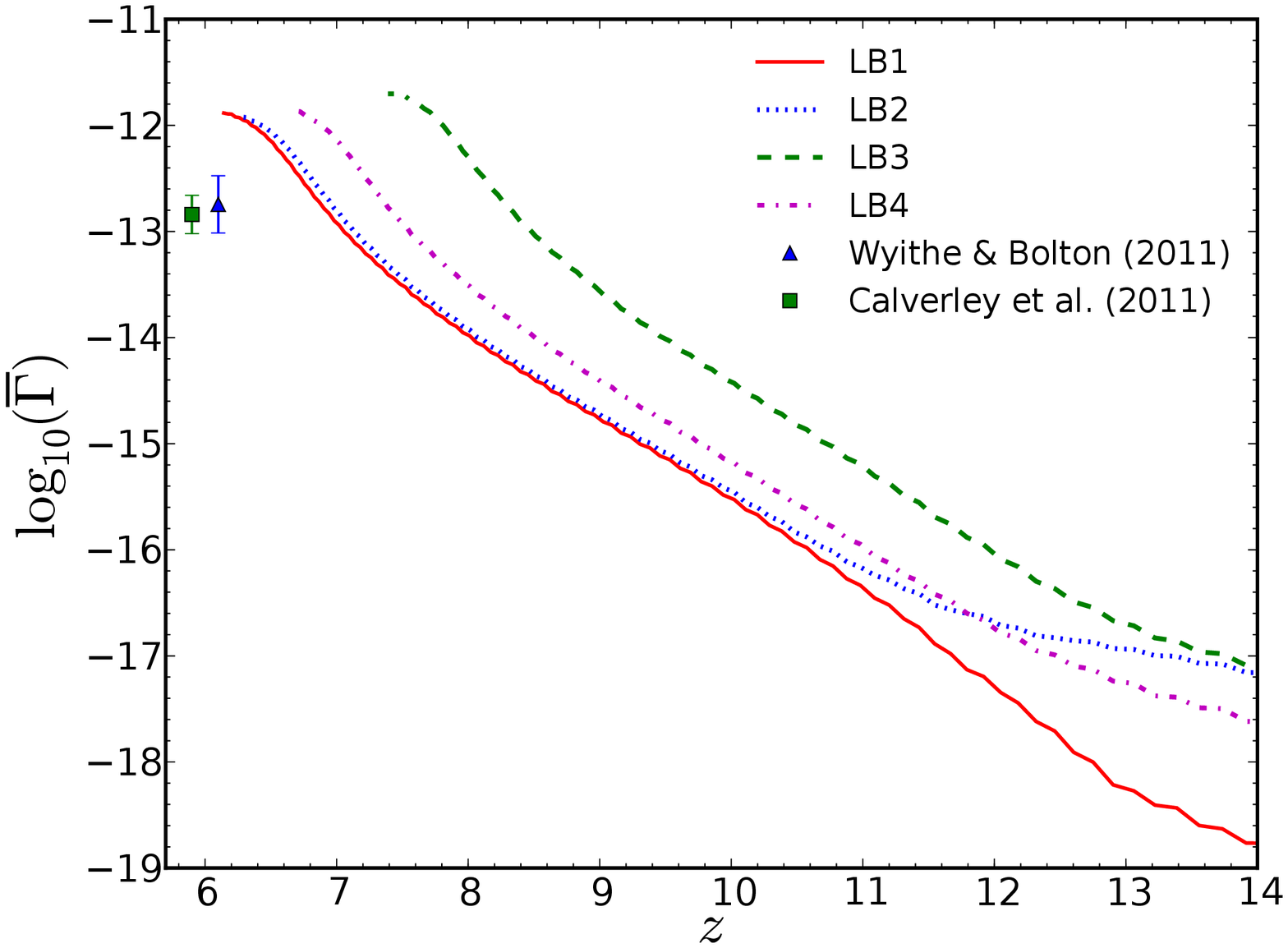}
\vspace{-0.2in} 
\hspace{-0.25in} 
\vspace{+1.1cm}
\caption{The four source models in the 244\,$h^{-1}$~Mpc box compared to observational constraints. \emph{Left:} The volume-weighted mean neutral fraction of hydrogen compared to observational inferences from \Lya forest transmission (squares) \citep{Fan06}, dark \Lya forest pixels (triangles) \citep{McGr11a,McGr15a}, quasar near zones (circles) \citep{Schr13a}, GRB damping wing absorption (diamonds)\citep{McQu08a,Chor13a}, decline in \Lya emitters (hexagons) \citep{Ota08a,Ouch10a}, and \Lya clustering (pentagons) \citep{Ouch10a}, following the discussion in \citep{Robe15a}. \emph{Middle:} The integrated electron-scattering optical depth compared to the \emph{Planck}TT+lowP+lensing+BAO 2015 results (dashed horizontal line) and the 1$\sigma$  error interval (shaded region) \citep{Plan15a}. \emph{Right:} The mean volume-weighted hydrogen photoionization rate compared to the observational constraint of \citet{Wyit11a} (hexagon).
\label{fig:obs}}
\end{center}
\end{figure*}

The current observational constraints on the timing and duration of reionization are still not tight. The main observables include the integrated electron-scattering optical depth derived from the cosmic microwave background polarization power spectra, which suggest an extended process \citep[e.g.][]{Robe15a}, and observations of the galaxies and intergalactic medium towards the end of reionization, which indicate that it ended around redshift $z\sim6$ \citep[e.g.][]{McGr15a}. Our models yield a range of results for these quantities, generally consistent with these constraints (Fig.~\ref{fig:obs} and Table~\ref{tab:summary}). The left panel of Fig~\ref{fig:obs} shows the volume-weighted mean neutral fraction of hydrogen, $x_{\rm \ion{H}{i}}^{\rm v}$, from a variety of observations. The most well-known results for $x_{\rm \ion{H}{i}}^{\rm v}$ are from measurements of the effective optical depth evolution of the \Lya forest (including higher-order transitions, if available) along many lines of sight to high-redshift quasars in \citet{Fan06}, represented by squares and shortened to \Lya forest transmission. Interpreting the transmission as a neutral fraction requires significant modelling, so the resultant neutral fraction is somewhat uncertain \citep{Mesi10}. Nearly model-independent upper limits on the neutral fraction come from the fraction of dark pixels in the \Lya forest, shown as triangles \citep{McGr11a,McGr15a}. Gamma-ray burst (GRB) damping wings, though rare, provide upper limits in this redshift range (diamonds) \citep{McQu08a,Chor13a}. The size of near zones around quasars give some indication of the minimum neutral fraction (circle), but these measurements are dependent on uncertain intrinsic quasar properties \citep{Bolt11a,Schr13a}. \Lya emitters \citep{Ota08a,Ouch10a} and clustering \citep{Ouch10a} provide further constraints, shown as hexagons and pentagons, respectively. Our late reionization models (LB1, LB2 and LB4) agree well with the observed fast rise in the neutral hydrogen fraction observed at $z\sim6-7$. The early reionization model (LB3) does not agree, which due to its numerous and weakly suppressed sources leads to an earlier end of reionization. However, if we tune down the assumed source efficiencies in LB3, bringing the neutral fraction evolution into agreement with that data set is straightforward. 

The integrated electron-scattering optical depth from the CMB last scattering surface to the present era, $\tau_{\rm es}$, measured from our simulations is listed in Table~\ref{tab:summary} and plotted in the middle panel of Fig.~\ref{fig:obs}. Current constraints from \emph{Planck}TT+lowP+lensing+BAO data give $\tau_{\rm es} = 0.066\pm0.013$ \citep{Plan15a}, shown as the shaded region in Fig.~\ref{fig:obs}. Conversely to the end-of-reionization data, the simulated $\tau_{\rm es}$ is highest for the LB3 model, making it most in agreement the central observed $\tau_{\rm es}$ value. The late reionization models correspond to lower values, albeit still in agreement within $1\sigma$ (LB2 and LB4) and within $2\sigma$ (LB1). We note that these data do not independently constrain the exact start, finish, or duration of reionization. An earlier beginning to reionization generally gives a larger $\tau_{\rm es}$, since the early Universe is denser and larger densities amplify $\tau_{\rm es}$. The very beginning of reionization is likely driven by minihaloes \citep{Ahn12a}, which form much earlier than the LMACHs and HMACHs that we consider here. Since we do not include any contribution from minihaloes, we expect our results to be $\sim\!0.02$ too low compared to those cases with very active star formation inside minihaloes \citep{Ahn12a}. More massive haloes dominate the later stages of reionization, which are the focus of this work.

Finally, the right panel of Fig.~\ref{fig:obs} shows the volume-averaged hydrogen photoionization rate, $\Gamma$. Our late reionization simulations all predict $\Gamma\sim10^{-12}\,\mathrm{s}^{-1}$ at $z=6$, while the observations (hexagon) find a slightly lower value of $\Gamma_{\rm obs}=10^{-13}-10^{-12.4}$ \citep{Wyit11a,Calv11a}. This discrepancy might be resolved by, for example, including small-scale gas clumping, which is not done in the simulations presented here. This clumping delays the late stages of reionization, while not having very significant effect on the optical depth (Mao et al., in prep.). On the other hand, the early reionization model LB3 finds significantly higher value for the photoionization rate and likely can be excluded with the current efficiency parameters. Once again, this model can be reconciled with the observational data by tuning down the assumed ionizing photon efficiencies. 

\subsection{Reionization history}
\label{sec:reion_hist}

The mean global reionization histories derived from our simulations can be characterised by several basic parameters, as detailed in Table~\ref{tab:summary}. The first of these parameters is the end of the reionization epoch, $z_{\rm reion}$, which we customarily define as the time when the mass-weighted ionized fraction of the gas, $x_{\rm m}$, first surpasses 99 per cent. This value also quantifies the overall duration of reionization, since the start of reionization is determined by when the first resolved haloes form in our simulations, which is fixed by structure formation alone. The second global parameter is $\tau_{\rm es}$, as discussed in the previous section. Finally, the redshifts at which $x_{\rm m}$ reaches 10 per cent, 50 per cent, and 90 per cent are also listed, which correspond to the early, middle, and late stages of reionization. The middle redshift, when 50 per cent of the gas mass is ionized for the first time, is of particular interest for observations, since it is a good, if rather rough, indicator of the epoch when the ionization fluctuations reach a maximum \citep[e.g.][]{Mell06b}. This maximum corresponds to the maximum of many observables, such as the redshifted 21-cm fluctuations.

\begin{figure*}
\includegraphics[width=3.2in]{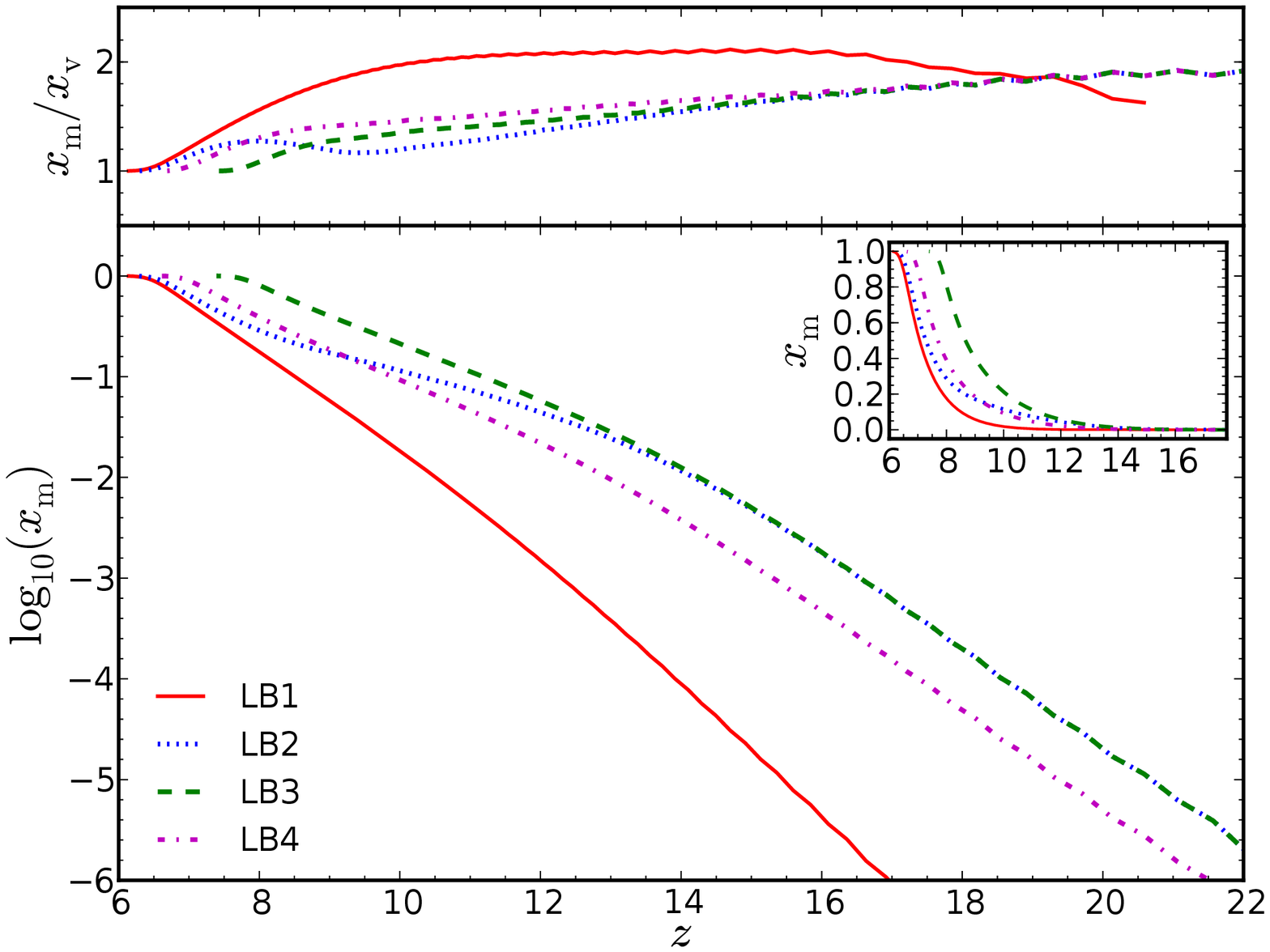}
\includegraphics[width=3.2in]{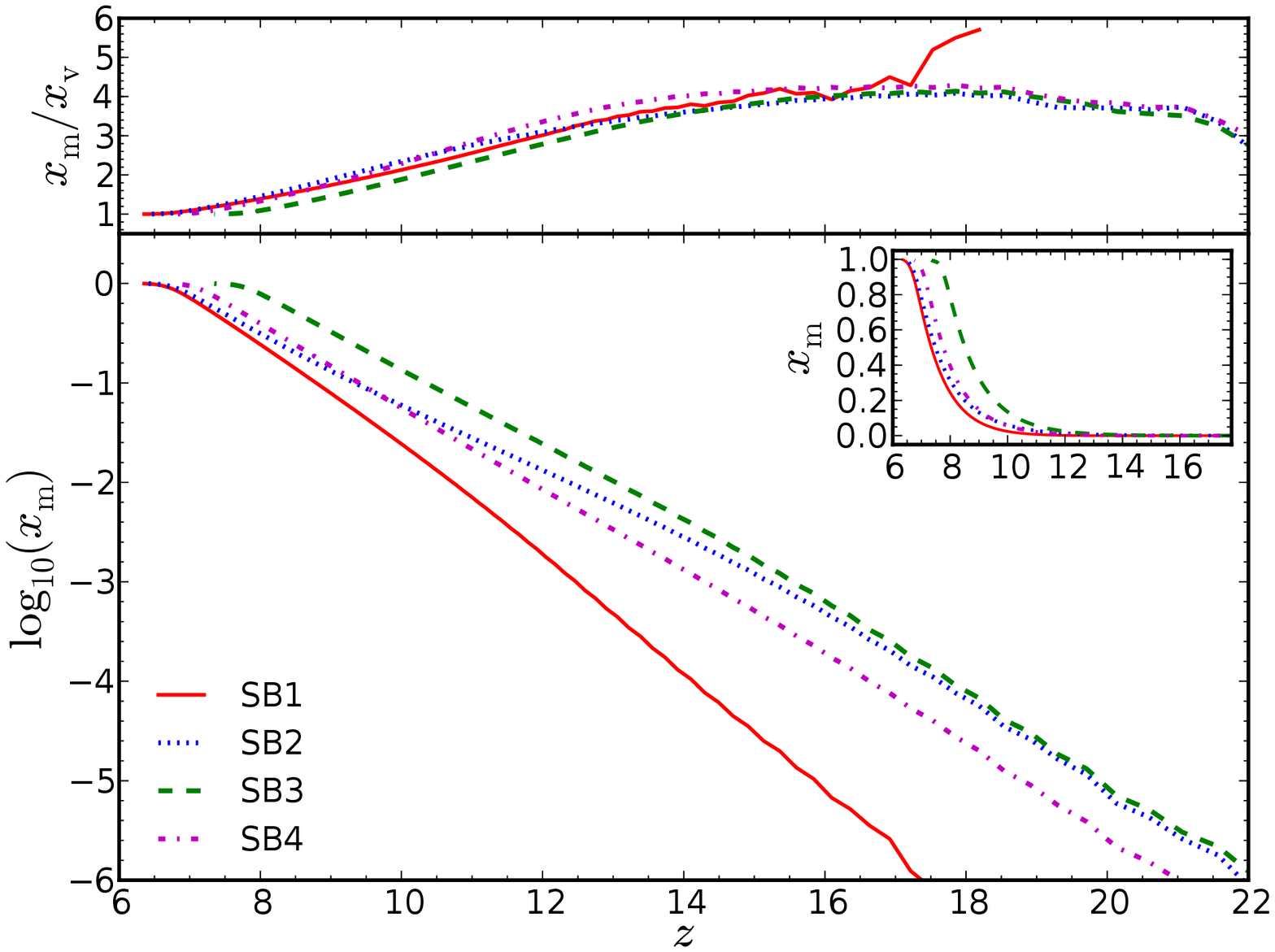}
\caption{Redshift evolution of the mass-weighted ionized fraction (lower panels) and the corresponding ratios of mass-weighted and volume-weighted ionized fractions (top panels), which are equal to the mean density of the ionized regions in units of the mean density of the Universe. \emph{Left:} the $244\,h^{-1}$~Mpc box shows the evolution for source models LB1 (solid), LB2 (dotted), LB3 (dashed), and LB4 (dot-dashed) (bottom panel). \emph{Right} for the $47\,h^{-1}$~Mpc box, SB1 (solid), SB2 (dotted), SB3 (dashed), and SB4 (dot-dashed) are displayed. \emph{Insets:} the same reionization histories in linear scale, as opposed to logarithmic.
\label{fig:fracs}}
\end{figure*}

The variation in the progression of reionization from source model differences is demonstrated in the globally averaged reionization histories as a function of redshift, shown in Fig.~\ref{fig:fracs}. Predictably, reionization starts significantly earlier in models where LMACHs are present, as LMACHs form earlier. The first HMACHs in the $244\,h^{-1}$~Mpc volume form at $z\sim21$, well after the first LMACHs. Accordingly, the SB1/LB1 HMACH-only models start reionizing with a significant delay with respect to the other models. Cases with high-efficiency LMACHs (LB2 and LB3) naturally reionize faster than the low-efficiency one (LB4). Initially, the method of LMACH suppression, either the full, aggressive one (LB2) or the partial one (LB3), makes essentially no difference in the global history, because the exponential growth of the halo collapsed fraction drives the exponential rise in the ionized fraction. However, once the ionized fraction reaches a few per cent, these two models begin to depart, as the LMACH suppression becomes more pronounced. The ionized fraction in the LB3 continues to grow quickly, with a change in slope due to the decreasing efficiency of the LMACHs. In contrast, LB2 results in a considerable slowdown and flattening of the reionization history until the HMACHs become dominant at $z\sim9-10$, after which the exponential growth resumes. In the global reionization history, the gradual, mass-dependent suppression model (LB4) follows the same trends as LB3, but with some delay due to its lower-efficiency LMACHs. Accordingly, the ionization fraction in LB4 overtakes the one in LB2 at redshifts just below $z\sim10$, as the lack of full LMACH suppression compensates for their lower efficiencies. The end of reionization and $z_{\rm reion}$ is dictated by the surviving sources and their efficiencies in each case, with little influence from the previous history, which is related to a process we refer to as self-regulation \citep{Ilie07a}. Consequently, LB1 and LB2 reach the end of reionization at approximately the same time, since only HMACHs remain at late times in either case. In contrast, LMACHs survive, albeit at lower efficiencies, in LB3 and LB4 and, thus, still contribute significantly to the entire evolution, leading to an earlier completion of reionization. The effective efficiency of LMACHs in LB4 is lower (though, increasing over time due to growing average source mass) than in LB4, slowing reionization. 

Reionization is fastest in models LB1 and LB3 and relatively slow and extended in LB2 and LB4. Accordingly, in the former cases, detecting an all-sky `global step' in the 21-cm emission due to the relatively fast transition of the IGM from fully neutral to ionized will be easier \citep{Shav99a}. However, the EoR is fairly extended \emph{all} cases, so such a measurement remains very difficult.

The smaller, $47\,h^{-1}$~Mpc volumes are based on a higher underlying $N$-body resolution (eliminating the need for sub-grid halo modelling) \emph{and} a higher radiative transfer grid resolution ($953\,h^{-1}$~kpc cells vs. $183\,h^{-1}$~kpc). The main drawback is that the volume is 1/140th of the $244\,h^{-1}$~Mpc boxes. The reionization process starts later in smaller volumes, because the earliest sources are very rare and are statistically unlikely to exist at very high redshift. For the same reason, the transition between LMACH-dominated and HMACH-dominated evolution is quicker and less pronounced. Regardless of these underlying dissimilarities, the overall trends in the global reionization histories discussed above remain the same for both simulation sizes, indicating the overall robustness of the results. 

The resolution also plays a small role in whether a source can ionize its cell and immediate surroundings, which is most evident in the fully suppressed model as SB2 appears depressed compared to LB2 (dotted lines on the right and left, respectively). Though a minor effect overall, the higher resolution of the smaller box means a smaller cell is more easily ionized, suppressing more sources for a given threshold for suppression. For a fixed simulation volume, increasing the RT grid resolution for the $47\,h^{-1}$~Mpc box does not have an appreciable effect on the reionization history or the number of photons emitted, indicating full convergence. For the larger, $244\,h^{-1}$~Mpc volume, the RT grid resolution has very little effect on the reionization history in case LB1, but LB3 slightly delays (by $\Delta z<0.5$) reionization, particularly in the intermediate stages where the LMACH suppression is more prominent. We do not show these comparisons due to their similarity, but the main change from higher resolution to lower is a decrease in suppression of LMACHs.

In contrast to the reionization histories, the ratio of mass-weighted to volume-weighted ionized fraction, which indicates the character (inside-out or outside-in) of the reionization process \citep{Ilie06a}, mostly shows only minor variations between models. The only exception here is model LB1 (HMACH-only), where the ionized regions are comparatively more overdense. However, this difference is largely due to numerical resolution, rather than a physical effect, as we discuss below. For the larger box, the mass-weighted over volume ionized fraction is always lower in the suppression models than in the HMACH-only model, indicating that reionization has less pronounced inside-out character. In other words, ionized regions are less correlated with the highest density peaks in models with LMACHs, since reionization is driven by wider range of sources, including low-mass, less-biased ones. The gradual suppression model LB4 is somewhat higher than the other two suppression models once reionization is underway ($x_{\rm m}\gsim0.01$). This model is more biased, because the largest LMACH sources are more strongly clustered and have higher efficiency on average compared to the LMACHs in the fully or partially suppressed models. 

For the $47\,h^{-1}$~Mpc box (right) in Fig.~\ref{fig:fracs}, we can see that $x_{\rm m}/x_{\rm v}$ is nearly converged. The higher-resolution run (SB2\_HR, not shown) differs in the ratio only slightly, indicating a robust inside-out nature of all models. In the context of the small, high-resolution boxes versus the large, low-resolution boxes, the high-resolution ratios take a somewhat different shape: rising initially, then falling towards unity at the end of reionization, by definition. The values reached are significantly higher, due to the better radiative transfer grid resolution that amplifies the inside-out nature of the process. For the higher resolution runs in the larger volume (L1\_HR and L3\_HR, not shown), $x_{\rm m}/x_{\rm v}$ peaks at higher values for the same reason. As noted above, only the HMACH-only model at low resolution achieves a similar shape to the results of the smaller boxes, and at high resolution, the ratio exceeds three. The partially suppressed, high-resolution model has a flatter shape, indicating a lack of convergence. Whether a source can ionize its own cell and immediate surroundings is resolution dependent; therefore, models with suppression generally require higher resolution than models with straightforward sources with a constant mass-to-light ratio.

In summary, the different LMACH suppression models result in significant variations in the duration and shape of the reionization history, even for same underlying cosmological structures and same efficiencies for HMACHs. For all models, HMACHs dominate during the late stages of reionization, which are the focus of this work. However, the inside-out nature of the process, in the sense of denser structures being ionized earlier on average, remains robust and roughly independent of the source suppression model, depending somewhat on the resolution. 

\begin{figure*}
\includegraphics[width=3.2in]{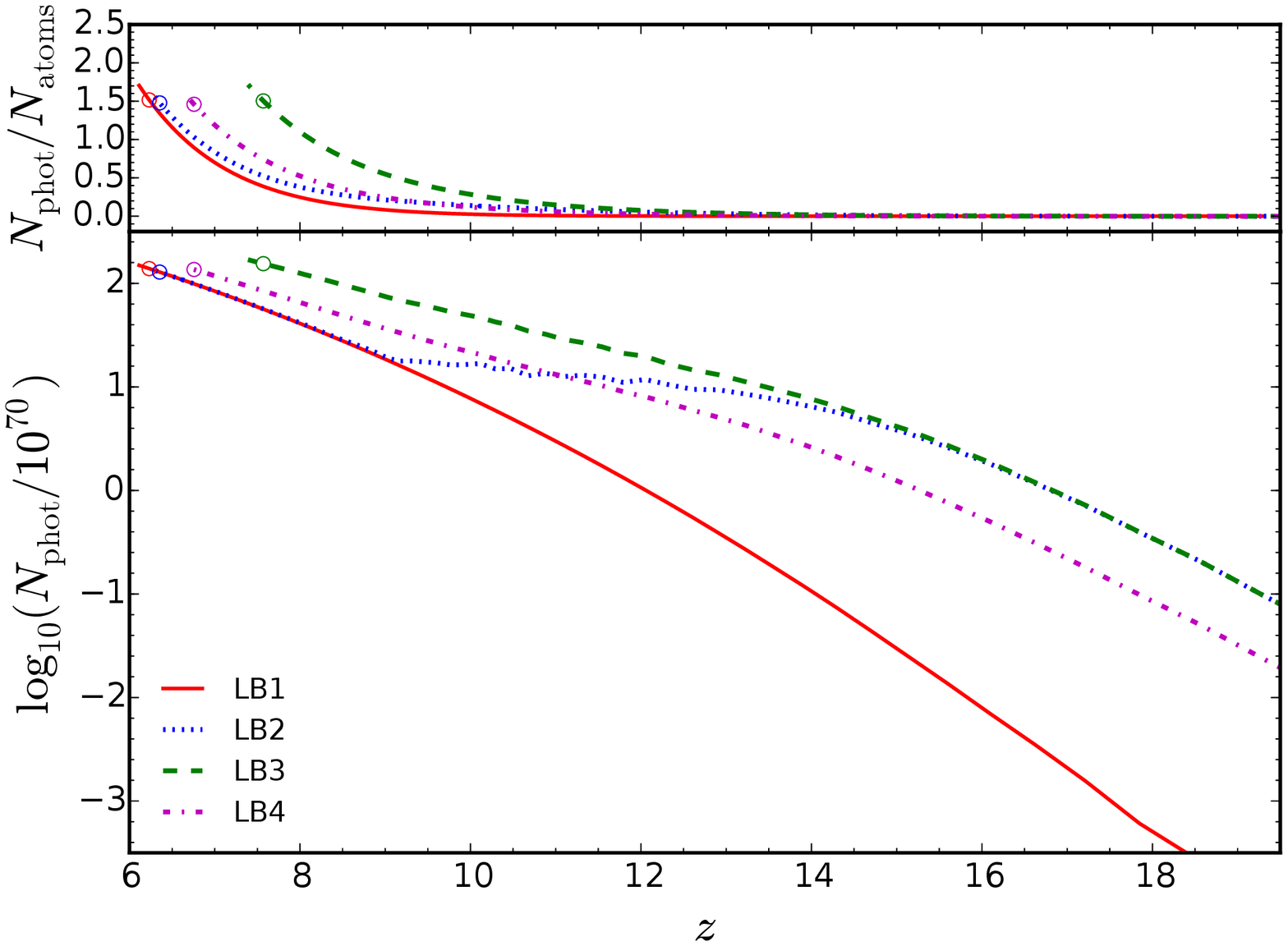}
\includegraphics[width=3.2in]{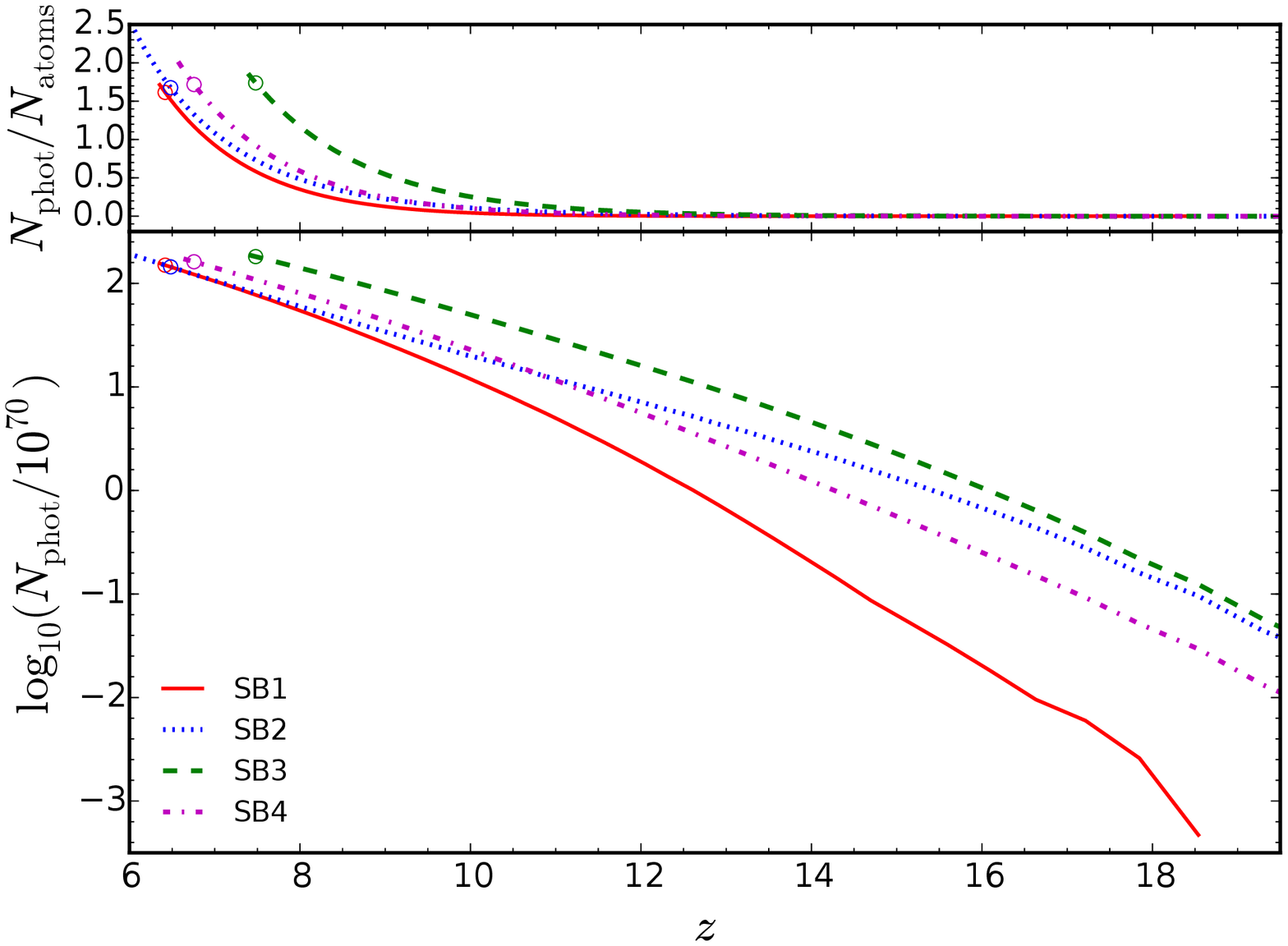}
\caption{\label{fig:source_lum} Number of ionizing photons emitted by all active sources in the computational volume per time-step renormalized to a ($100\,h^{-1}$\,Mpc)$^3$ volume (bottom panels) and cumulative number of photons per total gas atom released into the IGM (top panels). Shown are the 244\,$h^{-1}$~Mpc box (left) with LB1 (solid), LB2 (dotted), LB3 (dashed), and LB4 (dot-dashed) and 47\,$h^{-1}$~Mpc box (right) with SB1 (solid), SB2 (dotted), SB3 (dashed), and SB4 (dot-dashed). The open circles indicate $z_{\rm reion}$.}
\end{figure*}

These reionization histories are a direct consequence of the overall number of ionizing photons being emitted by all active sources, shown in Fig.~\ref{fig:source_lum}. For ease of comparison, the number of photons emitted by both the large and small boxes are renormalized to a $100\,h^{-1}$\,Mpc$^3$ volume. In the case LB1, where only HMACHs contribute, the number of photons emitted per time-step simply rises proportionally to the halo collapsed fraction, which is roughly exponentially. In all cases with LMACHs, reionization begins earlier, and all models initially have similar slopes. In fact, cases LB2 and LB3 are nearly identical until sufficient reionization occurs to produce significant self-regulation, around $x_{\rm m} = 0.20$. In the full-suppression case LB2, the initial exponential rise is halted around redshift $z\sim15$ and increases slowly (and moderately non-monotonically) until $z\sim9$, where high-mass, non-suppressible sources become dominant and the low-mass sources become highly suppressed. Therefore, similarly to our earlier results in \citet{Ilie07a}, the late phase of reionization and the end of the epoch, $z_{\rm reion}$, are dominated by HMACHs, while the LMACHs dominate the early phase of reionization and provide a significant boost to $\tau_{\rm es}$. 

Similarly to the reionization histories above, the resolution and box-size effects play a minor role, making the main trends in the evolution of the number of ionizing photons robust. The HMACH-only model is minimally affected by resolution, with the high-resolution flux appearing nearly identical to the low-resolution case. The models with suppression are somewhat affected by resolution, particularly through the sub-grid modelling of the LMACHs (in the large boxes) and the lower RT grid resolution. The combined effect is more significant suppression in the higher-resolution case, since the cells are smaller and therefore easier to ionize, which suppresses the LMACHs. However, once the ionization fraction grows sufficiently, the amount of suppression in the low-resolution case reaches and then surpasses the high-resolution case, since the sub-grid LMACHs are more strongly clustered than the resolved ones \citep{Ahn15a}. With more nearby sources increasing the ionizing radiation experienced by an LMACH, the source cell is ionized more easily, resulting in earlier suppression and the majority of ionizing photons being produced by HMACHs at late times. All models require just under two photons per atom to reach end of reionization, independent of resolution and simulation volume. Therefore, increasing the resolution by a factor of 13 from 0.98 to $0.078 h^{-1}$\,Mpc resolves more gas clumping, but does not substantially increase the impact of recombinations on reionization. Gas clumping at much smaller scales is needed to increase the number of photons per atom to higher values (Mao et al., in prep). The LLS, only partly included here, may also increase the number of photons required to complete reionization \citep{Shuk15a}.

\subsection{Ionization morphology}
\label{sec:morph}

\begin{figure*}
  \begin{center}
    \includegraphics[width=1.7in]{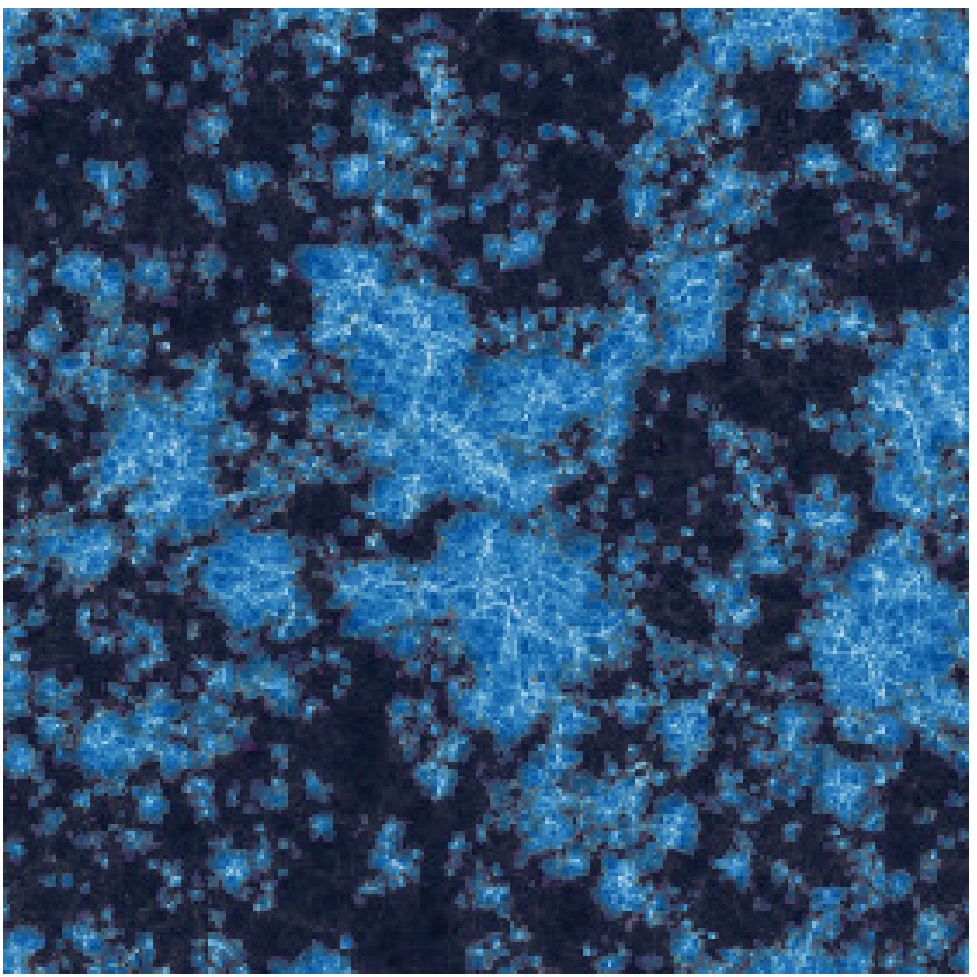}
    \includegraphics[width=1.7in]{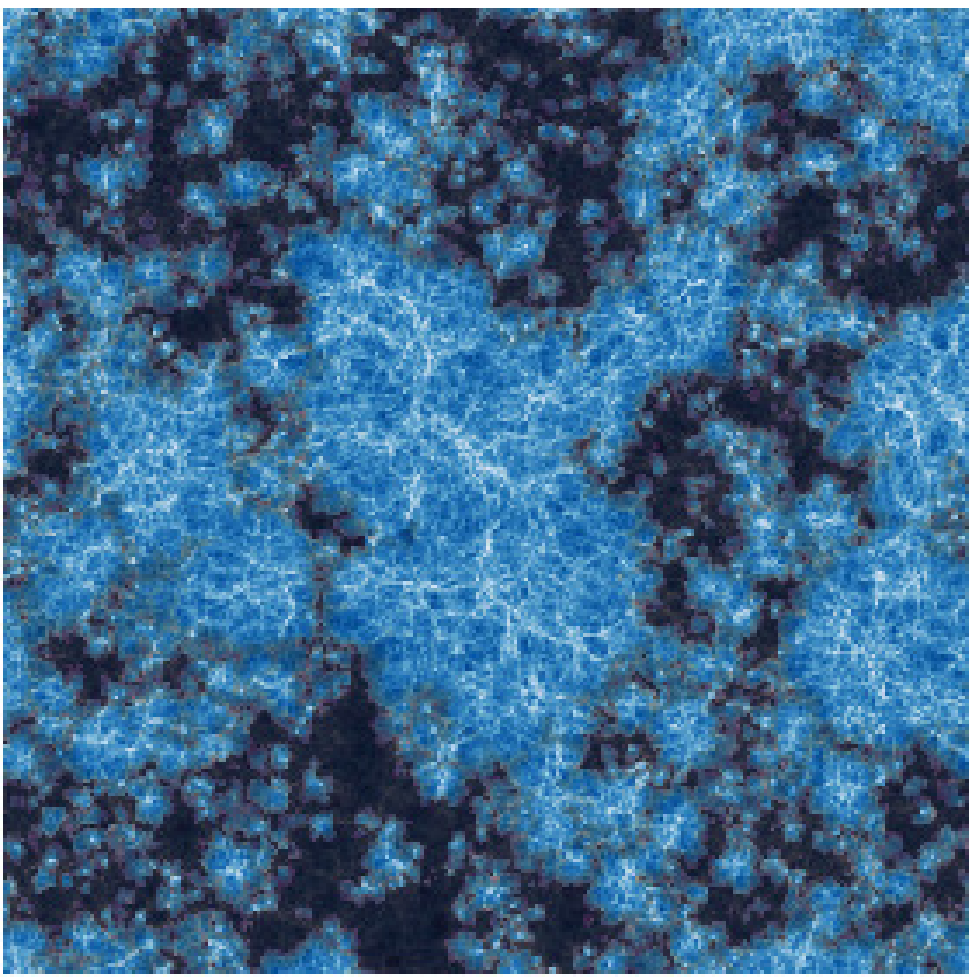}
    \includegraphics[width=1.7in]{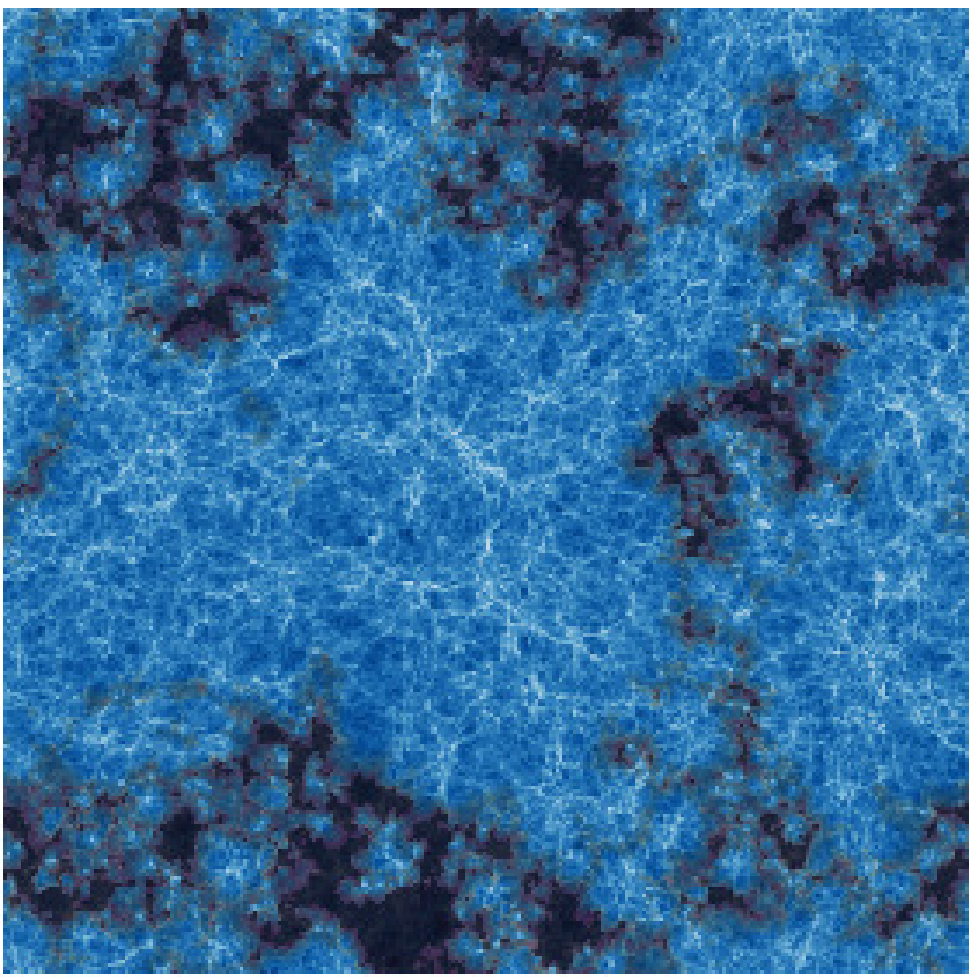}
    \includegraphics[width=1.7in]{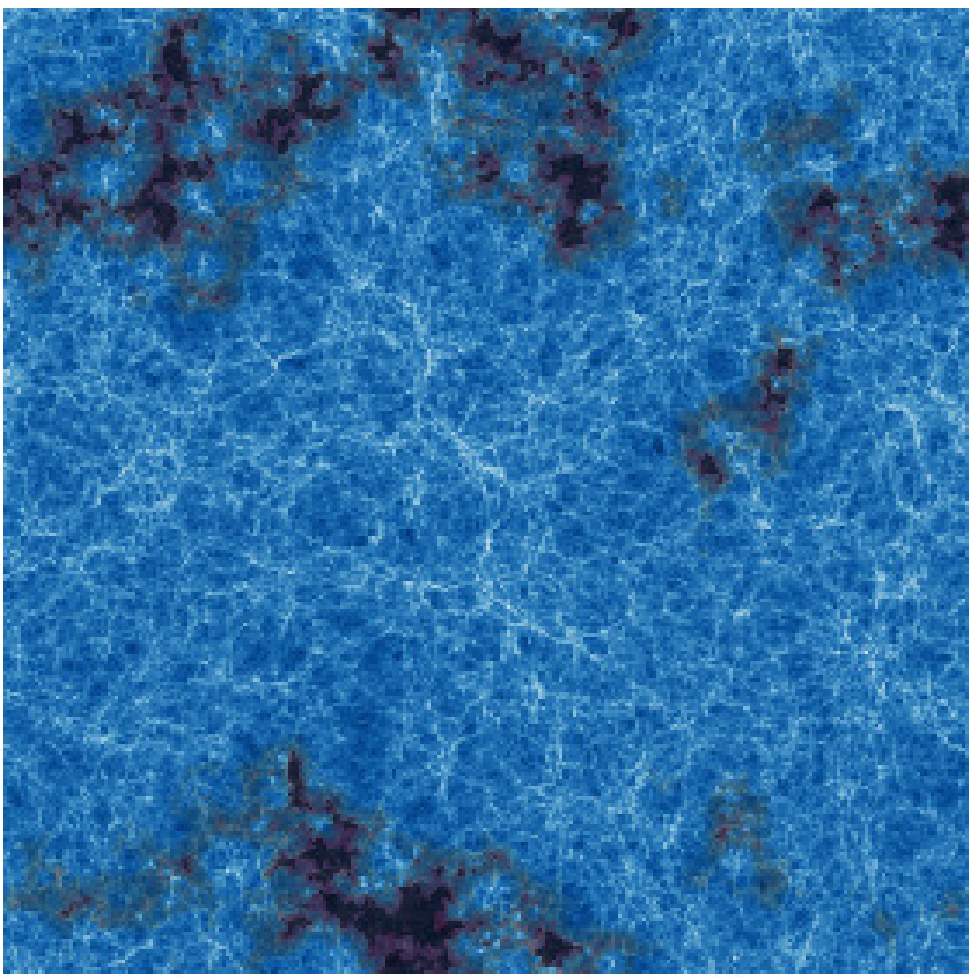}
       \vspace{-0.3cm}
  \end{center}
  \caption{Spatial slices of the ionized and neutral gas density from our radiative transfer simulation SB2\_{HR} at box-averaged by mass ionized fraction $x_{\rm m}= 0.3, 0.5, 0.7,$ and 0.9 from left to right. The density field is shown in blue, with lighter shades corresponding to denser regions and vice versa, and overlaid with the ionization field, where dark is neutral and light is fully ionized.
    \label{fig:images_gradual}}
\end{figure*}

In Fig.~\ref{fig:images_gradual}, we illustrate the evolution of the reionization geometry at several key stages of the process, corresponding to mass-weighted ionized fractions of $x_{\rm m}= 0.3, 0.5, 0.7,$ and 0.9 from left to right. We use a small-box, high-resolution simulation here, specifically SB2\_HR, which allows for better discrimination of any differences between models as small-scale structure is more discernible. Note that these smaller volumes are missing the large-scale density modes, which introduce additional large-scale fluctuations \citep{Ilie14a}. Even the new, mass-dependent source suppression model contains the basic features of source models considered in previous work \citep[e.g.][]{Ilie12a}, which we will explore in detail. Initially, a large number of fairly small, Mpc-size \ion{H}{ii} regions form. These regions are strongly clustered on small scales, following the clustering of the sources. Locally, these small \ion{H}{ii} regions quickly start merging into larger ones, with sizes between few and $\sim\!10$~Mpc across. We note that, of course, these are 2D cuts of the ionization field and that \ion{H}{ii} regions can, and do, have different sizes depending on the direction considered, as quantified e.g. in \citet{Ilie08b}. Significant large-scale percolation of the \ion{H}{ii} regions only occurs when the universe reaches $\sim\!50$ per cent ionization by mass, at which point, many ionized regions reach sizes of tens of Mpc and become connected by bridges to other nearby, large ionized regions of similar size. At the same time, different regions of similar size still remain neutral. The \ion{H}{ii} regions continue percolating up to still larger scales, and by $x_{\rm m}=0.7$, some reach tens of Mpc across, with significant neutral regions remaining between them. These large ionized and neutral regions both reflect the large-scale fluctuations of the underlying density field, as the densest regions are also sources. Finally, when the mass is 90 per cent ionized, most \ion{H}{ii} regions have percolated into one, though significant neutral regions remain even in this late phase.

\begin{figure*}
  \begin{center}
    \includegraphics[width=1.7in]{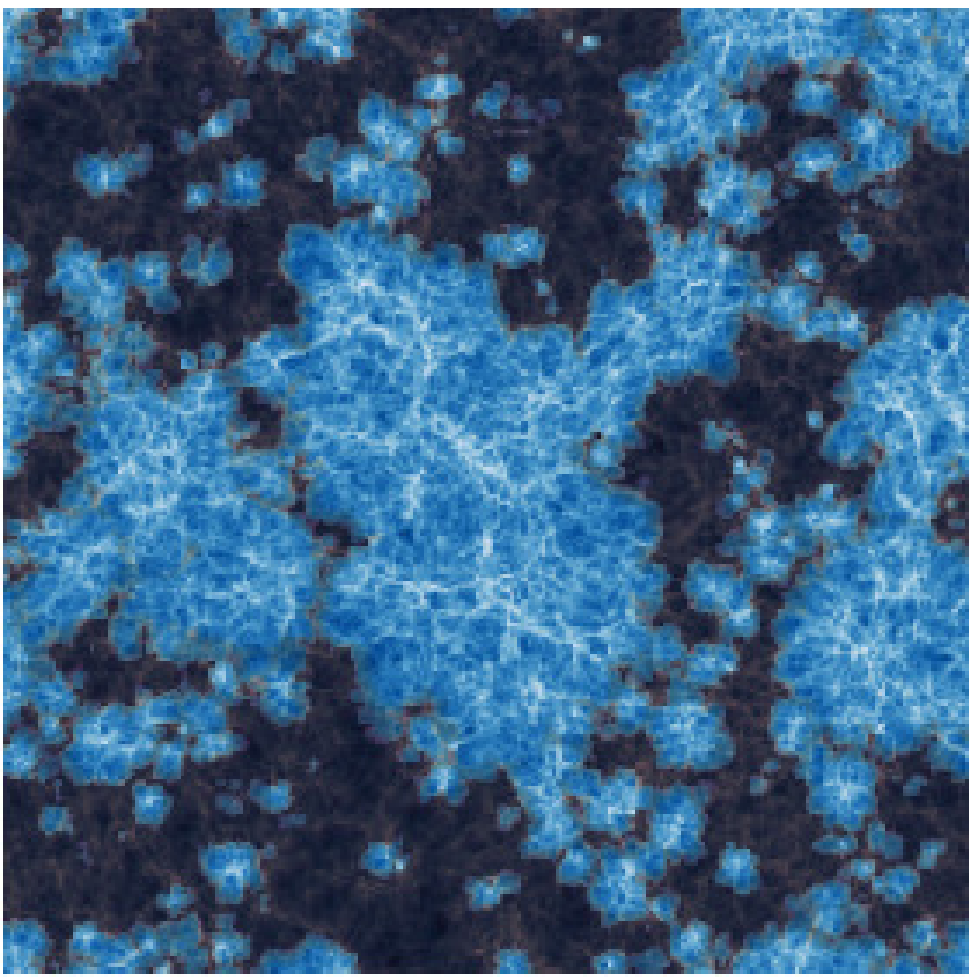}
    \includegraphics[width=1.7in]{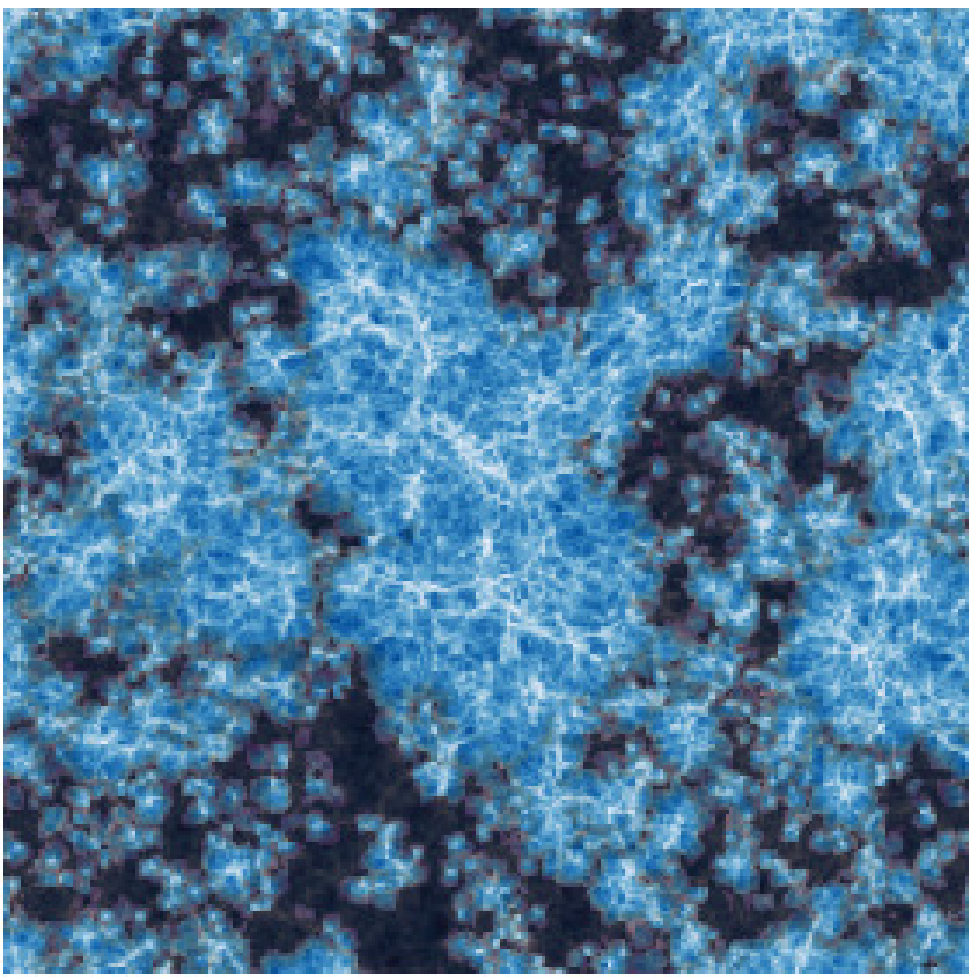}
    \includegraphics[width=1.7in]{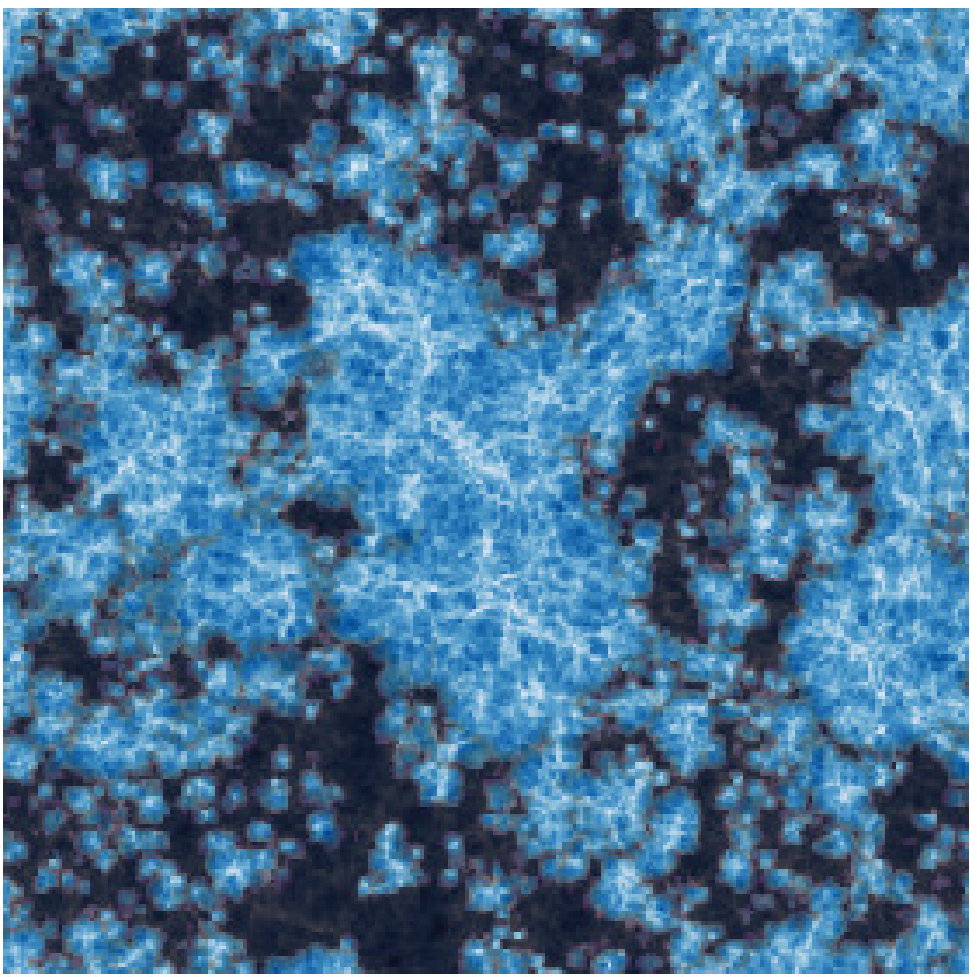}
    \includegraphics[width=1.7in]{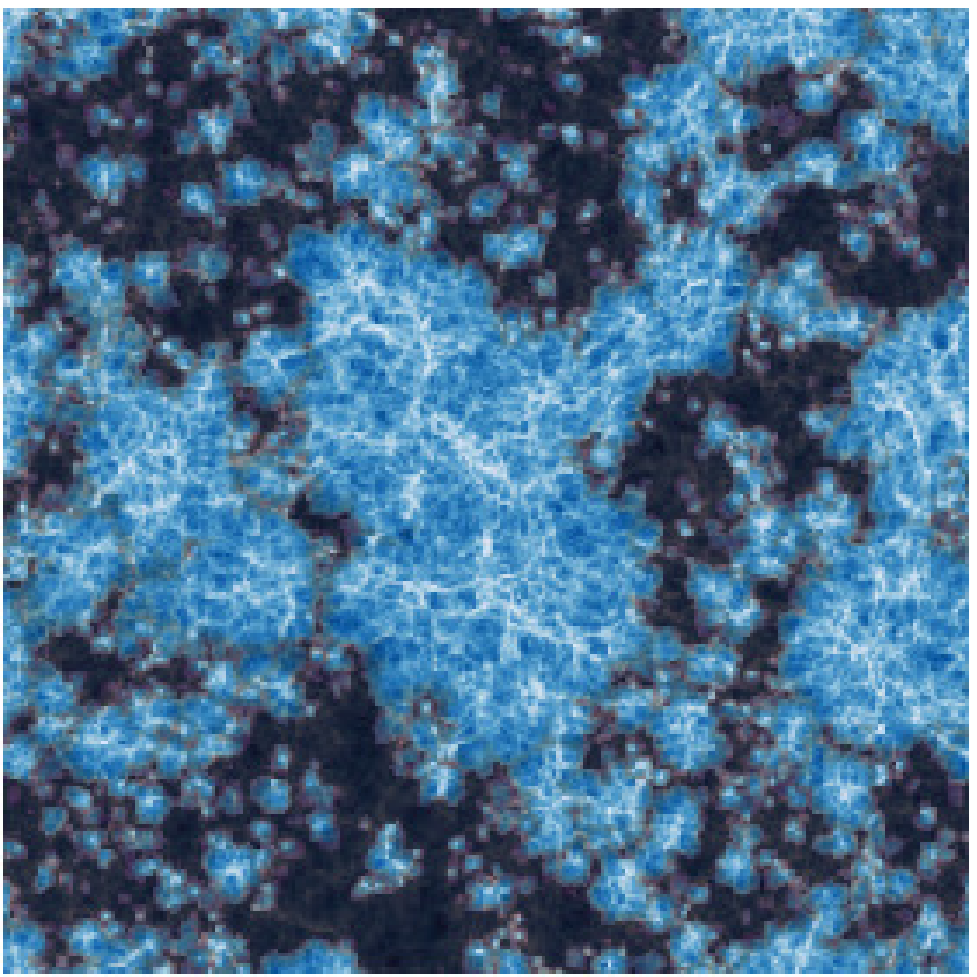}
       \vspace{-0.3cm}
  \end{center}
  \caption{Spatial slices of the ionized and neutral gas density from our 47\,$h^{-1}$~Mpc box. Models SB1, SB2, SB3, and SB4 (from left to right) are shown at the same mass-weighted ionized fraction, $x_{\rm m}\approx0.5$. The density field is shown in blue, with lighter shades corresponding to denser regions and vice versa, and overlaid with the ionization field, where dark is neutral and light is fully ionized.    \label{fig:images_47Mpc}}
\end{figure*}

\begin{figure*}
  \begin{center}
    \includegraphics[width=2.2in]{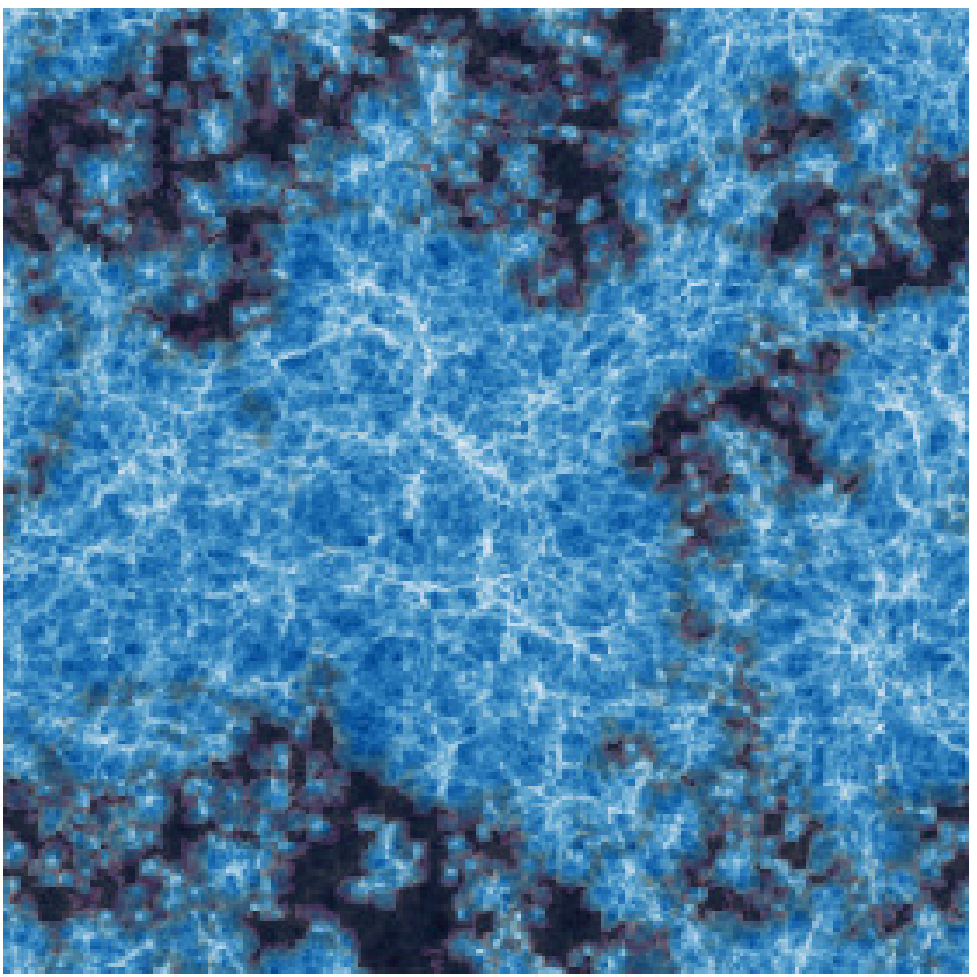}
    \vspace{-0.0in}
    \hspace{-0.055in}
    \includegraphics[width=2.2in]{xy306_ion_47Mpc_8.2S_612_7.059.eps}\\
    \includegraphics[width=2.2in]{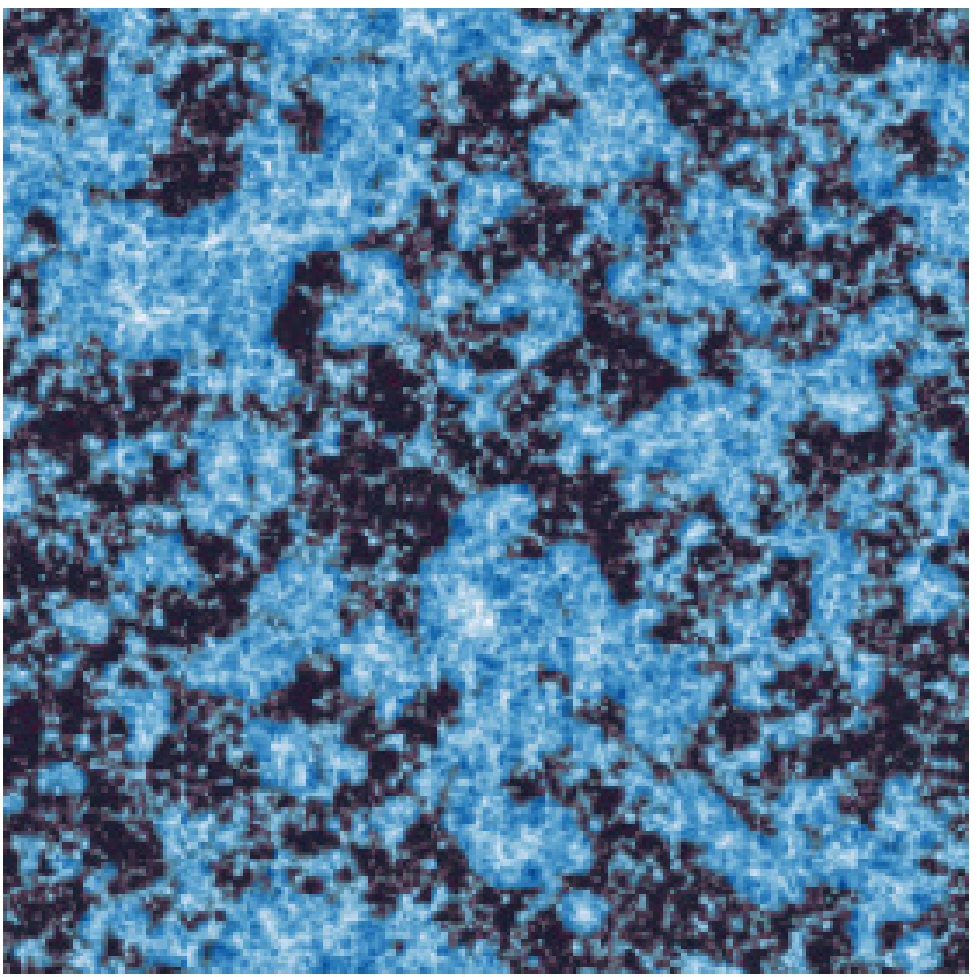}
    \hspace{-0.055in}
    \includegraphics[width=2.2in]{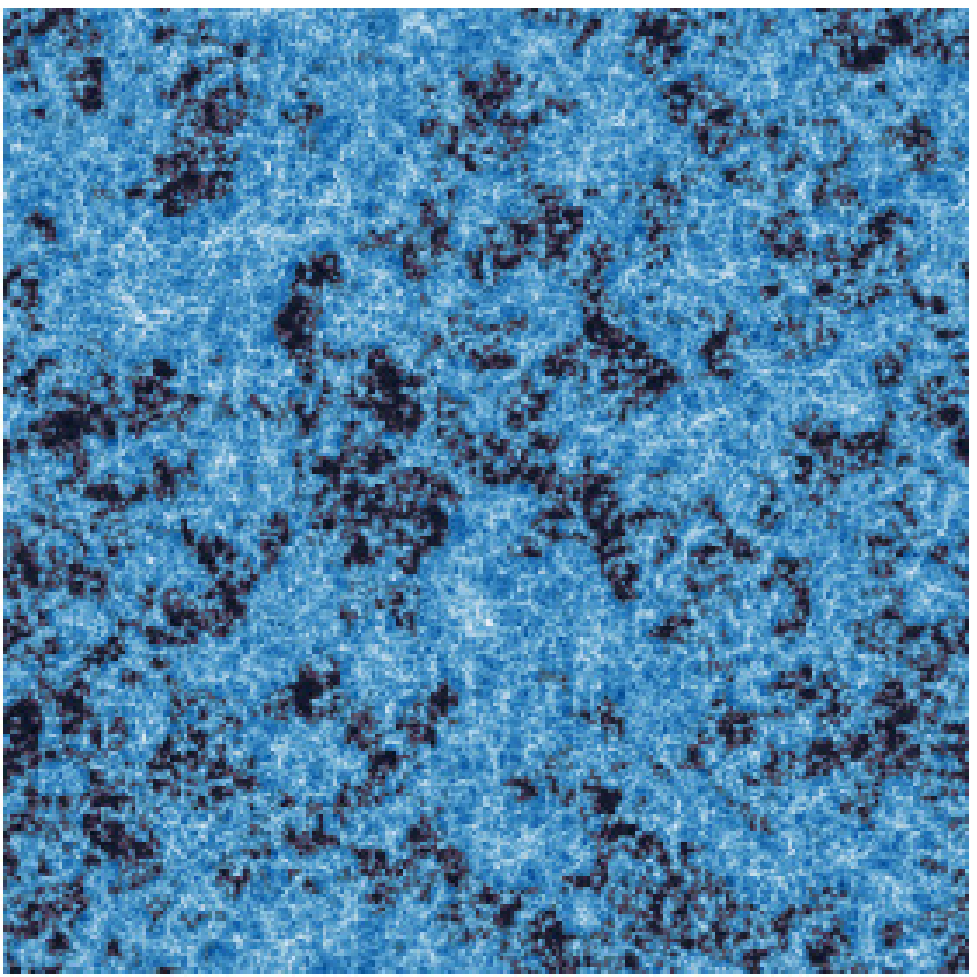}
  \end{center}
  \caption{Spatial slices of the ionized and neutral gas density from our radiative transfer simulations with box sizes 47\,$h^{-1}$~Mpc (upper panels) and 244\,$h^{-1}$~Mpc (lower panels), all at mass-averaged ionized fraction $x_{\rm m}\sim0.70$. The density field is shown in blue, with lighter shades corresponding to denser regions and vice versa, and overlaid with the ionization field, where dark is neutral and light is fully ionized. The left panels are low resolution, 306$^3$ for 47\,$h^{-1}$~Mpc and 250$^3$ for 244\,$h^{-1}$~Mpc, and the right panels are high resolution, 612$^3$ for  47~Mpc and 500$^3$ for 244\,$h^{-1}$~Mpc. Shown are cases SB2, SB2\_HR, LB3, and LB3\_HR (left to right and top to bottom).
    \label{fig:images_res}}
\end{figure*}

Direct comparison of all four simulations at the same ionized fraction illustrates the differences in morphology caused by the various LMACH suppression models, shown as SB1, SB2, SB3, and SB4 from left to right in Fig.~\ref{fig:images_res} at $x_{\rm m}\approx0.50$. In all cases, the large-scale structures of the ionization field strongly correlate with the underlying distribution of density and clustered haloes and are, thus, quite similar. There are significant differences in the smaller scale structures among the range of simulations. Naturally, the HMACH-only SB1 has larger, smoother ionized patches and few small-scale ones. The aggressive suppression case (SB2) has more widespread relic \ion{H}{ii} regions, where the local sources have switched off, compared to SB3 and SB4, where most LMACHs remain active, albeit at a lower emissivity. Cases SB2 and SB3 have much more fine, small-scale structures compared to SB1 and (to a lesser extent) SB4. Finally, we compare different RT grid resolutions for the same source models in Fig.~\ref{fig:images_res}. Apart from (obviously) much sharper images in the high-resolution cases, the reionization morphology is largely the same. Although especially true for the smaller, $47\,h^{-1}$~Mpc volume, the two distributions are quite close in both volumes. Clearly, more small-scale structure is revealed at higher resolution, but that is likely too small to make a difference for the first generation of observations, which will have relatively low resolution. At least visually, the overall differences are small between the four source models and depend weakly on the RT resolution. We quantify these differences in more detail below.

\begin{figure*}
\begin{center} \vspace{+0.1in}  
\includegraphics[height=1.7in]{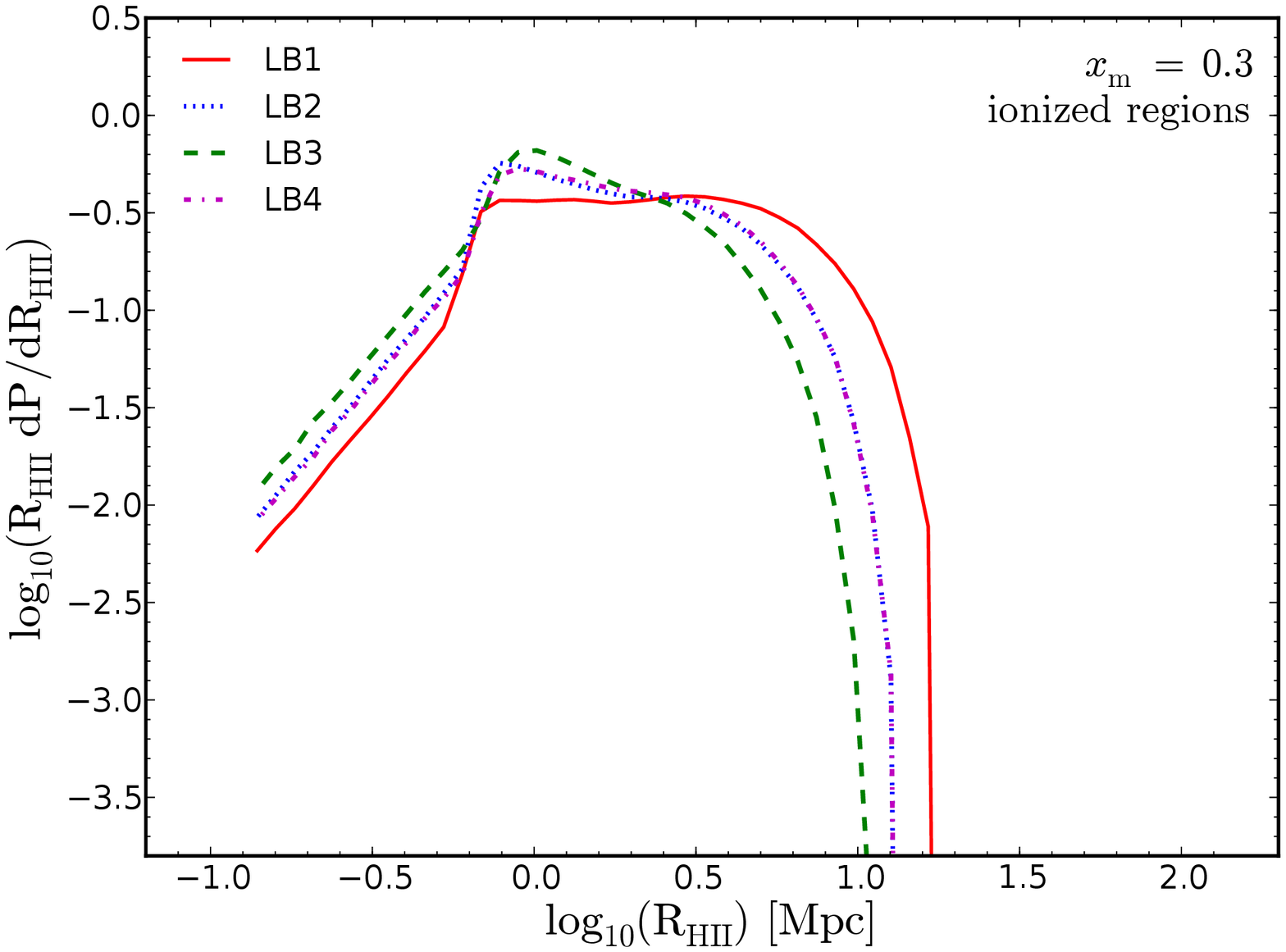}
\vspace{-0.2in} 
\includegraphics[height=1.7in]{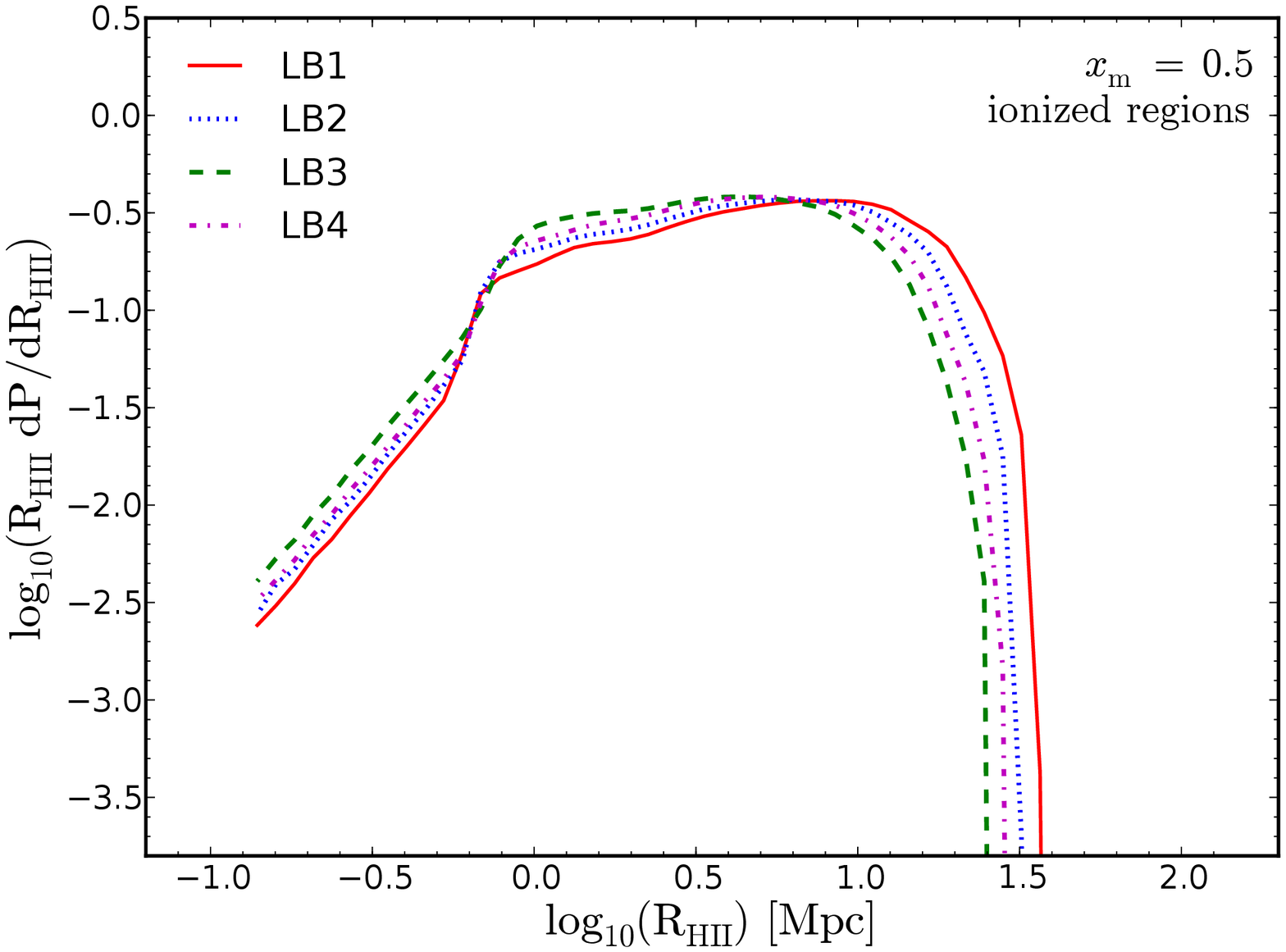}
\vspace{-0.2in} 
\includegraphics[height=1.7in]{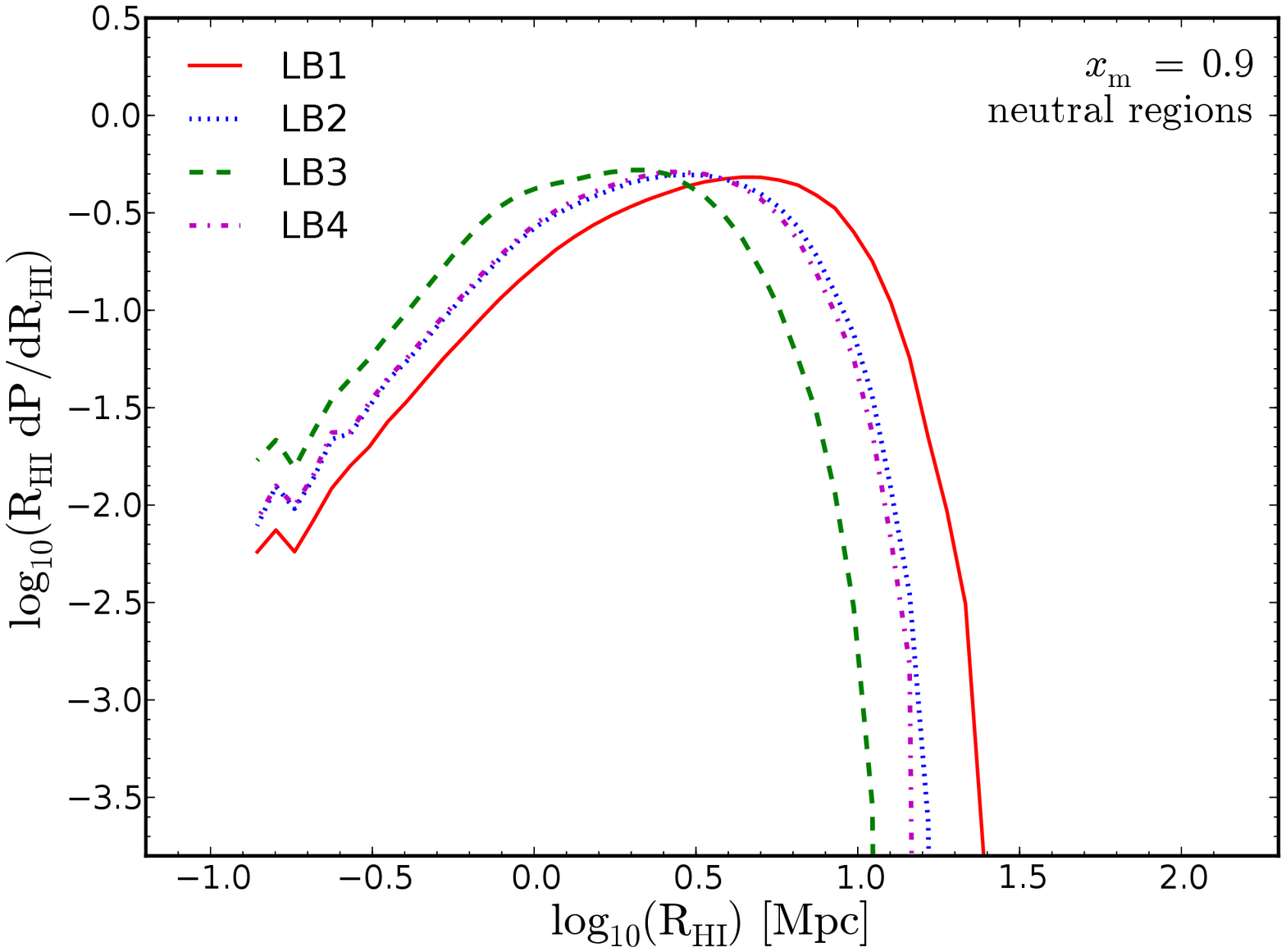} 
\vspace{+0.8cm}
\caption{
\label{fig:R_dist}
Size distributions of ionized or neutral regions for the $244\,h^{-1}$~Mpc box. Distributions are shown at different stages of the reionization process with ionized fraction by mass as $x_{\rm m} = 0.3$ (\ion{H}{ii} regions), 0.5 (\ion{H}{ii} regions), and 0.9 (\ion{H}{i} regions) from left to right. The LB1 (solid), LB2 (dotted), LB3 (dashed), and LB4 (dot-dashed) simulations are represented.}
\end{center}
\end{figure*}

A more quantitative measure of the size distributions of the ionized regions, based on the spherical average method \citep[SPA,][]{Zahn07a,McQu07a}, supports the qualitative conclusions drawn from the slices. Fig.~\ref{fig:R_dist} shows the probability distributions for the radius of ionized regions, $R_{\rm \ion{H}{ii}}$, at $x_{\rm m}$ = 0.3 and 0.5 (left and middle panel, respectively) for LB1 (solid), LB2 (dotted), LB3 (dashed), and LB4 (dot-dashed). To investigate the sizes of ionized regions, we rely on the $244\,h^{-1}$Mpc volume, since small simulation volumes severely constrain the abundance and sizes of large \ion{H}{ii} regions \citep{Ilie14a}. The distributions reflect both the suppression mechanism and the epoch at which the corresponding reionization stage is reached. As expected, the size of the ionized bubbles grows during reionization, starting mainly at the Mpc scale for $x_{\rm m}=0.3$ (left). At this stage, LB1 has the flattest distribution, whereas the models with LMACHs produce majority small bubbles. By the midpoint ($x_{\rm m}=0.5$, middle), ionized regions of at least $\sim\!10$~Mpc begin to emerge for all source models. Here, LB3 has the most numerous and uniform source, yielding the flattest distribution. As more ionized regions merge together, large bubbles of $\gtrsim\!10$~Mpc begin to dominate. Throughout reionization, the partially suppressed model (LB3) always has smaller bubbles on average, since the smallest, abundant sources are never fully suppressed. Conversely, HMACH-only model (LB1) has the largest bubbles on average.

As expected from visual observation of the spatial slices, LB1 and LB3 are at the two extremes during the early stages of reionization ($x_{\rm m}=0.3$, left), with distributions skewed towards very large patches for the former and small patches for the latter. This behaviour reflects the size of the sources, with large sources -- that cannot be suppressed -- creating large bubbles from emitting more photons. Conversely, highly efficient, small sources create small bubbles, are then suppressed, and just maintain the ionized region. The other two cases, LB2 and LB4 show almost identical distributions at this time, intermediate between the two extremes. Around 50 per cent ionized (middle panel), the bubble sizes for all models have grown, and the distributions for all models have become increasingly similar. LB2 is becoming dominated by the large sources that drive LB1, narrowing the gap between the distributions from early times. By $x_{\rm m} = 0.7$ (not shown here), the distributions have nearly converged for all models with log$_{10}(R_{\rm \ion{H}{ii}}^{\rm max})$ ranging from $\sim\!1.1 - 1.4$. 

The rightmost plot of Fig.~\ref{fig:R_dist} shows the probability distributions for the radius of neutral islands, $R_{\rm \ion{H}{i}}$, at $x_{\rm m} = 0.9$, since, at this late time, the ionized patches have all topologically merged and only the neutral islands are distinct. As before, LB3 is the most uniform with the smallest neutral regions, and LB1 is the most stochastic with the the largest neutral regions. The remaining models (LB2 and LB4) are very similar at this point. The neutral regions are also more Gaussian as compared to the ionized regions, especially in the large-$R_{\rm \ion{H}{i}}$ tail.

\subsection{21-cm background} 
\label{sec:21cm}

\subsubsection{Calculating redshifted 21-cm emission}

The differential brightness temperature of the redshifted 21-cm emission with respect to the CMB is determined by the density of neutral hydrogen, $\rho_{\rm \ion{H}{i}}$, and its spin temperature, $T_{\rm S}$, and is given by \citep{Fiel59a}:
\ba
 \delta T_{\rm b}&=&\frac{T_{\rm S} - T_{\rm CMB}}{1+z}(1-e^{-\tau})\nonumber\\
&\approx&
\frac{T_{\rm S} - T_{\rm CMB}}{1+z}
\frac{3\lambda_0^3A_{10}T_*n_{\rm \ion{H}{i}}(z)}{32\pi T_{\rm S} H(z)}. 
\label{eq:dT0}
\ea
Here, $T_{\rm CMB}$ is the temperature of the CMB radiation at that time, $\tau$ is the corresponding 21-cm optical depth (assumed to be small when writing equation~\ref{eq:dT0}), $\lambda_0=21.16$~cm is the rest-frame wavelength of the 21-cm line, $A_{10}=2.85\times10^{-15}\,\rm s^{-1}$ is the Einstein A-coefficient, and $T_*=0.068$~K corresponds to the energy difference between the two levels. The mean number density of neutral hydrogen, $n_{\rm \ion{H}{i}}(z)$, at redshift, $z$, is:
\ba
\langle n_{\rm \ion{H}{i}} \rangle(z)&=&
\frac{\Omega_{\rm b}\rho_{\rm crit,0}}{\mu_{\rm H}m_{\rm p}}(1+z)^3\nonumber\\
&=&1.909\times10^{-7}\rm cm^{-3}\left(\frac{\Omega_{\rm b}}{0.042}\right)(1+z)^3,
\ea
with $\mu_{\rm H}=1.22$ is the corresponding mean molecular weight (assuming 24 per cent He abundance), and $H(z)$ is the redshift-dependent Hubble constant,
\ba
  H(z)&=&
H_0[\Omega_{\rm m}(1+z)^3+\Omega_{\rm k}(1+z)^2+\Omega_\Lambda]^{1/2}
                      \nonumber\\  
&=&H_0E(z)\approx H_0\Omega_{\rm m}^{1/2}(1+z)^{3/2}.
\ea
Here, $H_0$ is the Hubble constant at present, and the last approximation in the above equation is valid for $z\gg 1$. 

Throughout this work, we assume that $T_{\rm S} \gg T_{\rm CMB}$, i.e., that all of the neutral IGM gas is Ly-$\alpha$-pumped by the background of UV radiation below 13.6~eV from early sources and heated well above the CMB temperature (due to, e.g., a small amount of X-ray heating). Therefore, the 21-cm line is seen in emission. These assumptions are generally well-justified, except possibly at the earliest times \citep[see e.g.][and references therein]{Furl06a}. In the high-$T_{\rm S}$ limit, equation~(\ref{eq:dT0}) becomes
\ba
\delta T_{\rm b}&=&{28.5\,\rm mK}\left(\frac{1+z}{10}\right)^{1/2}(1+\delta)\nonumber\\
& &\times\left(\frac{\Omega_{\rm b}}{0.042}\frac{h}{0.73}\right)\left(\frac{0.24}{\Omega_{\rm m}}\right)^{1/2},
\label{eq:dT}
\ea
where $1+\delta={n_{\rm \ion{H}{i}}}/{ \langle n_{\rm H} \rangle}$ is the density of the neutral hydrogen in units of the mean gas density.

Since our simulations take place in real space, we need to transform our data to redshift space, where the observations of lines occur. If the redshift is caused only by the Hubble expansion, then the redshift space position, $\mathbf{s}$, of some emitter will be the same as its comoving real space position, $\mathbf{r}$. However, if there is also a peculiar velocity along the line of sight, $v_{\parallel}$, then an emitter at position $\mathbf{r}$ in real space will be shifted to a position $\mathbf{s}$ in redshift space:
\be
	\mathbf{s} = \mathbf{r} + \frac{1 + z_{\mathrm{obs}}}{H(z_{\mathrm{obs}})} v_{\parallel} (t, \mathbf{r}) \hat{r},
	\label{eq:reddist}
\ee
where $1+z_{\mathrm{obs}} = (1+z_{\mathrm{cos}})(1-v_{\parallel}/c)^{-1}$, $z_{\mathrm{obs}}$ is the observed redshift, and $z_{\mathrm{cos}}$ is the cosmological redshift \citep[e.g.][]{Mao12a}. In other words, an emitter with a peculiar velocity away from the observer (i.e., $v_{\parallel}>0$) will be more redshifted than one with no velocity and will appear to be farther away than is really the case, and vice versa. \cite{Mao12a} describe several ways to calculate the redshift-space signal from a real-space simulation volume with brightness temperature and velocity information. Here, we use a slightly different method, introduced in \citet{Jens13a}, which splits each cell along the line-of-sight into $n$ sub-cells, each with a brightness temperature $\delta T (\mathbf{r})/n$. We then interpolate the velocity and density fields on to the sub-cells, move them around according to equation~(\ref{eq:reddist}), and re-grid to the original resolution. This scheme is valid only in the optically thin and high-$T_{\rm s}$ case, when equation~(\ref{eq:dT}) holds and each parcel of gas can be treated as an independent emitter of 21-cm radiation. For this paper, we use $40$ sub-cells, which is converged to less than one per cent.

\begin{figure*}  
\includegraphics[height=4in]{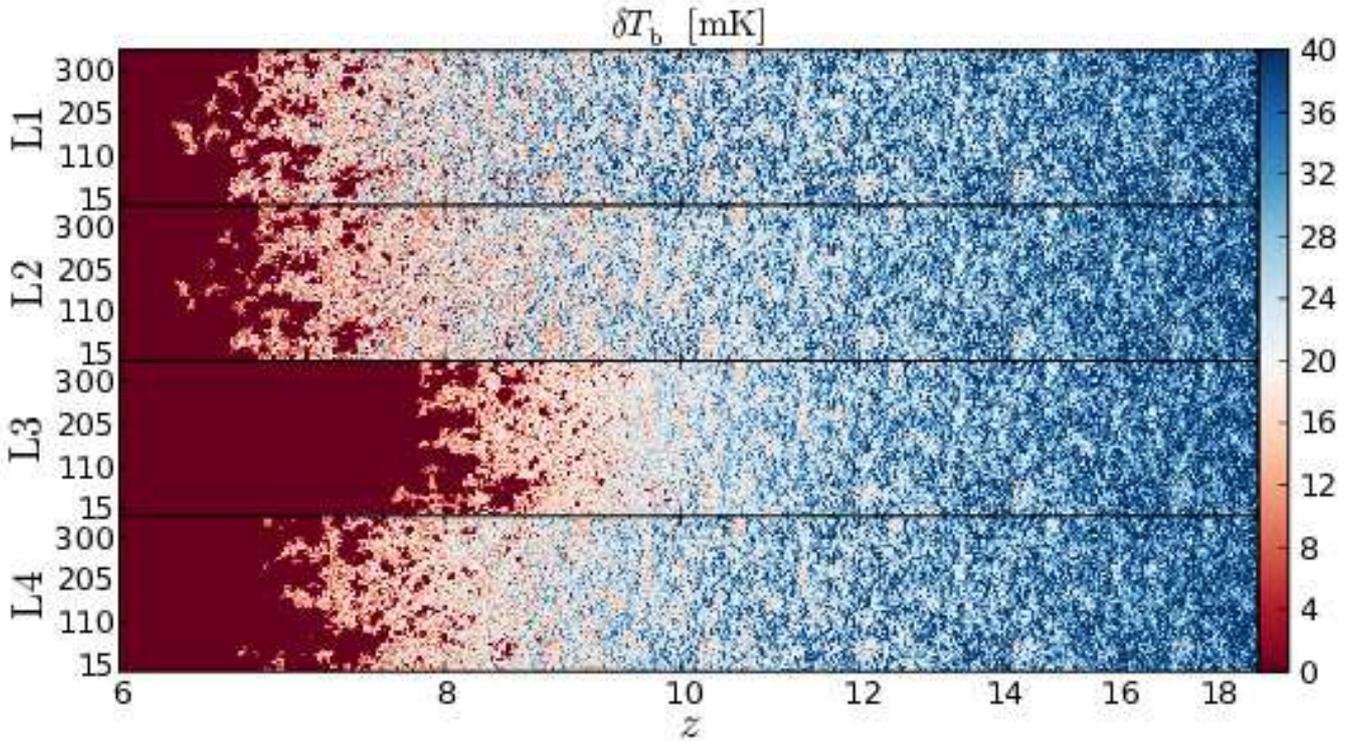}
\caption{
\label{fig:nu_box_raw}
Position-redshift slices from our 244 $h^{-1}$ Mpc boxes. These slices illustrate the large-scale geometry of reionization and the significant local variations in reionization history as seen in the redshifted 21-cm line. From top to bottom, the images show the differential brightness temperature (colour scale at right) at the full grid resolution in linear scale for LB1, LB2, LB3, and LB4. The spatial (vertical) scale is comoving Mpc.}
\vspace{-0.5cm}
\end{figure*}

In Fig.~\ref{fig:nu_box_raw}, we show the position-frequency slices cut through the simulated image cube. The vertical scale is the spatial dimension, and the horizontal is the observed frequency. Images are of the differential brightness temperature (colour scale at right) for simulations LB1, LB2, LB3, and LB4 from top to bottom, continuously interpolated in frequency and including the redshift-space distortions. At low frequency (high redshift), all \ion{H}{ii} regions are small and mostly isolated, though the exact redshift where this is no longer true depends on the reionization history and, therefore, on the source model. As these bubbles begin to merge, larger structures ($\sim\!10$~Mpc) begin to form, culminating in hundreds of such bubbles that all merge together towards the end of reionization. The intervening period, when the 21-cm fluctuations peak, varies significantly in duration between models. This period is most extended in model LB2, due to the combination of an early start and aggressive LMACH suppression exclusive to that model. Conversely, LB1 and LB3, where all sources are always active, have reionization proceed relatively fast. Finally, LB4 is intermediate between these two extremes. In that case, reionization starts early, but the lowest-mass sources, which initially dominate the photon budget, are quickly suppressed and form only small ionized patches. Only when the larger sources become more common do the \ion{H}{ii} regions become larger. 

\begin{figure*}
\includegraphics[height=4in]{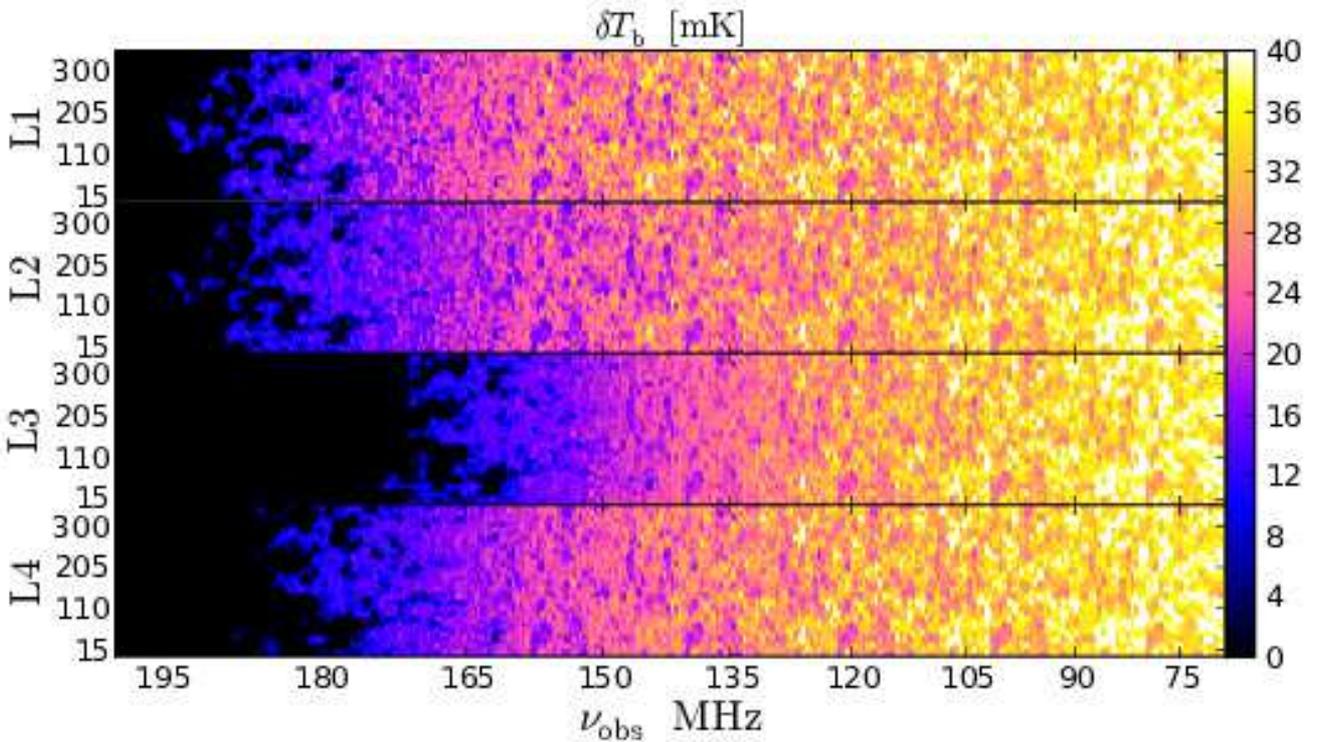}\caption{
\label{fig:nu_box_smooth}
Position-frequency slices from our 244 $h^{-1}$ Mpc boxes. These slices illustrate the large-scale geometry of reionization and the significant local variations in reionization history as seen at redshifted 21-cm line with a realistic 3~arcmin (Gaussian FWHM) beam size and 0.44~MHz (top-hat) bandwidth filter. From top to bottom, the images show the smoothed differential brightness temperature in linear scale for LB1, LB2, LB3, and LB4. The spatial (vertical) scale is comoving Mpc.} 
\vspace{-0.5cm}
\end{figure*}

Fig.~\ref{fig:nu_box_smooth} indicates how these fluctuations will be seen as a function of observed frequency, $\nu_{\rm obs}$, at resolution similar to the that of the first generation experiments, such as LOFAR. The same volume and simulations are shown, but the entire image cube is smoothed with a 3~arcmin Gaussian beam and a 0.44~MHz (top-hat) bandwidth filter. The early, small-scale structure is effectively erased given probable noise levels and foreground signals for the current experiments (to be presented in detail in a companion paper), but might become detectable in future, more sensitive experiments, such as SKA. However, the large-scale patches remain clearly visible even with this relatively large smoothing. The smoothing also somewhat diminishes the visual distinction between models, although the overall trends and features remain the same.

\subsubsection{Mean and rms}

\begin{figure*}
\includegraphics[width=3.2in]{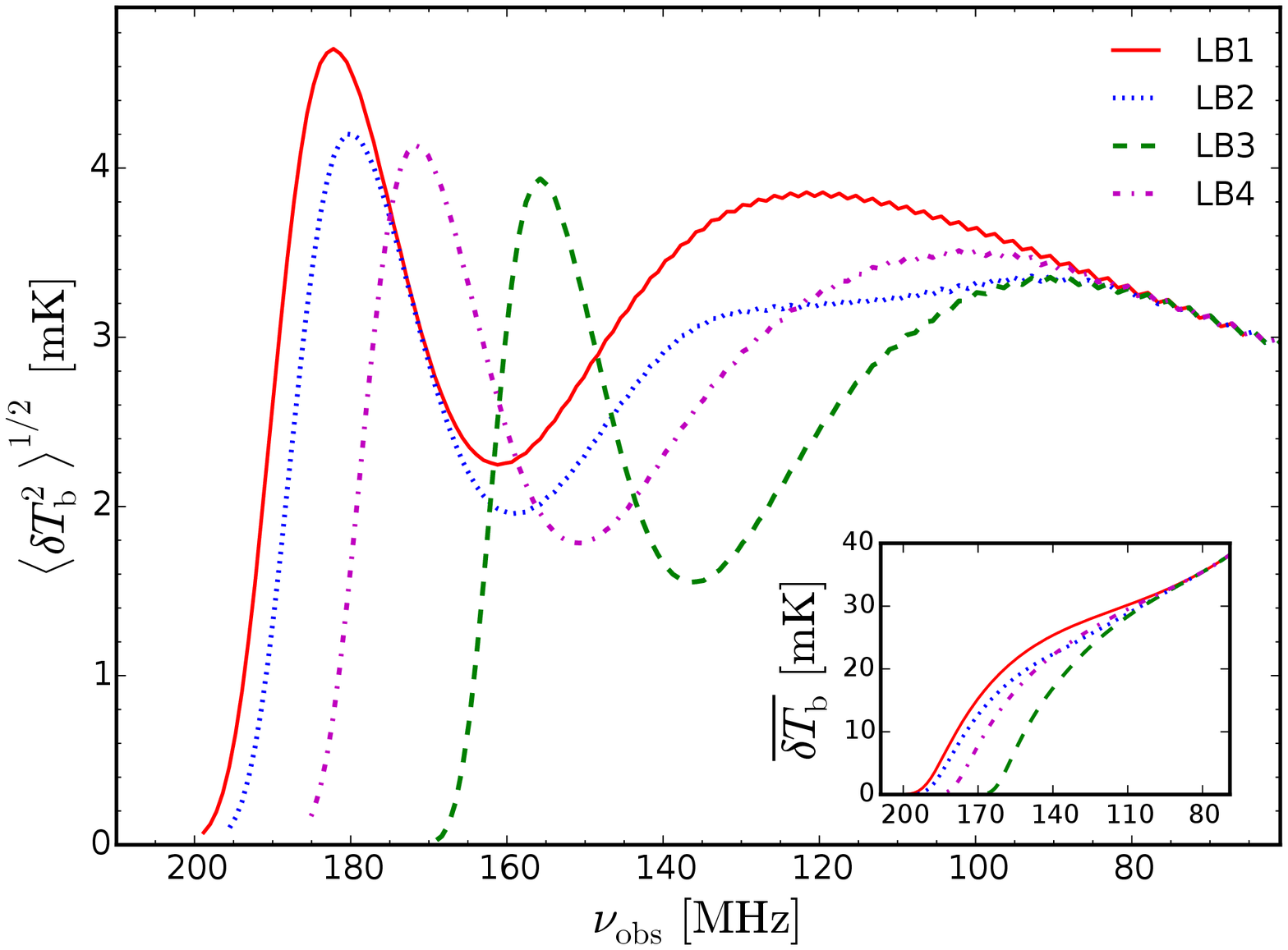}
\includegraphics[width=3.2in]{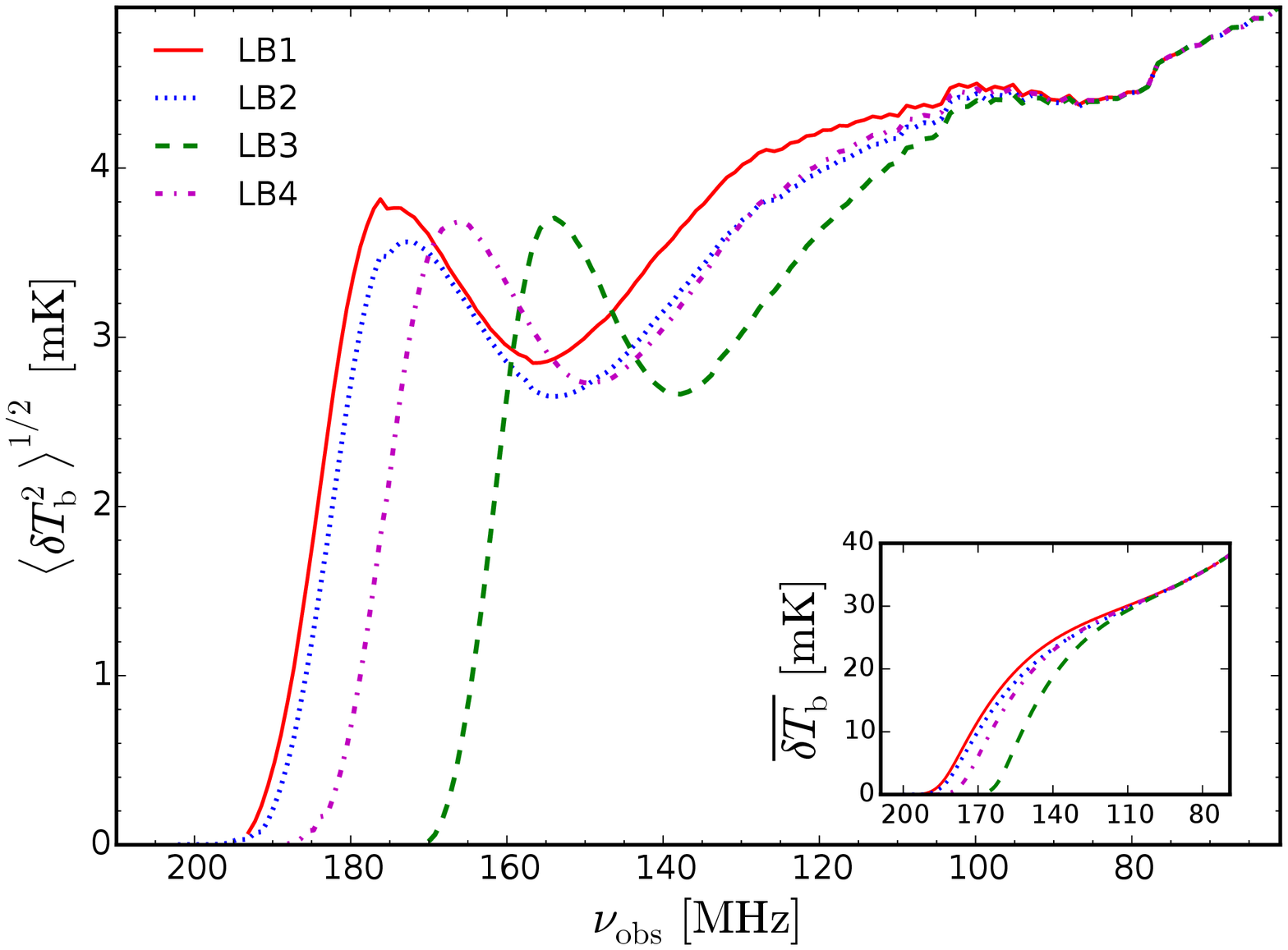}
\caption{The evolution of the rms and mean (inset) of the 21-cm background. \emph{Left:} for the $244\,h^{-1}$~Mpc box, four models are shown, LB1 (solid), LB2 (dotted), LB3 (dashed), and LB4 (dot-dashed). \emph{Right:} for the $47\,h^{-1}$~Mpc box, four models are shown, SB1(solid), SB2 (dotted), SB3 (dashed), and SB4 (dot-dashed).
\label{fig:mean_rms}}
\end{figure*}

The evolution of the mean differential brightness temperature, $\overline{\delta T_{\rm b}}$, as a function of observed frequency for all low-resolution cases is shown in Fig.~\ref{fig:mean_rms} (insets) with the $244\,h^{-1}$~Mpc ($47\,h^{-1}$~Mpc) on the left (right). In all cases, the evolution is gradual over time. With regards to experiments looking for rapid changes in the 21-cm signal as the Universe reionizes \citep{Shav99a,Bowm10a}, this behaviour means that all the suppression models are difficult to detect and likely impossible to distinguish at this aggregate level. The HMACH-only and fully suppressed model show a steeper drop in signal at late times than the gradual and partially suppressed models, but the effect is weak.

\begin{figure*}
\includegraphics[height=1.7in]{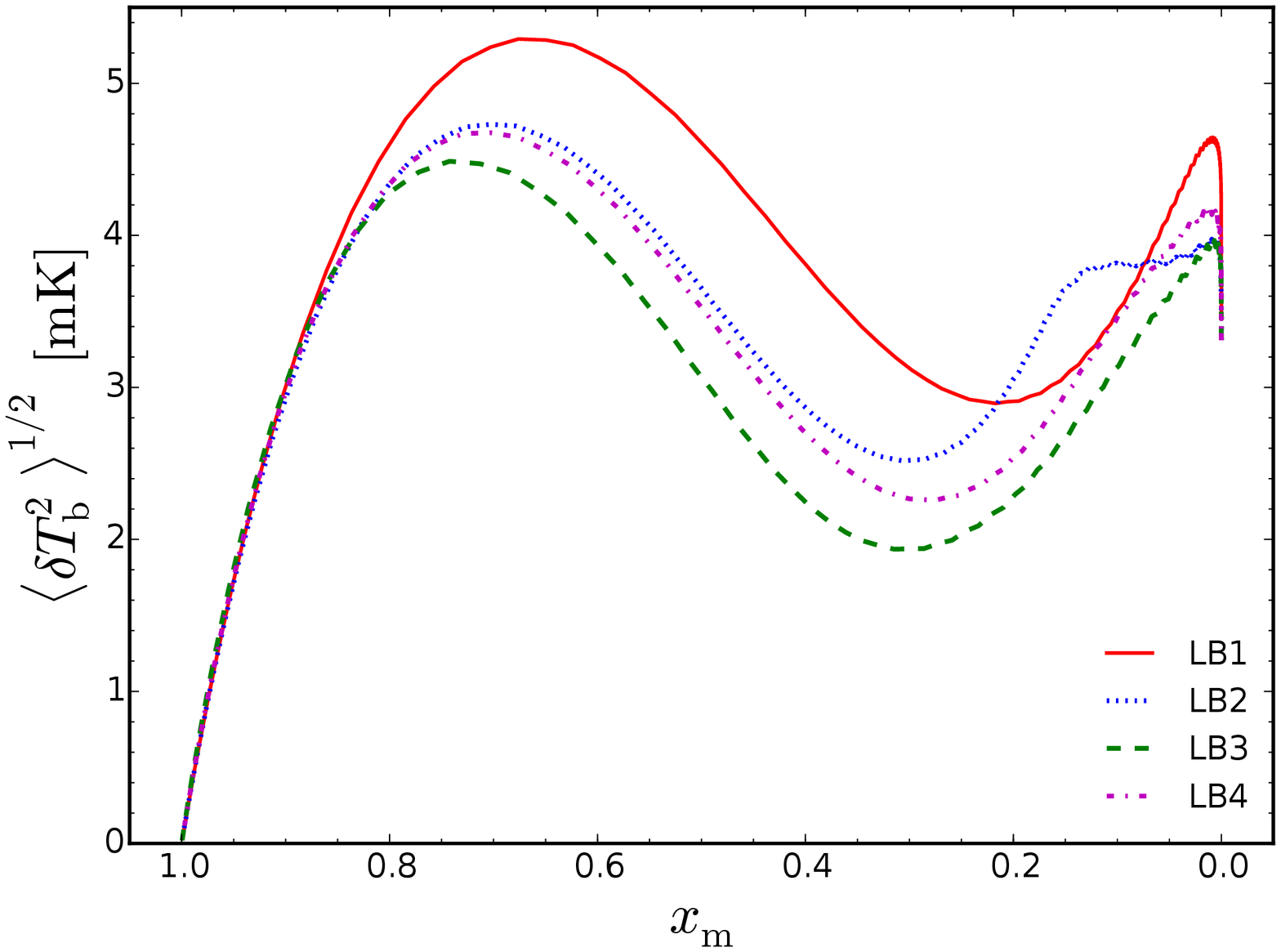}
\includegraphics[height=1.7in]{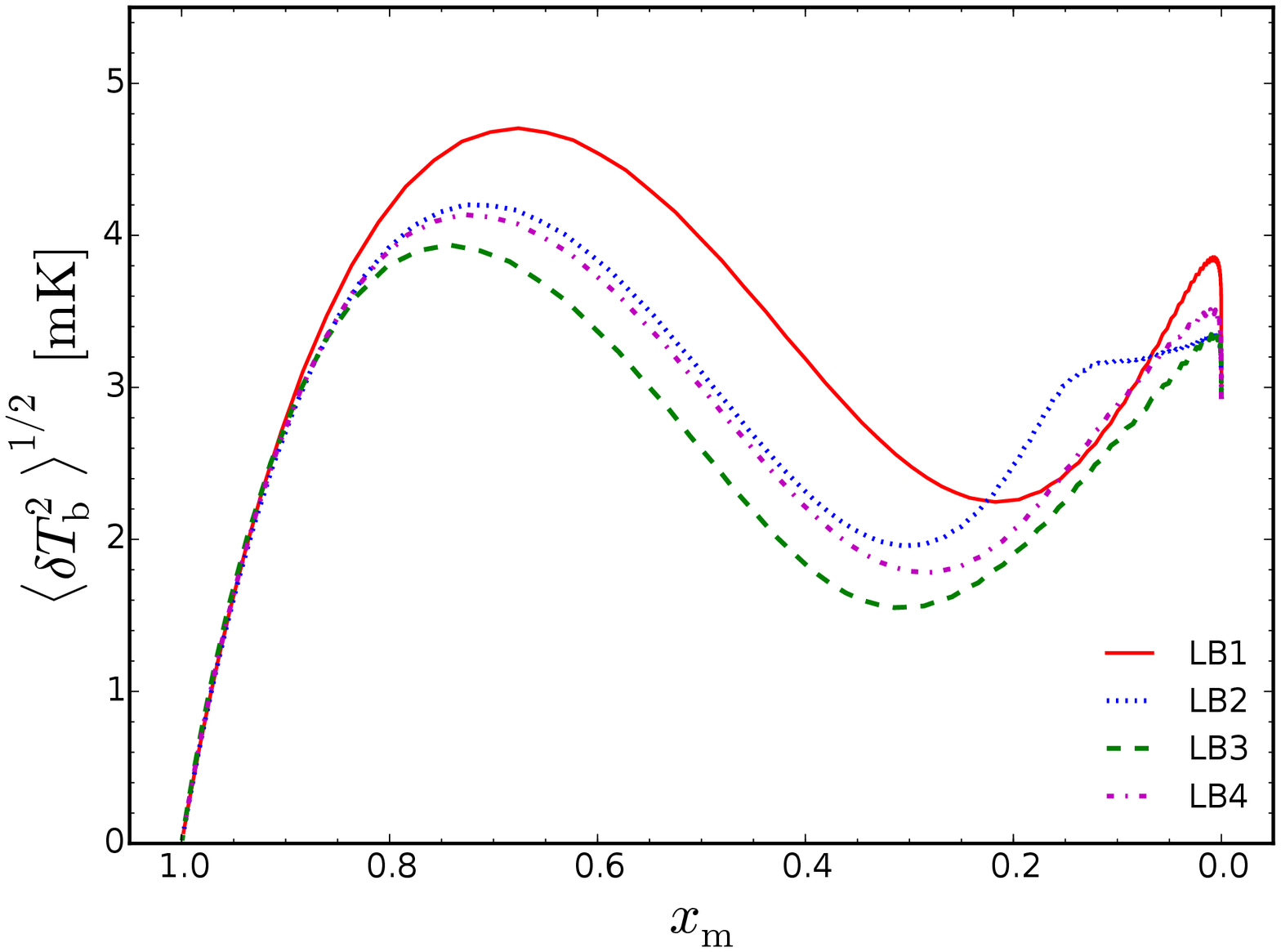}
\includegraphics[height=1.7in]{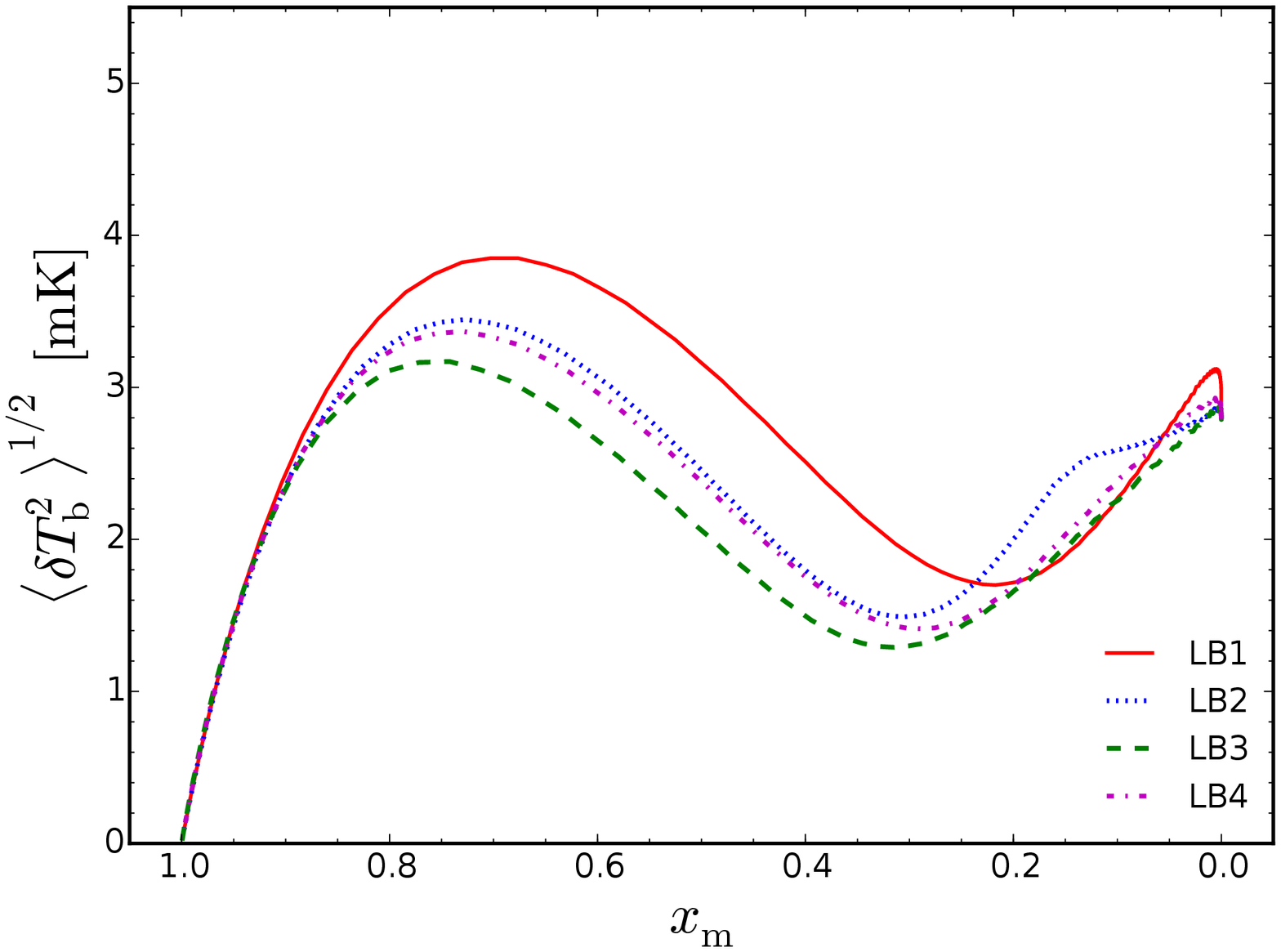}
\caption{The evolution of the rms fluctuations in the 21-cm background versus mean mass-weighted ionized fraction for different instrument realizations. The left, middle, and right panels are smoothed with a Gaussian beam of size 2, 3, and 5~arcmin and bandwidth 0.29, 0.44, and 0.73~MHz, respectively, for the 244$h^{-1}$~Mpc box and all source models LB1 (solid), LB2 (dotted), LB3 (dashed), and LB4 (dot-dashed).
\label{fig:dTrms_xi}}
\end{figure*}

Comparing the root-mean-square (rms) of the fluctuations in $\delta T_{\rm b}$ with respect to the mean ($\langle \delta T_b^2 \rangle$) averaged over a LOFAR-like beam and bandwidth (3~arcmin Gaussian and 0.44~MHz bandwidth filter) shown in Fig.~\ref{fig:mean_rms}, we see that the overall evolution follows similar paths in all cases with some variations. Since very little gas is ionized at early times, the 21-cm fluctuations track the underlying density fluctuations. First, consider the larger box in the left panel. As reionization progresses, the rms curves begin to diverge from being purely density-driven, with LB2 and LB3 diverging the earliest at $\nu_{\rm obs} > 85$~MHz. The higher efficiency of the LMACHs drive this behaviour by more rapidly ionizing the universe as compared to the other models. The mass-dependent suppression model (LB4) deviates from the density fluctuations later at $\nu_{\rm obs} > 95$~MHz, since most LMACHs  have very low efficiency after suppression. Of course, the rms for the model lacking LMACHs entirely (LB1) turns over the latest at $\nu_{\rm obs} \sim110$~MHz. Essentially, this feature gives no information about the effects of source suppression, just the mean source efficiency and the type of sources. As the highest density peaks are ionized and the the mean $\delta T_{\rm b}$ decreases, the rms curves dip, because the \ion{H}{ii} regions are still smaller than the smoothing scale and, therefore, do not contribute to the fluctuations. As reionization proceeds further, the size of the \ion{H}{ii} regions increases, eventually outgrowing the smoothing size and producing the peak in the signal. The position of this peak is largely dictated by the reionization history and the typical bubble size as compared to the beam size; hence, the fastest ionizing model (LB3), which makes large ionizing patches more quickly, peaks first at 155~MHz with LB4 following at 173~MHz. LB1 and LB2 share nearly identical sources at the end of reionization, because LMACHs are completely or nearly (respectively) suppressed. These models peak latest at $\nu_{\rm obs} \sim 185$~MHz. Larger fluctuations result from a later global reionization, benefitting from greater density fluctuations. The peak of LB2 is lower than that of LB1, because more reionization occurred earlier, front-loading the fluctuations. The peak height depends on the typical ionized bubble size at maximum fluctuations and how this size compares to the smoothing scale.

Although the rms shapes are similar across the simulations, the fully suppressed model (LB2) differs the most. For 110~MHz $< \nu_{\rm obs} < 135$~MHz, the signal is flatter. Specifically, the high-$\nu_{\rm obs}$ is significantly less distinct, aligning with the peak in LB3, and the signal does not begin dropping until $\sim\!140$~MHz. The aggressive suppression of LMACHs in this model limits the growth of ionized bubbles early on as compared to the other suppression models. 

Comparing these fluctuations versus $x_{\rm m}$ (Fig.~\ref{fig:dTrms_xi}) removes the dependence on the reionization timing from the comparison of the models (see the middle panel for the same smoothing as above). The peak position occurs later in the reionization process in models with more numerous and brighter LMACHs, with LB4 peaking the latest. The trough has similar behaviour, with LB1 bottoming out the earliest. The full and partial suppression models look very similar in the early universe, dominated by high-efficiency LMACHs, so the trough position is nearly identical for LB2 and LB3 around $x_{\rm m}\sim0.3$. Since LB4 has low-efficiency LMACHs, the trough appears between the high-efficiency LMACH models and the HMACH-only model. As above, the largest differences in shape occur during the early stages of reionization before $x_{\rm m} \sim 0.2$. The flattening of the signal in LB2 is more pronounced, with all the other models showing a steep initial drop in the magnitude of fluctuations. The RT grid resolution has essentially no effect in SB2 vs. SB2\_HR and LB1 vs. LB1\_HR and a minor effect in LB3 vs. LB3\_HR, where the peak and trough values change by up to 10 per cent, but their position in frequency remains the same (not shown). This consistency demonstrates the robustness of our results to changes in RT grid resolution.

In Fig.~\ref{fig:dTrms_xi}, we also compare various levels of smoothing with a 2, 3, and 5~arcmin Gaussian beam (left to right) and a corresponding bandwidth filter of 0.29, 0.44, and 0.73~MHz, respectively. The different levels of smoothing have only a mild effect on the 21-cm rms fluctuations. The overall shapes and relative levels for the different source models are robust to these toy models of smoothing. The larger beam size slightly decreases the peak fluctuations, from $\sim\!4.5-5.5$~mK for 2~arcmin to $\sim\!3.5-4.5$ for 5~arcmin beam. This larger smoothing also moves the peak to slightly later times (higher $x_{\rm m}$), since ionized patches grow over time and better match the larger beam and bandwidth sizes. 

\begin{figure*}
\includegraphics[height=1.7in]{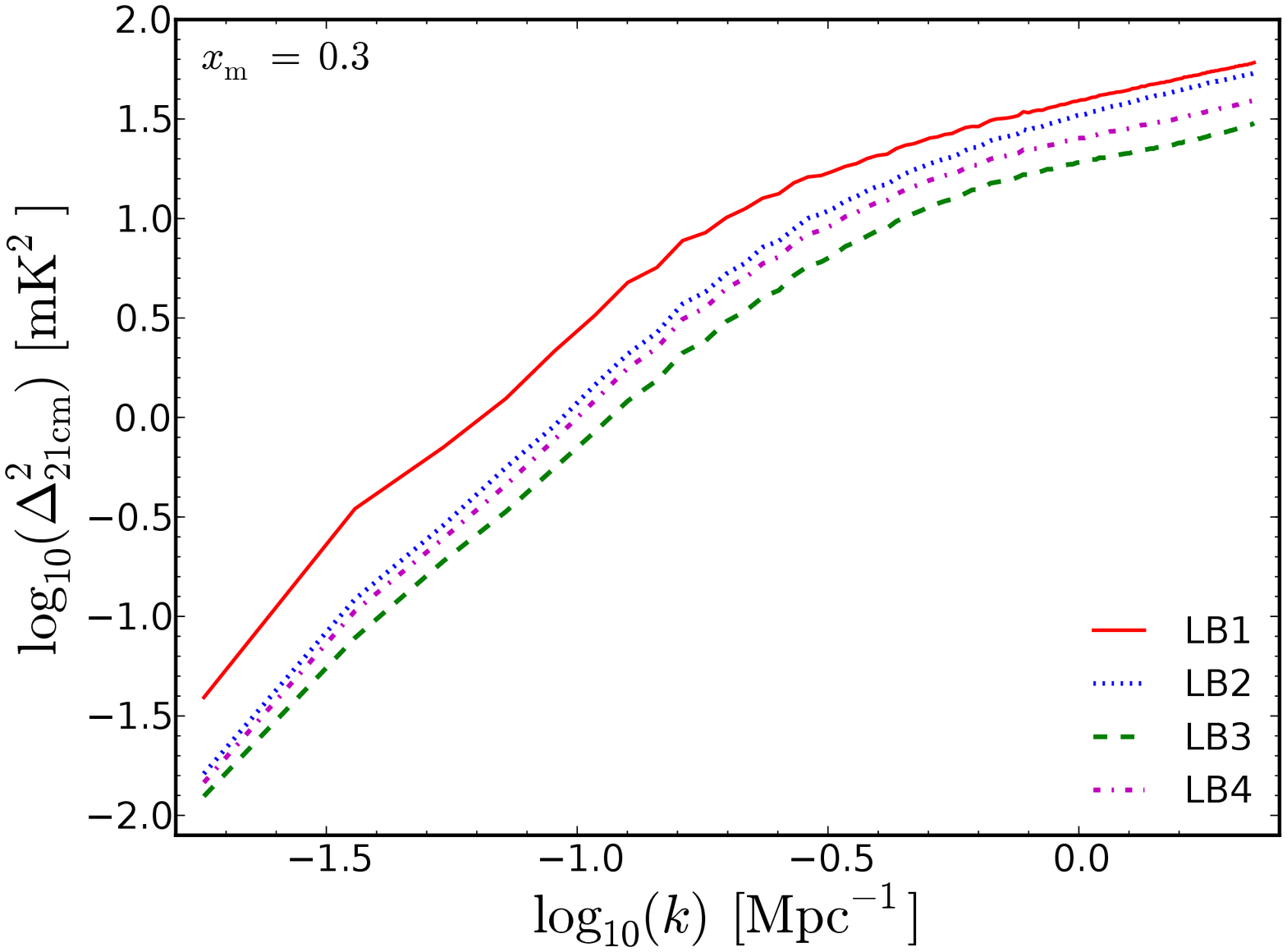}
\includegraphics[height=1.7in]{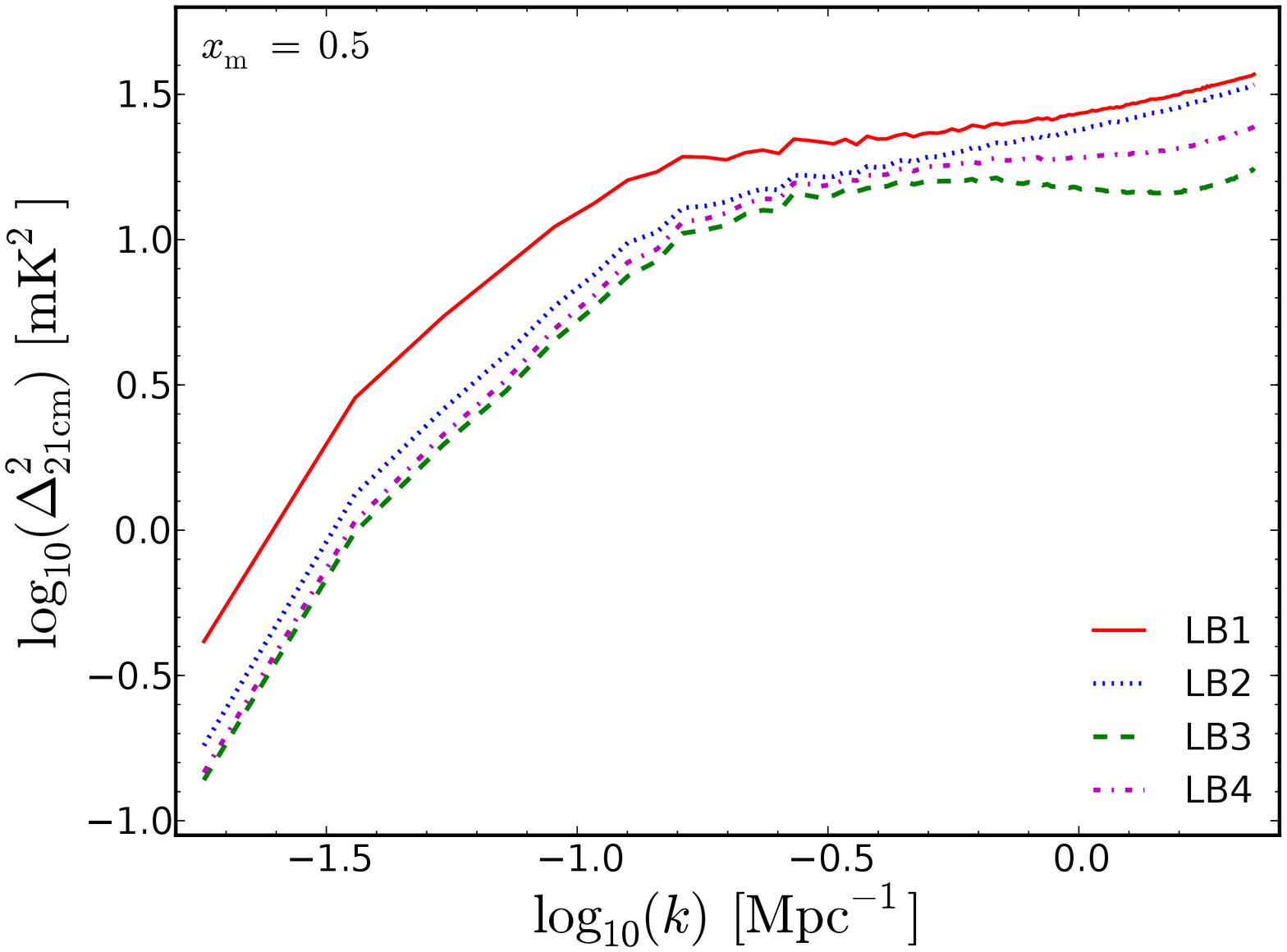}
\includegraphics[height=1.7in]{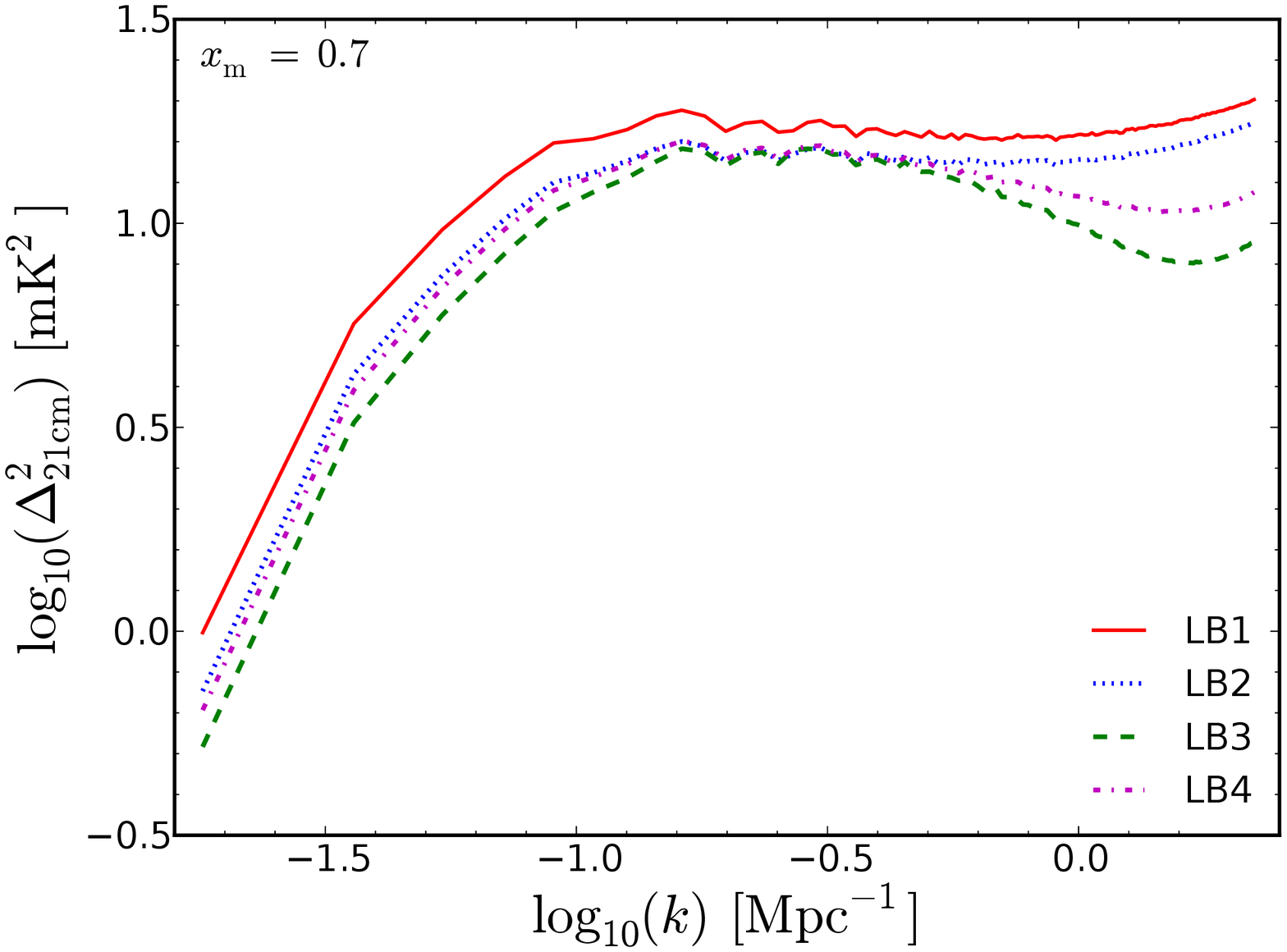}
\caption{The power spectra for the $244\,h^{-1}$~Mpc box and all source models at $x_{\rm m} = 0.3, 0.5,$ and 0.7 from left to right. The LB1 (solid), LB2 (dotted), LB3 (dashed), and LB4 (dot-chased) simulations are represented. These power spectra are calculated for coeval cubes, including redshift-space distortions.
\label{fig:244_ps}}
\end{figure*}

\subsubsection{Power spectra}

Another key statistical quantity is the autocorrelation power spectrum of the 21-cm differential brightness temperature fluctuations, referred to as the 21-cm power spectrum. The power spectrum, $P_{21}(\mathbf{k})$, is defined as:
\be
	\langle \widetilde{\delta T_{\rm b}^*}(\mathbf{k}) \widetilde{\delta T_{\rm b}}(\mathbf{k'}) \rangle \equiv (2 \upi)^3 P_{21}(\mathbf{k}) \delta_D^3(\mathbf{k}-\mathbf{k'}),
	\label{eq:ps_def}
\ee
where $\widetilde{\delta T_{\rm b}}$ is the Fourier transform of $\delta T_{\rm b}$ and $\delta_{\rm D}^3$ is the three-dimensional Dirac delta function. Throughout this work, we will use the (spatially) dimensionless power spectrum, $\Delta^2_{21}(k)$, where
\be
	\Delta^2_{21}(k) = \frac{k^3}{2 \upi^2} P_{21}(k)
	\label{eq:delta_21}
\ee
and has the units of mK$^2$. We follow the methodology in \cite{Mao12a} that includes redshift-space distortions.

In Fig.~\ref{fig:244_ps}, we compare $\Delta^2_{21}(k)$ for simulations LB1 (solid), LB2 (dotted), LB3 (dashed), and LB4 (dot-dashed) at $x_{\rm m} = 0.3$ (left), 0.5 (middle), and 0.7 (right). We do not discuss the power spectrum results from the smaller volume simulations, since these volumes are unable to represent the large-scale fluctuations that are important during reionization. At $x_{\rm m} = 0.3$ (left), the power is dominated by small scales, which is expected in the early stages of reionization when the ionized bubbles have yet to grow large and the density field is a significant contributor to the power spectrum. As reionization progresses (see $x_{\rm m} = 0.5$, middle), the power spectrum flattens with larger modes contributing significantly. These large scales come from the large ionized bubbles that are beginning to form with the density contributing mainly on small scales. At later times, the power spectrum weakens on small scales, as seen at $x_{\rm m} = 0.7$ (right panel), and LB3 and LB4 have more power on larger scales than the smallest. Here, $\Delta^2_{21}(k)$ is dominated by the ionization field contribution, and with more and more ionized regions overlapping, the small-scale structure diminishes.

Generically, the power is higher for models that reionize later, since the increased density fluctuations at later times increase the 21-cm fluctuations at a particular ionized fraction. Therefore, LB1 is the highest at all scales, and LB4 is the lowest, except when all LMACH models converge at the latest times (as seen on the right, $x_{\rm m}=0.7$). Since LB1 has only large sources, all ionized bubbles are larger on average, and LB1 has increased power on all scales, especially at large scales and at early times before significant overlap occurs. Distinguishing between the three suppression methods is easiest at late times and at small scales (high $k$). As seen in the left panel, a significant spread in the high-$k$ tail is present. More numerous sources, as in LB3 where LMACHs are never fully suppressed, create a more uniform ionization field, which suppresses small-scale power. Unfortunately, such small scales ($k>1\,h$Mpc$^{-1}$) are below the resolution of the current 21-cm experiments. At intermediate scales ($0.1<k<1\,h$Mpc$^{-1}$), all models yield largely identical power spectra, while at large scales differences are somewhat greater, but likely also difficult to detect. 

\subsubsection{PDFs and non-Gaussianity}

The 21-cm power spectra would fully characterise the emission field if the differential brightness distribution were purely Gaussian, which is manifestly not the case during reionzation \citep{Mell06b,Ilie08a,Hark09a,Watk14a}. The probability distribution functions (PDFs) of $\delta T_{\rm b}$ could be significantly non-Gaussian, particularly at the later stages of reionization \citep{Mell06b}, and their measured skewness can be used to discriminate between different reionization scenarios \citep{Hark09a}. The PDFs and their evolution could also be used to derive the reionization history of the IGM \citep{Ichi10a,Glus10a}.

\begin{figure*}
\begin{center} \vspace{+0.1in}  
\includegraphics[height=1.7in]{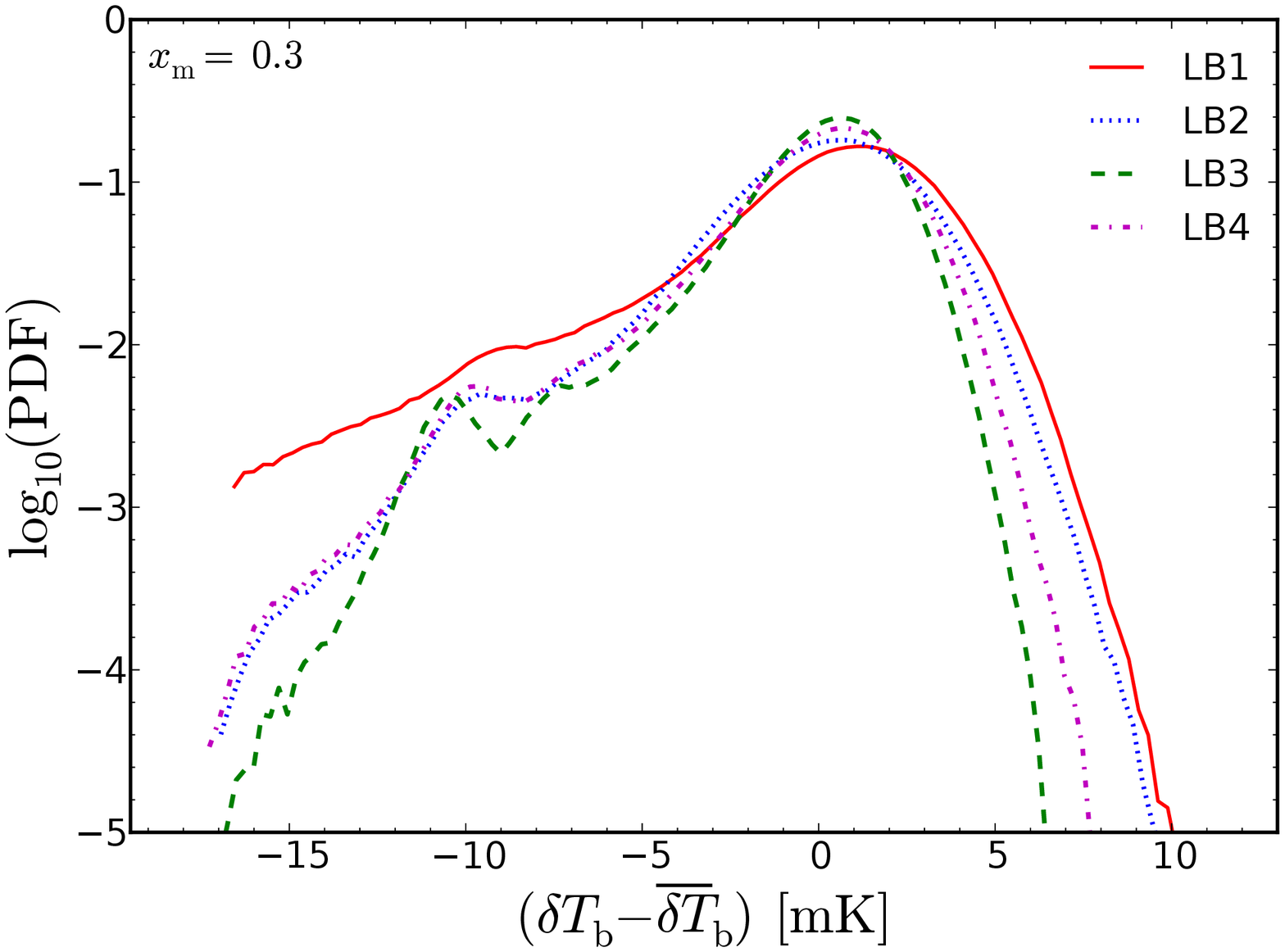}
\vspace{-0.2in} 
\includegraphics[height=1.7in]{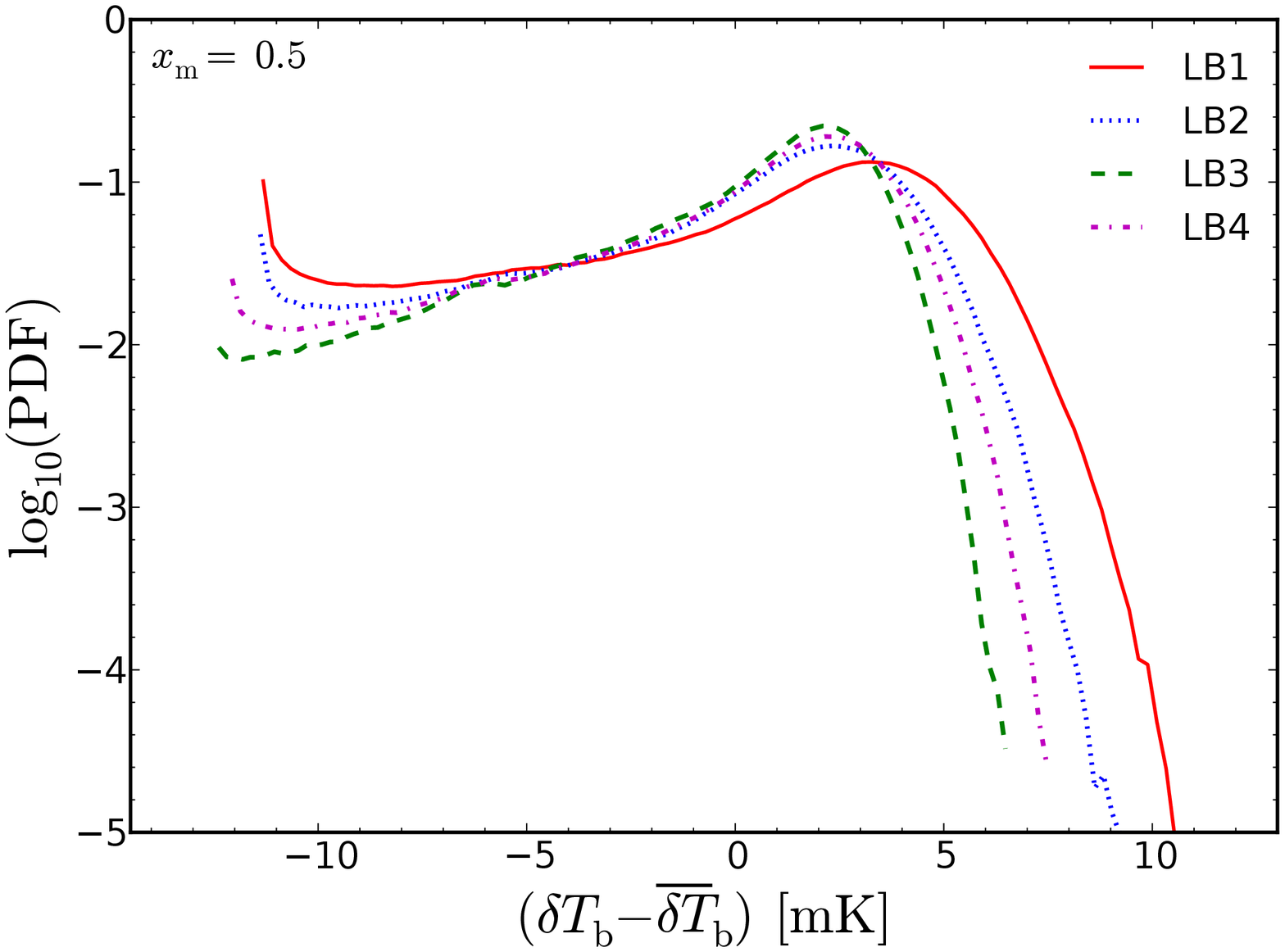}
\vspace{-0.2in} 
\includegraphics[height=1.7in]{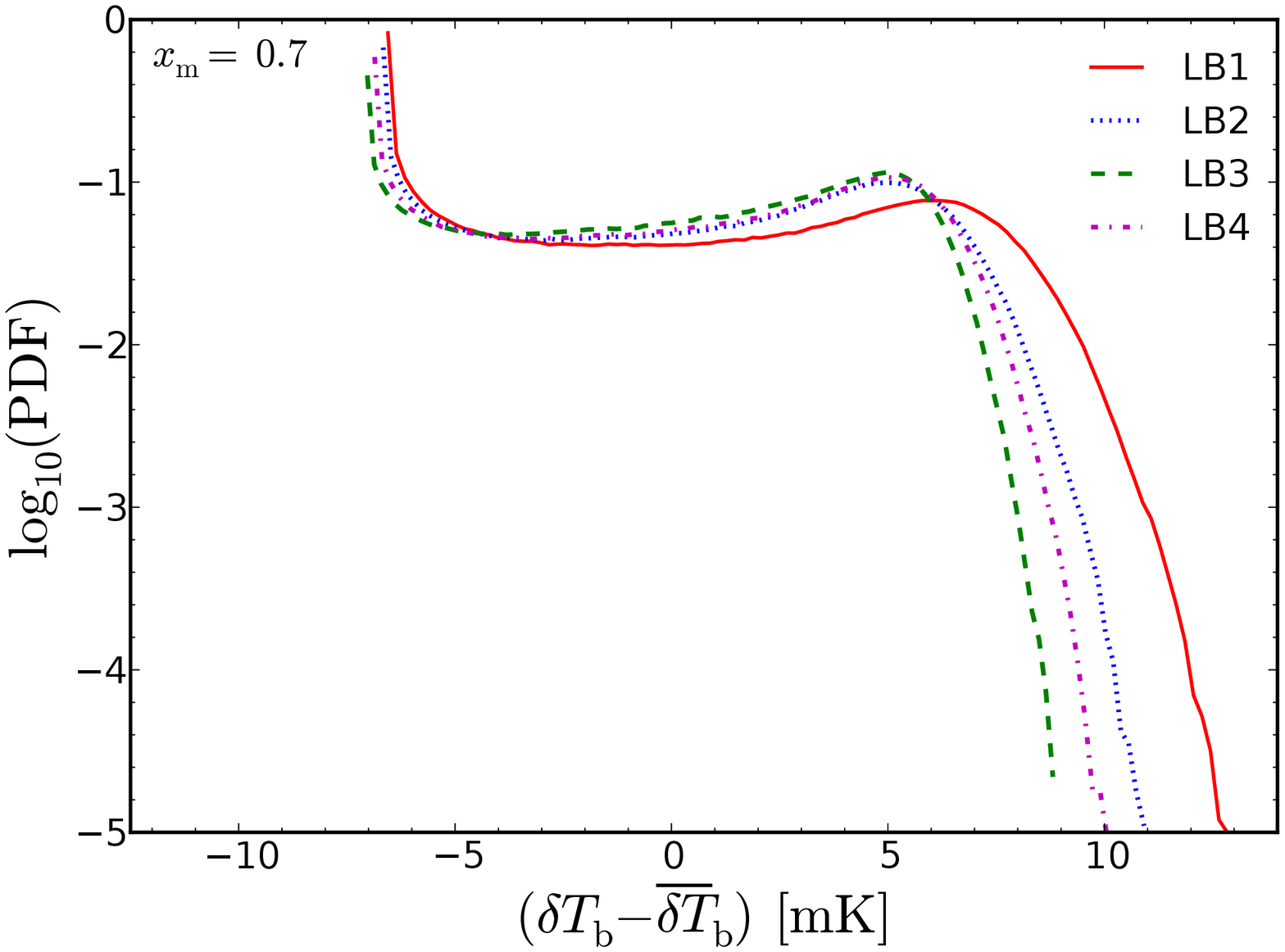} 
\vspace{+1cm}
\caption{
\label{fig:dT_PDF}
The probability distribution function for $\delta T_{\rm b}$ from our 244$\,h^{-1}$~Mpc simulations. Distributions are shown at different stages of the reionization process with ionized fraction by mass, $x_{\rm m} = 0.3, 0.5,$ and 0.7 from left to right. The LB1 (solid), LB2 (dotted), LB3 (dashed), and LB4 (dot-dashed) simulations are represented. In order to mimic the behaviour of an interferometer, we apply a Gaussian 3~arcmin beam size and 0.44~MHz (top-hat) bandwidth filter, and the mean signal has been subtracted. }
\end{center}
\end{figure*}

The 21-cm PDFs smoothed over a Gaussian 3~arcmin beam and 0.44~MHz (top-hat) bandwidth for all suppression models in our large box at three representative stages of reionization ($x_{\rm m}=0.3, 0.5,$ and 0.7, from left to right) are shown in Fig.~\ref{fig:dT_PDF}. Early on (see $x_{\rm m}=0.3$ on the left), the distributions are mostly following the underlying density field and are, therefore, the closest to Gaussian. Non-linear density evolution introduces non-Gaussianity that increases over time. Reionization itself introduces some non-Gaussianity at low $\delta T_{\rm b}$, as the first \ion{H}{ii} regions form around the highest density peaks, and moves the corresponding cells into the extreme left of the distributions. This effect slightly skews the distribution towards below-average (i.e., negative in $\delta T_{\rm b}-\overline{\delta T_{\rm b}}$) temperature values, since the low-density regions remain more neutral on average. The HMACH-only model (LB1) produces the largest skew and distribution width, because the high-mass sources reside in the densest regions that are strongly clustered as a consequence of the Gaussian density field statistics \citep[see, e.g., Figure~4 in][]{Ilie14a}. LB3 has the narrowest distribution, because the smallest sources, which are less biased and more common, are never fully suppressed. The remaining two models lie in between with full suppression, LB2, producing a wider distribution than the gradual suppression, LB4. 

As reionization progresses (see $x_{\rm m}=0.5$ in the middle), the hierarchy of the PDF width among the models remain, but stronger non-Gaussianity develops. The significant negative tail in the PDF is due to the ionized regions ($\delta T_{\rm b}-\overline{\delta T_{\rm b}}<0$). The remaining neutral regions are a mostly voids that have low, but positive $\delta T_{\rm b}-\overline{\delta T_{\rm b}}$. During the latest stages of reionization (see $x_{\rm m}=0.7$ on the right), the four simulations are most similar to each other, as most regions are fully ionized. As before, LB1, with the rarest sources and latest $z_{\rm reion}$, has the most high-$\delta T_{\rm b}$ cells and LB3 least, with the most uniform sources and earliest $z_{\rm reion}$. This effect is due to the uniformity of sources, as more uniform source make more uniform ionization fields, and partially due to the increased density fluctuations at later times, i.e., the density fluctuations are larger for L1 than L3 at the same ionized fraction. Therefore, the statistics of bright peaks, particularly at late times, provide a promising way to discriminate between the different suppression scenarios and to learn about the nature of the ionizing sources. For example, at $x_{\rm m}=0.7$, there are no 10~mK peaks in LB3, very few in LB4, and orders of magnitude more in LB2 and LB1. Although not shown here, the RT grid resolution again has essentially no effect in SB2 vs. SB2\_HR and LB1 vs. LB1\_HR and only a minor effect in LB3 vs. LB3\_HR. In the latter case, the PDF distribution exhibits more low-$\delta T_b$ pixels at early times and more high-$\delta T_b$ pixels at late times for the high-resolution grid.

\begin{figure*}
\includegraphics[width=3.2in]{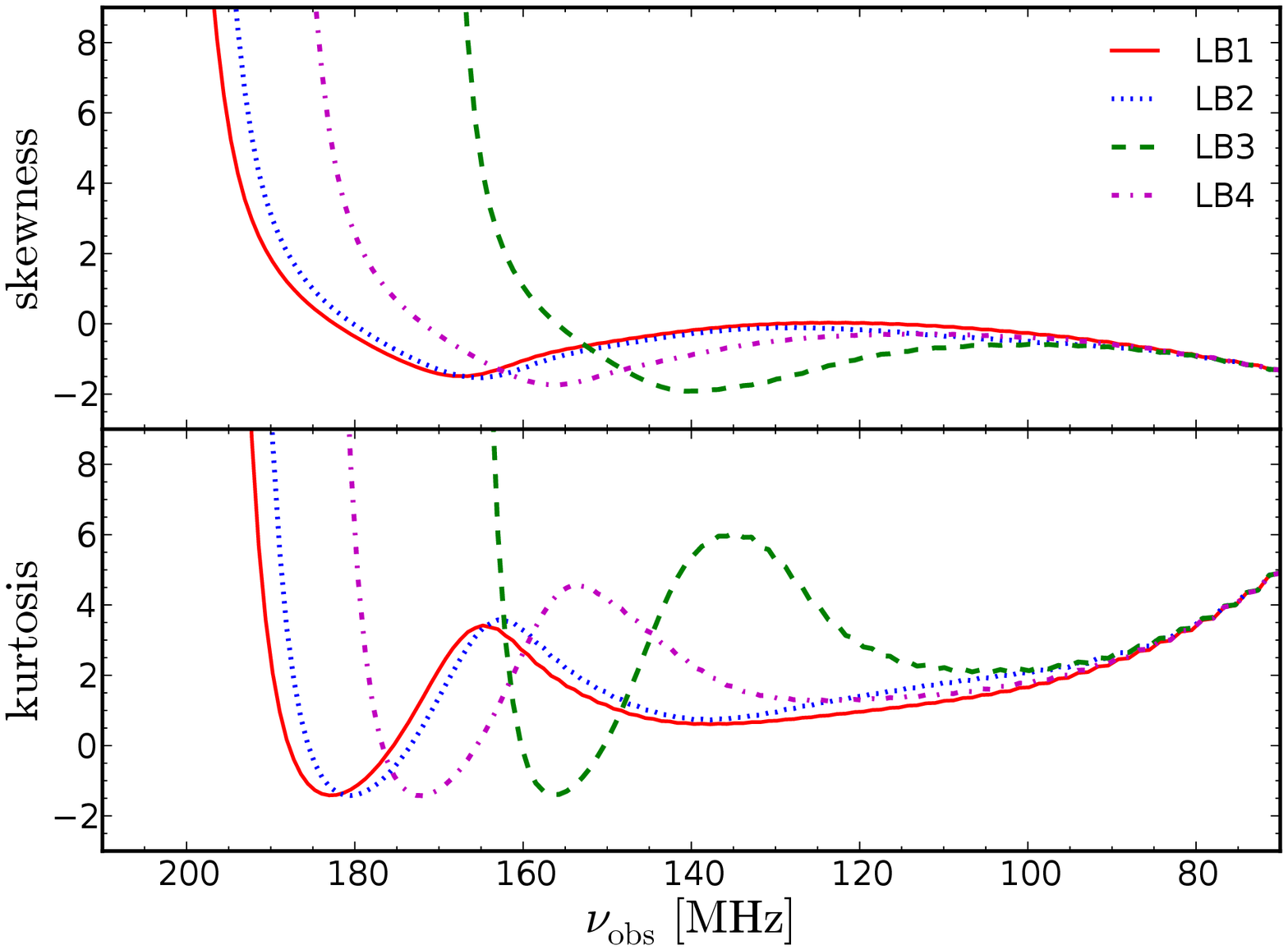}
\includegraphics[width=3.2in]{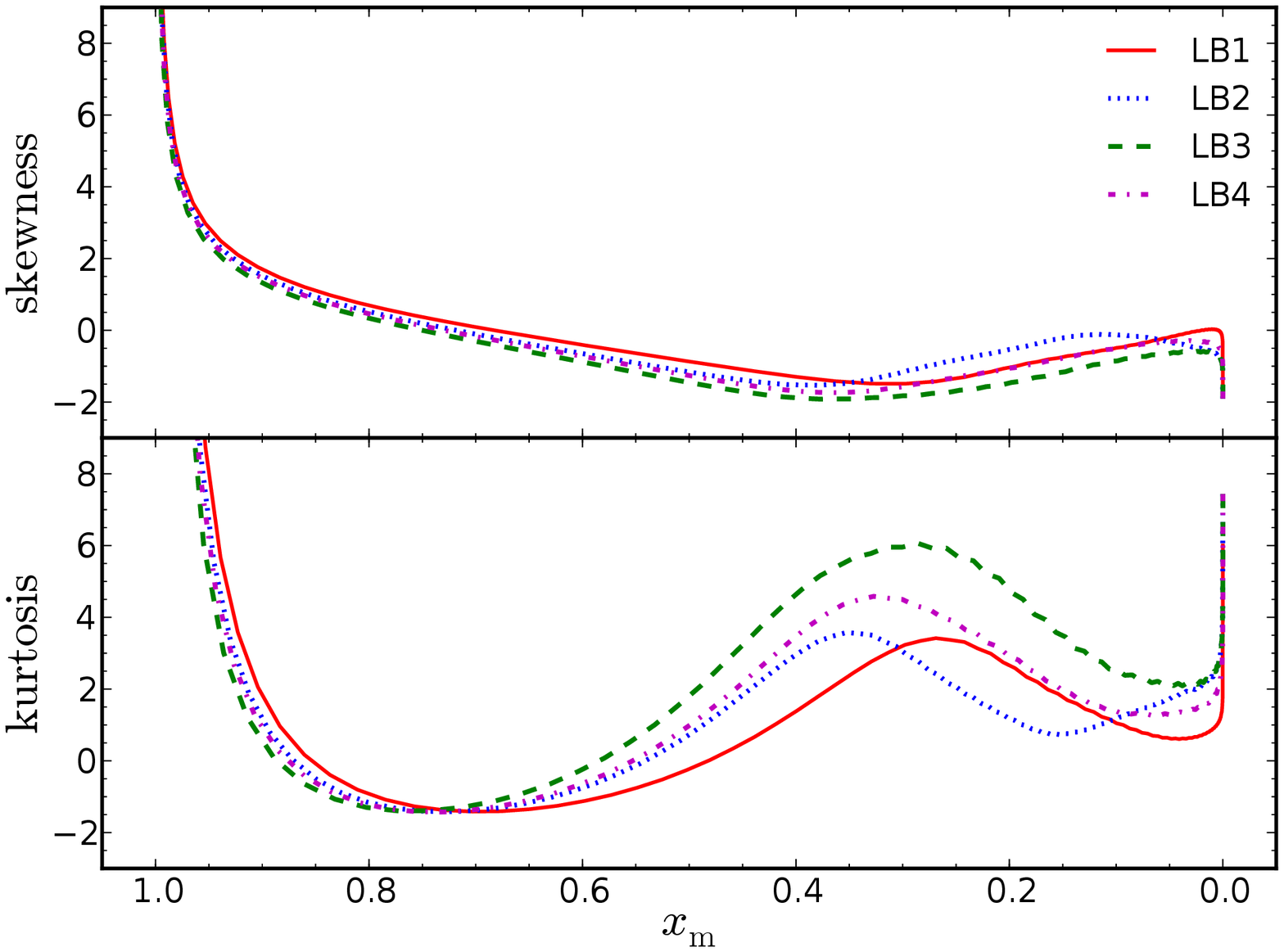}
\caption{The evolution of the skewness (top) and kurtosis (bottom) in the 21-cm PDFs for the 244\,$h^{-1}$~Mpc box and all source models. The models LB1 (solid), LB2 (dotted), LB3 (dashed), and LB4 (dot-dashed) are smoothed with a 3~arcmin (Gaussian FWHM) and bandwidth 0.44~MHz. The left (right) panel shows the evolution as a function of frequency (ionized fraction).
\label{fig:dTskewkurt}}
\end{figure*}

\begin{figure}
\includegraphics[width=3.2in]{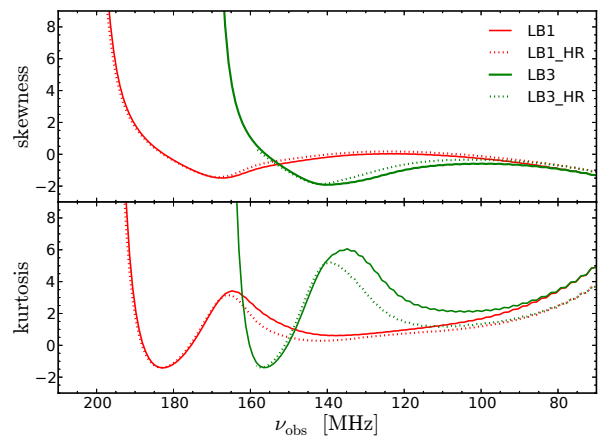}
\caption{The evolution of the skewness (top) and kurtosis (bottom) in the 21-cm PDFs for the 244\,$h^{-1}$~Mpc box for the high-resolution RT grid (dashed) vs. corresponding low-resolution models (solid) with a Gaussian beam size 3~arcmin and bandwidth 0.44~MHz. Here are two suppression models with HMACHs only (left trough, LB1) and partial suppression of LMACHs (right trough, LB2).
\label{fig:dTskewkurt_res}}
\end{figure}

The level of non-Gaussianity of the PDFs can be quantified to first and second order by the skewness and kurtosis, respectively. Fig.~\ref{fig:dTskewkurt} shows the evolution of the skewness (upper panel) and kurtosis (lower panel) vs. observed frequency (left) for simulations LB1 (solid), LB2 (dotted), LB3 (dashed), and LB4 (dot-dashed). As expected from the above discussion, the skewness is very large at high frequencies, or late times, and is very low and slightly negative at low frequencies, or early times. The major feature of the skewness is the dip at intermediate frequencies. The position of this feature is mainly determined by the timing of reionization, where the dip occurs earlier for faster reionization scenarios. When plotted against the ionization fraction all scenarios produce almost identical evolution (Fig.~\ref{fig:dTskewkurt}, right top panel), with the dip occurring at $x_{\rm m} \approx 0.35$. The depth of this trough is weakly dependent on the distribution of the ionizing sources, where the most uniform sources model (LB3) has deepest trough. Similarly, LB1 and LB2 are roughly the same at the frequency of this feature, because there are only HMACHs and many fully suppressed LMACHs, respectively. The variations between the models are very minor, however. Skewness also proves insensitive to the RT grid resolution, which can be seen in the upper panel of Fig.~\ref{fig:dTskewkurt_res}. The two large-volume simulations with high resolution available are shown, with L1 (solid), L1\_HR, (dotted), L3 (solid), and L3\_HR (dotted), where the HMACH-only models have a higher-frequency trough as compared to the partially suppressed LMACH models. The trough is narrower and shifted slightly to higher frequency for the high-resolution simulations, though the effect is minimal.

The kurtosis differentiates the models more than the skewness. We will focus on the two main features: the peak, which always occurs first and roughly coincides in frequency with the trough of the skewness, and the trough, which happens later and occurs at approximately the same frequency as the peak in the rms fluctuations. Accordingly, the frequency position of the peak (trough) depends mostly on the timing of reionization with earlier reionization producing a lower-frequency peak (trough). Also, the timing of reionization affects the size of density fluctuations at a given frequency. This effect plus the general uniformity of the sources determines the height of the peak, where LB3 (earliest and most uniform) is the highest and LB1 (latest and least uniform) is the lowest. For the trough, the kurtosis turns slightly negative. All models reach the same approximate depth at different frequencies, dependent on the speed of reionization, but at similar ionized fraction, $x_{\rm m}\sim0.7-0.8$. The RT grid resolution has no significant effect on the kurtosis evolution for case SB2 vs. SB2\_HR, but for the larger boxes the higher grid resolution shifts the peak to higher frequency (by few MHz), as shown in the lower panel of Fig.~\ref{fig:dTskewkurt_res}. Interestingly, this frequency shift also aligns the kurtosis peak and skewness trough more closely in frequency. This correlation suggests that measuring both quantities at the same time might serve as a check and validation of the measurements. These higher order statistics show the greatest promise for differentiating between possible models of suppression.

\section{Discussion}
\label{sec:summary}

We presented a suite of full radiative transfer simulations of the epoch of reionization designed to investigate the observational signatures of star-formation suppression in dwarf galaxies due to radiative feedback (or, possibly, mechanical feedback). We considered four different, physically motivated suppression models, all with a large and (comparatively) small box size and with two RT grid resolutions. We investigated mainly the large-scale effects of reionization and addressed the robustness of our results to numerical effects, namely simulation volume and grid resolution. Specifically, we sought to discover which observational signatures are most sensitive to the method of radiative feedback and which are the most robust. We primarily focused on the redshifted 21-cm signatures that can probe the full reionization history and the detailed morphology thereof. The 21-cm signals can provide a wealth of information, including the mean history, rms evolution, power spectra, PDFs, imaging, and higher-order statistics. 

We find that the morphology of reionization and the overall shape of observational features addressed here are generally insensitive to source suppression model, box size, and resolution. The exact timing of reionization varies among the suppression models, where the degree of survival for small sources determines how quickly reionization progresses. Despite these differences, the evolution of the 21-cm signal is very similar. The same is true for the evolutions of the fluctuations in the signal, where the characteristic rms peak and trough locations depend on the reionization history.

A closer examination does reveal important differences, however, especially in the higher-order statistics. The small-scale power spectrum is affected by the typical mass of the dominant ionizing sources. More aggressive suppression of LMACHs (low-mass sources) effectively removes them and, therefore, cause differences in the power spectrum at small scales between the suppression models, especially at late times. Similar differences can be seen in the PDFs of the differential brightness temperature. Here, larger (and more rare) sources cause more high-$\delta T_{\rm b}$ regions to form. Therefore, the HMACH-only model (LB1) will always have a brighter tail, and the least-suppressed LMACH model (LB3) will have the narrowest distribution. The kurtosis of the 21-cm PDF distribution shows significant variation between the models, while the skewness is quite insensitive and has a largely universal shape. Using the fact that the trough in the kurtosis approximately corresponds to peak in the rms and the peak roughly coincides with the trough in the skewness can be a useful check on the measurements of these quantities, and these features contain information on the characteristic size of ionized regions, the beam size, and timing of reionization.

As expected, the smaller box size misses some of the large-scale structure and rare, bright sources found in the larger box size, which creates differences between the two volumes. Since we are mainly focussing on observables and large-scale structures, we mainly presented results for the larger volume. By definition, higher resolution simulations capture smaller structure better, as can be readily seen in the simulation slices in Fig.~\ref{fig:images_res}. Although the overall shape of the observables remain robust to RT resolution, we found that the models of suppression were sensitive to the resolution, leading to small-scale differences that are mainly present during the intermediate stages of reionization. During early times, there are fewer (or dimmer, depending on the source model) sources in high-resolution cases, because a source can more easily ionize its own smaller cell and introduce suppression of ionizing radiation. Towards the end of reionization, all models are dominated by large halos that are not affected by suppression, so the resolution matters less at these late times.

This study does not cover every possible influence that a source model might have on a signature of the EoR. We are only considering the possible signatures of nature of LMACH suppression, with all other parameters (e.g., source efficiencies and thresholds for suppression) being equal. In previous studies, we considered varying the ionizing efficiency of sources \citep{Mell06a,Ilie07a,Ilie08a}, the typical mass and the very nature of the active sources \citep{Ilie12a,Ahn12a,Grif13a}, and how large a simulation volume is sufficient to derive the various quantities \citep{Ilie14a}. Ongoing observations that constrain galaxy formation and small-scale, high-resolution hydrodynamical simulations of galaxy evolution will help to further limit the available freedom of these other quantities.

These simulations are also necessarily a simplified version of the early Universe, since all relevant scales cannot be simulated simultaneously. In particular, we do not consider the small-scale, unresolved IGM structure, which includes gas clumping, and only include a simplified model for the Lyman-limit systems. These structures should have the greatest impact at latest times, particularly in suppressing the large-scale fluctuations \citep[][and Mao et al., 2015, in prep.]{Soba14a,Shuk15a}. In addition to self-shielding regions in the IGM that may remain neutral, galaxies may also hold dense regions that act as photon sinks. These effects reduce the overall contrast between ionized and neutral regions, suppressing fluctuations generally and preventing the late-time rise in the skewness \citep{Watk15a}. Throughout this work we also assume that $T_{\rm S} \gg T_{\rm CMB}$ and ignore any early heating from X-ray sources. These effects should be important early on in reionization \citep[e.g.][]{Venk01a,Furl06a,Seme07a,Prit07a,Sant10a,Mesi13a}, and we plan investigate these effects in a future paper. Due to the resolution of our $N$-body simulations, we do not consider sources below $10^8~\msun$, which should be important during the early stages of reionization \citep{Ahn12a} and may contribute a significant fraction of the total ionizing photons \citep{Wise14a,Paar15a}.

Although the current generation of radio interferometer experiments may not be able to detect the differences resulting from the various suppression methods presented here \citep[see e.g.][]{Pati14a}, we present which 21-cm signal features are robust to this physical uncertainty. Any comparisons to observations must take into account the details of the instrument, which is beyond the scope of this paper. Future experiments may indeed have the sensitivity to distinguish between our models presented here. None the less, this suite of simulations will aid in the interpretation of any upcoming 21-cm measurements.

\section{Acknowledgements}
We thank Hannes Jensen for sharing \textsc{\small c2raytools} and Yi Mao for providing us with his power spectrum code. We acknowledge PRACE for awarding us computational time under PRACE4LOFAR grants 2012061089 and 2014102339 and access to resource Curie based in France at CEA and to resource SuperMUC at LRZ. This work was supported by the Science and Technology Facilities Council [grant numbers ST/F002858/1 and ST/I000976/1] and the Southeast Physics Network (SEPNet). GM was supported in part by Swedish Research Council grant 60336701. KA was supported by NRF-2012K1A3A7A03049606 and NRF-2014R1A1A2059811. P.R.S. was supported in part by U.S. NSF grant AST-1009799, NASA grant NNX11AE09G, NASA/JPL grant RSA Nos. 1492788 and 1515294, and supercomputer resources from NSF XSEDE grant TG-AST090005 and the Texas Advanced Computing Center (TACC) at the University of Texas at Austin. Some of the numerical computations were done on the Apollo cluster at The University of Sussex.

\bibliographystyle{mnras} 
\bibliography{sources_ms.bbl}

\end{document}